\documentclass[letterpaper,twocolumn,10pt]{article}
\usepackage{paper}

\usepackage{tikz}
\usepackage{amssymb}
\usepackage{amsmath}
\usepackage{multicol}
\usepackage{caption}
\usepackage{subcaption}
\usepackage{tabulary}
\usepackage{multirow}
\usepackage{xcolor}
\usepackage{tikz}
\usepackage{enumitem}
\usepackage{graphicx}
\usepackage{longtable}

\newcommand{\noteng}[1]{\textcolor{red}{NG: #1}}
\newcommand{\notegb}[1]{\textcolor{blue}{GB: #1}}

\newcommand{\new}[1]{\textcolor{brown}{#1}}

\begin{document}

\date{}

\title{A Comparative Audit of Privacy Policies from \\Healthcare Organizations in USA, UK and India}

\def\plainauthor{Author name(s) for PDF metadata. Don't forget to anonymize for submission!}

\author{
{\rm Gunjan Balde}\\
 IIT Kharagpur
 \and
 {\rm Aryendra Singh}\\
  IIT Kharagpur
   \and
 {\rm Niloy Ganguly}\\
  IIT Kharagpur
   \and
 {\rm Mainack Mondal}\\
  IIT Kharagpur
} %

\maketitle

\begin{abstract}
Data privacy in healthcare is of paramount importance (and thus regulated using laws like HIPAA) due to the highly sensitive nature of patient data. To that end, healthcare organizations mention how they collect/process/store/share this data (i.e., data practices) via their privacy policies. Thus there is a need to audit these policies and check compliance with respective laws. This paper addresses this need and presents a large-scale data-driven study to audit privacy policies from healthcare organizations in three countries---USA, UK, and India. 

We developed a three-stage novel \textit{workflow} for our audit. First, we collected the privacy policies of thousands of healthcare organizations in these countries and cleaned this privacy policy data using a clustering-based mixed-method technique. We identified data practices regarding users' private medical data (medical history) and site privacy (cookie, logs) in these policies. Second, we adopted a summarization-based technique to uncover exact broad data practices across countries and notice important differences. Finally, we evaluated the cross-country data practices using the lens of legal compliance (with legal expert feedback) and grounded in the theory of Contextual Integrity (CI). Alarmingly, we identified six themes of non-alignment (observed in 21.8\% of data practices studied in India) pointed out by our legal experts. Furthermore, there are four \textit{potential violations} according to case verdicts from Indian Courts as pointed out by our legal experts. We conclude this paper by discussing the utility of our auditing workflow and the implication of our findings for different stakeholders.

\end{abstract}

\section{Introduction}\label{sec:intro}

\noindent Today, healthcare organizations like hospitals around the world collect highly sensitive patient data. This data is used for multiple purposes---keeping track of patients, their treatment, and even health informatics via health-data mining~\cite{tekieh-2015-healthdatamining,goncalves-2015-qualityoflife,xue-2018-health-data}. Due to the sensitive and personal nature of this data, regulating bodies (e.g. governments) have drafted and enforced laws on how to store and utilize the patient data (e.g., their medical records, financial information, and even passwords) while maintaining the privacy and security of this data. These laws include GDPR~\footnote{General Data Protection Regulation} article 35 in EU, Data Protection Act 2018 in the UK, HIPAA, 1996~\footnote{Health Insurance Portability and Accountability Act of 1996} in the USA, and Information Technology Rules, 2011 and more recent proposal of DISHA~\footnote{Digital Information Security in Healthcare Act} in India~\cite{gdpr-article35-2021,DPA-2018,hipaa-law,IT_ACT_2011,DISHA_ACT_2018}. Naturally, healthcare organizations should comply with applicable laws for handling sensitive patient data and mention their data practices (how they collect/process/store/share this data) in official \textit{privacy policies}. Thus, auditing privacy policies should uncover data practices of healthcare organizations.

\begin{figure*} [t]
    \centering
    \includegraphics[width=0.65\textwidth,scale=0.5]{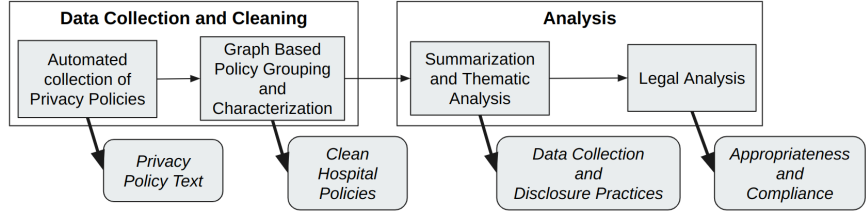}
    \caption{Workflow for auditing privacy policies of healthcare organizations in a country with outputs for each step. }
    \label{fig:work_flow}
\end{figure*}

In general, mining privacy policies is an active research domain where researchers use machine learning as well as natural language processing techniques for finding privacy violations / non-compliance with these laws for different industries. However, we noted that, although a large body of work on privacy policy primarily focused on services like shopping sites or technology companies~\cite{zimmeck2019maps,zimmeck2017automated,oltramari2018privonto,libert2018automated}, mining healthcare organizations' privacy policies have attracted relatively scant attention. %
One reason might be that in many western countries these privacy policies are strictly required to follow specific templates which already ensure legal compliance, making them less interesting to analyze. However, beyond the privacy policy of a single organization, there may be differences in broad practices across countries or  regions that may be interesting to analyze. %
Side by side, a country like India which has a still-nascent approach to enforce privacy laws necessitates auditing the privacy policies put forward by the Indian hospitals at scale and checking legal compliance. %
Hence there is a need for systematically auditing privacy policies of healthcare organizations in a country to help various stakeholders to take further action. %

To that end, in this paper, we investigate the following questions: \textit{How can we systematically audit healthcare privacy policies at scale across different countries to uncover data practices? Are these practices appropriate and compliant with applicable laws?} We answered these questions by (i) developing a novel auditing workflow and (ii) using this workflow on the privacy policies from USA, UK and India.

\textbf{Our novel mixed-method auditing \textit{workflow}}: Our workflow (Figure \ref{fig:work_flow}) is a three-stage methodology to collect privacy policies of healthcare organizations of a country, extract the data practices from the policies and analyze the policies. 
In the \textit{first} stage of our \textit{workflow}, we use a semi-automated process to {\it collect} the privacy policies declared by hospitals on their websites (a large majority of the countries across the globe mandates such declaration). This is then followed by a clustering-based data cleaning stage, which yields hospital-specific privacy policies.  %

In \textit{second} stage of workflow, we used a {\it segmentation-summarization} approach to select the most important sentences belonging to a particular data practice category and create a `template' that readily lends itself to auditing via comparative thematic analysis. In the \textit{final stage} we analyze the data practices from policies of healthcare organizations using the theory of Contextual Integrity (CI) and through the lens of legal compliance (using expert feedback). Our workflow provides both high-level and in-depth exploration of data practices revealed by privacy practices of healthcare organizations in different countries.

\vspace{1mm}
\noindent\textbf{Auditing outcome from applying our novel \textit{workflow} on USA, UK and India:} We audited privacy policies of healthcare organizations from three different countries, {\bf USA, UK}, and {\bf India} that have all their health information in English and have different health laws. 
In this direction, we make the following contributions.

\begin{itemize}[noitemsep,topsep=0pt,parsep=1pt,partopsep=0pt,leftmargin=5.5mm]
    
    \item  Using our workflow we put forward a first-ever large-scale  representative dataset of privacy policies from hospitals of three countries: India (377), UK (183), and USA (1,505) (Section \ref{sec:RQ1}).
    
    \item Through the graph-clustering-based data cleaning phase of our workflow, our audit shows that for USA roughly 50\% of the policies are extremely similar to each other and also discussed the practices in a similar fashion as use-cases described in the body of HIPAA. In contrast, in India, few policies actually discussed the handling of users’ sensitive medical information (Section \ref{sec:RQ2}).

    \item Finally, we found and validated (using feedback from a team of legal experts)  potential \textit{non-compliance} in the data practices (i.e., how they collect/process/store/share health-related data) of Indian hospitals' privacy policies. In fact, our audit found six main reasons of non-alignment in Indian privacy policies which were present in 21.8\% data practices of which four were identified as \textit{potential} violations (Section~\ref{sec:RQ4}). %

\end{itemize}

\noindent\textbf{Ethical considerations:} In this study we automatically collected publicly available privacy policies from healthcare organizations in line with earlier work~\cite{zimmeck2019maps,zimmeck2017automated,oltramari2018privonto}. The only human-subjects study phase was the expert feedback. 
Although our institute lacks a traditional IRB, we extensively discussed among ourselves and with other researchers to minimize any harm in our protocol. During our feedback phase by legal experts (Section~\ref{sec:RQ4}) we were extremely careful about ethically collecting data and followed the best practices prescribed in the Menlo report~\cite{kenneally2012menlo}. Specifically, to protect privacy of our legal experts we removed all personally identifiable information (e.g., name or email id) before storing and analyzing the annotated data and qualitative feedback. Also the experts had the option to abort any time from giving feedback on legal compliance. Lastly, we only report aggregate results without revealing any individual or organization. 

Next, we start with discussing the necessary related work and then move to describing our auditing workflow as well as findings of our audit after applying the workflow in practice.

\section{Related Work}\label{sec:rel_works}

\noindent We present the prior work along three dimensions---mining privacy policies, leveraging natural language processing (NLP) for consumer privacy protection, and leveraging CI (Contextual Integrity) theory.%

\vspace{1mm}

\noindent \textbf{Privacy policy mining.} Prior studies focused on automated mining privacy policies, often with the goal to uncover potential privacy violations. These earlier studies range from mining  privacy policies of mobile applications from Google Play store to mining generic privacy policies~\cite{zimmeck2019maps,zimmeck2017automated,oltramari2018privonto}. Some of them even proposed to improve readability and structure using automated learning methods~\cite{gopinathautomatic,liao2020measuring}. 
In general, these studies often collected privacy policy data from the most popular websites---Using Google trends (to get websites against most popular queries) or ~\url{Alexa.com} %
to identify top domains~\cite{oltramari2018privonto,bannihatti2020finding,liu2014step,sathyendra2017identifying}. They identified privacy policies of these domains by collecting hyperlinks with ``privacy policy'' or ``terms of service'' mentioned in the associated text or even looked for specific keywords like ``privacy'' in the URL itself~\cite{libert2018automated,bannihatti2020finding,sathyendra2017identifying}. We build on these studies, but our work is quite distinct---we focused on privacy policies from \textit{USA, UK, and Indian Healthcare Organizations} which is distinctly separate from top (often shopping or technology-focused) websites. Furthermore, we study variation of privacy policies across different countries and audit their legal compliance (e.g., selling customer information without consent). 

\vspace{1mm}

\noindent \textbf{NLP for consumer privacy.} Ravichander et al, 2021~\cite{ravichander_breaking_2021}, discuss  {\em how NLP can benefit consumer privacy}. Furthermore, Wilson et al.~\cite{OPP_115} have identified data practices from policies and put forward OPP-115---a dataset of 115 privacy policies with 3,792 annotated segments. These segments were annotated across 10 data privacy categories like first party collection,  data retention etc. Earlier works also extracted opt-out choices present from policy text using NLP techniques~\cite{sathyendra_helping_nodate,bannihatti2020finding}.
Works of Zimmeck et al~\cite{zimmeck2017automated,zimmeck2019maps} employed NLP techniques to analyze privacy policies of mobile applications  to identify potential legal non-compliance. Other works utilized NLP techniques to carry out the readability of privacy policies using NLP techniques~\cite{massey-analysis,10.1007/978-3-642-39371-6_8} and explored possibility of question answering over privacy policies to allow consumers to selectively query privacy policies for issues that are important to them~\cite{ravichander-etal-2019-question,ahmad-etal-2020-policyqa}. Some studies even   utilized  summarization techniques to construct summaries to help the users easily understand the practices described in the privacy policy~\cite{zaeem2018privacycheck,keymanesh_toward_2020,tomuro_automatic_2016}. We leveraged these earlier studies on classification of data practices and summarization techniques.

\vspace{1mm}

\noindent \textbf{Theory of Contextual Integrity (CI).} Contextual Integrity~\cite{nissenbaum2004privacy} theory provides a systematic framework to study privacy norms and expectations. CI defines privacy as appropriate flows of information. Each information flow consists of five parameters: subject, sender, recipient, information type (or attribute), and transmission principle. Appropriate information flows conform to the socially acceptable values of these parameters. Earlier work demonstrates that we can infer privacy norms (i.e., rules regulating acceptable information flow) by measuring the acceptability of different information flows (created with a varying combination of CI parameter values) ~\cite{apthorpe2018discovering,apthorpe2019evaluating,shvartzshnaider2019going,MM_Deletion_Privacy}. For example, users in general might be comfortable when a healthcare organization (sender) discloses user’s medical records (attribute) to the doctor in-charge (recipient) to monitor the health status (transmission principle). However, they might be uncomfortable if the recipient is a health insurance provider. Our work aims to understand appropriateness  of the privacy policy data practices and check their alignment with law. Thus we build on prior work and choose CI as a suitable framework for our purpose.

Now, we present our first stage of auditing workflow for collecting and analyzing privacy policies from healthcare organizations.

\if{0}
\begin{figure*}
	\centering
	\begin{minipage}{0.64\columnwidth}
		\centering
		\includegraphics[width=\textwidth]{Chart/Words_Dist.eps}
		\label{label1}
	\end{minipage}%
	\begin{minipage}{0.64\columnwidth}
		\centering
		\includegraphics[width=\textwidth]{Chart/Sent_Dist.eps}
		\label{label2}
	\end{minipage}
	\begin{minipage}{0.64\columnwidth}
		\centering
		\includegraphics[width=\textwidth]{Chart/Flesch_Dist.eps}
		\label{label2}
	\end{minipage}
	\caption{CDF plots for (Left) Number of words in the policies , (Middle) Number of sentences in the policies , and (right)the Readability scores for India, the UK, and USA}
	\label{fig:cdf_countries}
\end{figure*}
\fi

\if{0}
\section{Outline of Our Auditing Workflow}\label{sec:work_flow}

\begin{figure*} [!t]
    \centering
    \includegraphics[width=\textwidth]{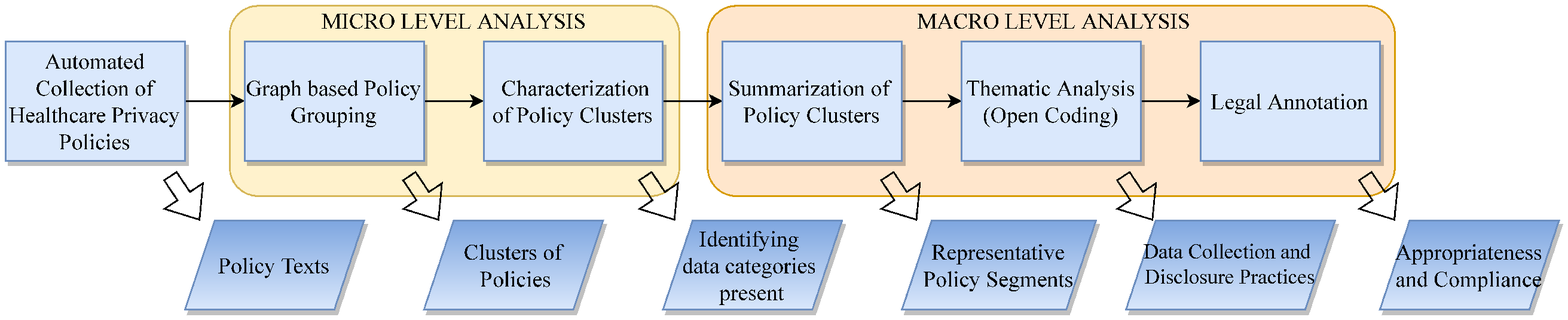}
    \caption{Workflow for auditing privacy policies of healthcare organizations in USA, UK and India with outputs at each step. }
    \label{fig:work_flow}
\end{figure*}

In this work we created a novel \textit{workflow} to collect and audit the privacy policies of healthcare organizations across three countries---USA, UK and India. These three countries conduct their official procedures predominantly in English (enabling comparison) and have different laws regulating their healthcare domain. The workflow is presented in Figure~\ref{fig:work_flow}. Key components of our workflow are:

\noindent\textbf{(i). Data Collection and Cleaning.} Our workflow begins with automated collection of privacy policies from each of these countries. We follow a three-step approach to extract policies from healthcare organizations with online presence (Sec.~\ref{sec:RQ1}). Intuitively, many healthcare privacy policy text in a country might be similar (following same regulations). So, we used a graph-based clustering approach to cluster together similar policies and identify the set of clean hospital-specifc policies.

\noindent\textbf{(ii). Summarization and thematic analysis.} Next using the \textit{clean} policy clusters, we create and analyze \textit{representative} policy templates for each country. 
We follow a segmentation-summarization based approach to systematically create these templates from a privacy policy cluster. %
We then conduct a thematic analysis of these templates to understand the data collection and disclosure practices being adapted in general by healthcare organizations (Sec.~\ref{sec:RQ3}). 

\noindent\textbf{(iii). Appropriateness and compliance.} We finally 
created a user-study to analyze appropriateness of these identified data practices using CI-theory and consulted legal experts to validate our findings. In effect,  we audited legal compliance of existing data practices and finally analyze the reasons for potential vagueness as well as non-compliance (Sec.~\ref{sec:RQ4}).

\fi
\section{Data Collection and Cleaning}\label{sec:RQ1}

We start with the first phase of our workflow---automatically collecting the privacy policies of healthcare organizations (in this work) from USA, UK and India. We will discuss our approach to collect a set of privacy policies, and then discuss how we obtain clean hospital-specific privacy policies.

\subsection{Approach to Collect Privacy Policy Text At-scale}

We follow the following three-step method for each country to systematically collect privacy policies from healthcare organizations and extract privacy policy text:

\noindent{\bf Step-I:} We collect the list of healthcare organization (or \textit{Hospital}) urls using an automated approach.
    
\noindent{\bf Step-II:} For every collected url, we visit the landing page and find hyperlinks on the page %
which potentially are privacy policy urls. 
Specifically in the `a' tag of a url we checked for words like {\em privacy policy, privacy notice, terms and conditions} (in line with ~\cite{srinath-etal-2021-privacy}).
    
\noindent{\bf Step-III:} We parse the HTML text present on privacy policy urls to finally extract privacy policy text (using  Trafilatura~\cite{barbaresi-2021-trafilatura}). We apply these steps for USA, UK and Indian hospitals.

\subsubsection{Collecting Country-Specific Privacy Policies}

For each country, we adopted different ways for obtaining the urls of the Healthcare organizations for {\bf (Step-I)}. %

\vspace{1mm}
\noindent{\bf USA.} We used an official list of USA Hospitals~\cite{USA-dir} that lists hospitals in the USA along with the State, type of facility, ratings etc., for a total of 5299 hospitals. We used a simple google query search using the format: $<Hospital\_Name+State>$ and extracted the top url result~\cite{Google-dir}. We finally had 2511 hospital urls for a total of 4333 hospitals. Using this set after processing through Step-II, we get 1828 URLs of which 1735 were unique URLs for a total of 3250 hospitals. For remaining 583 urls for which we could not find a url in an automated fashion, we  manually checked and found 88 more privacy policy urls. After processing these policy urls using Step-III, we collected total 1,505 unique privacy policies (several hospitals had identical privacy policies).%
\if{0}
\begin{table}[]
    \centering
    \begin{tabular}{|r|r|}
    Error Type & Count \\ \hline
    Max Retries & 2264 \\
    No URL Found & 1148 \\
    406 &  43 \\
    Duplicate URL & 1157 \\
    403 & 401 \\
    404 & 40 \\
    503 & 21 \\
    500 & 5 \\
    400 & 1 \\ \hline
    Total & 5080
    \end{tabular}
    \caption{Error Distribution in USA}
    \label{tab:USA_Error}
\end{table}

\begin{table}[]
    \centering
    \begin{tabular}{|p{4cm}|r|}
    Keywords used from the text and url of $<a>$ tag & Count \\ \hline
    Privacy but not Website & 1139 \\
    Privacy and Notice & 629 \\
    privacy in url & 39 \\
    Patient and Privacy & 133 \\
    Notice and HIPAA & 79 \\
    Terms and Condition & 24 \\ \hline
    Total & 2043 
    \end{tabular}
    \caption{Policies obtained using keywords}
    \label{tab:USA_PolicyDist}
\end{table}
\fi

\vspace{1mm}

\noindent{\bf UK.} 
For the UK, NHS (National Health Service) Trusts are the main entities responsible for functioning of the  healthcare organizations. Using a list of NHS Trusts~\cite{UK-dir}
we obtained 220  urls for web pages on this list and parsed the %
webpages to obtain privacy policies. We thus obtain a final set of 183 unique privacy policies after applying Step-II and III of our collection process.

\vspace{1mm}
\noindent{\bf India.} 
We used the MedIndia’s Hospital Directory~\cite{IN-dir}
of 31,522 hospitals. We extracted the name of the hospitals along with the city from this directory. We then used the following query on Google: $<Hospital\_Name + City>$, and extracted the top url using the googlesearch package~\cite{Google-dir}. %
We only consider the resulting urls if domains have non-empty overlap with the corresponding {\it Hospital\_Name} (we removed common words like `hospital', `nursing' etc. to preserve the specificity of hospital names). We also removed urls which contain obviously non-Indian domain names (e.g., `.co.uk' and `.au'). We finally end up with 6,015 unique urls from 7,662 unique hospitals. On processing these urls (using {\bf Steps-II and III}) we obtained 657 unique privacy policies. %

We noted that India lacked global repositories, which were the primary and a credible source for UK and USA. So we manually verified each of the 657 policies whether they correctly mapped to the corresponding hospitals, and in fact, were a healthcare organization policy. On manual verification, we found several  urls corresponding to either government (or private) directories mentioning these hospitals or news articles which we discarded. Finally, we identified 370 unique privacy policies from 3,120 hospitals.

\begin{table}[t]
    \footnotesize
    \centering
    \begin{tabular}{r|rr}
    \hline
    \textbf{Type of Facility} & \multicolumn{2}{c}{\textbf{Percentage}} \\ 
    & Initial & Clean\\\hline
    \multicolumn{3}{c}{\textbf{USA}} \\ \hline
    Acute Care Hospitals & 66.18 & 64.12\\
    Critical Access Hospitals & 22.56 & 23.62\\
    Psychiatric & 8.25 & 7.76\\
    Children & 2.06  & 1.98 \\
    Acute Care - Department of Defense &  0.95 & 2.52 \\
    \hline
    \multicolumn{3}{c}{\textbf{India}} \\ \hline
     Clinic & 30.27 & 35.95 \\
     Multi-Specialty Hospital & 19.92 & 15.58 \\
     Pathology lab and Diagnostic Centre & 17.36 & 16.46 \\
     Diagnostic and Treatment center & 8.90 & 16.32 \\
     Super-Speciality Hospital & 6.69 & 3.18 \\
     Medical College/Institute/Hospital & 5.16 & 7.49 \\
     Speciality Hospital & 1.73 & 1.29 \\
     Others & 9.97 & 3.73 \\
     
    \hline
    \end{tabular}
    \caption{Type of Facilities covered in USA and India}
    \label{tab:facility_desc}
\end{table}

\subsubsection{ Coverage of our Collection Strategy} 

Since we searched privacy policy urls on only the landing-page, we might have missed privacy policy urls not present on the landing pages. So, we manually looked into landing pages from India and USA that our algorithm discarded due to not finding a privacy policy (all urls from UK had a privacy policy). We found that in only 2 cases of India (out of 100) the discarded website had a privacy policy---however, in both the cases the policy is simple terms of service, not related to healthcare data. For USA we initially discarded 583 landing pages; upon manual inspection we discovered 88 privacy policies from them. We added these 88 policies to our dataset and arrived at 1,505 policies in total. %
Thus, 1,505 privacy policies for USA, 183 for UK and 370 for India constituted our final dataset. Next, we discuss the representativeness of these policies from healthcare organizations.
\if{0}
\begin{table}[t]
    \centering
    \resizebox{\columnwidth}{!}{
    
    \begin{tabular}{|l|c|c|c|c|c|c|}
    \hline    
    Country (\#Policies) & \multicolumn{2}{c|}{\#Words} &\multicolumn{2}{c|}{\#Sentences}&\multicolumn{2}{c|}{Readability Score} \\
       & Mean & Median & Mean & Median & Mean & Median \\
       \hline
       {\bf USA (1505)} & 2459.01 & 2401.0 & 96.93 & 84.0 & 51.15 &51.66  \\
       \hline
       {\bf UK (183)} & 1783.80 & 1040.0 & 61.0 & 42.0 &55.88 & 55.73\\
       \hline       
       {\bf India (370)} & 1125.21 & 747.0 & 40.47 & 32.0 &52.47 &52.58 \\
       \hline
    \end{tabular}}
    \caption{Distribution of Number of Words, Sentences and Flesch Reading Ease score for each country}
    \label{tab:stats_country}
\end{table}
\fi

\subsubsection{Representativeness of Collected Policies }\label{sec:representative} To judge the representativenenss of the policies obtained we looked into how many states were covered by the policies obtained. We found that for USA, 50 out of 50 states were covered, for UK we covered 183 out of 219 NHSs and for India we covered 24 out of 29 states. For USA and India, we also checked the diverse types of hospitals covered in our dataset (Table~\ref{tab:facility_desc}). In short, we covered diverse type of healthcare organizations across countries from various states, giving us confidence about representativeness of our dataset.

\subsubsection{Linguistic Properties of Privacy Policy Text}
We first investigated the distribution of the number of words, sentences, and Flesch Reading Ease Score---the result is shown in Table~\ref{tab:stats_country}. We did this analysis to understand the variation in the \textit{linguistic} complexity of the policies across countries and check for trivial (essential empty) policies that we might have \textit{accidentally} collected.The Indian privacy policies are on average 1125.21 words and 36.60 sentences long, whereas policies from the USA and the UK are 2459.01 and 1783.80 words long and are made up of 96.93 and 61.0 sentences~\footnote{Sentences obtained using NLTK sent\_tokenize module}. The word and sentence distribution was statistically significantly different (using Mann-Whitney U test) with a medium effect size (using Cohen's D) between India and US (p < 0.0001, Cohen's D = 0.49), and India and UK (p < 0.0001, Cohen's D = 0.43). The Flesch Reading Ease scores of all policies were between 50 to 60 which indicate that the documents are uniformly fairly difficult to read (and non-empty)---perhaps due to potentially complex privacy policies. 

\begin{table}[h]
    \centering
    \resizebox{\columnwidth}{!}{
    
    \begin{tabular}{|l|c|c|c|c|c|c|}
    \hline    
    Country (\#Policies) & \multicolumn{2}{c|}{\#Words} &\multicolumn{2}{c|}{\#Sentences}&\multicolumn{2}{c|}{Readability Score} \\
       & Mean & Median & Mean & Median & Mean & Median \\
       \hline
       {\bf USA (1505)} & 2549.01 & 2401.0 & 96.93 & 84.0 & 51.15 &51.66  \\
       \hline
       {\bf UK (183)} & 1783.80 & 1040.0 & 61.0 & 42.0 &55.88 & 55.73\\
       \hline       
       {\bf India (370)} & 1125.21 & 747.0 & 40.47 & 32.0 &52.47 &52.58 \\
       \hline
    \end{tabular}}
    \caption{Distribution of Number of Words, Sentences and Flesch Reading Ease score for each country}
    \label{tab:stats_country}
\end{table}

\subsubsection{Type of Information Mentioned in Policies: First Party Collection}

\noindent {\bf USA and UK.} We find for USA, out of 1,505 policies 435 (28.9\%) mention `PHI' (short for Personal Health Information), and 1179 (78.4\%) policies mention medical records, whereas 350 (25.3\%) policies mention `cookie', and only 43 (2.9\%) policies mention `logs'. For UK, out of 183 policies, 41 (22.4\%) policies mention `PHI', 71 (38.7\%) policies mention medical records, whereas 98 (53.5\%) policies mention `cookies' and only 12(6.5\%) policies mention `logs'. 

\noindent \textbf{India.} The Indian IT Rules, 2011~\cite{IT_ACT_2011} identifies `medical records and history' as one of the healthcare specific sensitive information. We checked (using keywords) if Indian policies talk about this information. We find that out of 370 policies, 46 (12.2\%) policies mention `medical history' and `medical records'. We also looked into how many policies talk about online data (cookies and logs) and found that 236 (62.6\%) policies mention `cookie', and 25 (6.6\%) mentions `logs' in them. Interestingly, potentially with different healthcare laws in place for USA and UK, we see that the mention of attributes relevant to patient information is much more frequent (78.4\% and 38.7\% respectively) in USA and UK than India (12.2\%) (Blocks 1 and 2 of Table~\ref{tab:dist_traits}).

\begin{table}[t]
    \centering 
    \footnotesize
    \resizebox{\columnwidth}{!}{
    \begin{tabular}{p{1.5cm}|p{2cm}|p{2cm}|p{2cm}}
        \textbf{Keywords} & {\bf USA (\#,\%)} & {\bf UK (\#,\%)} & {\bf India (\#,\%)} \\
        \hline
        \multicolumn{4}{c}{\textbf{Healthcare Specific Information}} \\ \hline
        medical records & 1179, 78.4\% & 71, 38.7\% &   46, 12.2\%\\ 
        PHI             & 435, 28.9\%  & 41, 22.4\% &   0, 0\%\\ \hline
        
        \multicolumn{4}{c}{\textbf{Online Information}} \\ \hline
        cookies & 350, 23.3\%  & 98, 53.5\% &  236, 62.6\%\\
        logs & 43, 2.9\% & 12, 6.5\%  & 25, 6.6\% \\ \hline

        \multicolumn{4}{c}{\textbf{Data Disclosure Practices}} \\ \hline                      
        share     & 1084, 72.1\% & 118, 64.5\% &  200, 53.1\%\\ 
        sell      & 374, 24.8\% & 19, 10.4\% & 93, 24.7\% \\
        disclose  & 806, 81.7\%  & 62, 33.8\%& 175, 43.8\% \\
        pass      & 12, 0.7\% & 71, 38.8\% & 4, 1.1\%\\ \hline
        
    \end{tabular}}
    \caption{Attributes corresponding to type of information as discussed in policies along with the count of policies they occur in and percentages.}
    \label{tab:dist_traits}
\end{table}

\subsubsection{Type of Information Mentioned in Policies: Data Disclosure Practices}
We identify how many policies actually mention the data disclosure practices (to disclose users' information) of these organizations. Specifically, we looked for mentions of  keywords {\em share, disclose, sell, provide, and pass}~\cite{ddp-kewords,GDPRContentAnalysis,comprehensiveKeyWords}. We find that 81.7\% and 64.5\% policies from the USA and UK mention data disclosure practices, whereas 53.1\% Indian policies mention such data disclosure practices (Block 3 of Table~\ref{tab:dist_traits}). Thus the fraction of policies that discuss such practices is almost similar (the most frequent practices are presented in Table~\ref{tab:snippets_freq}, Appendix \ref{append:ddp}). 

However, we noted that the data disclosure practices (i.e., the sentences) from USA and UK are detailed and diligently mentioned under what conditions and to whom is the data being disclosed. In contrast, in India, we noted that the policies just mentioned `We do not sell, trade or rent users personal identification information to others', where {\em others} seems indefinite---it doesn't explicitly mention the third parties and organizations (either govt. or private), and what {\em information } covered by these clauses. A potential reason is the right to `Accounting of Disclosures', which is exclusive to USA (mentioned in 245 policies)---consequently, the USA hospitals are required to give the patient an account of the disclosures it has made of the patient's data over a specific time frame (e.g. 6 months, 1 year). In Section~\ref{sec:RQ3} and~\ref{sec:RQ4}, we further explore these data practices and their legal compliance. Next, we present our methodolgy to obtain clean hospital policies.

\if{0}
\if{0}
\begin{table*}[t]
    \centering
    \begin{tabular}{p{4cm}|p{1cm}|p{4cm}|p{1cm}|p{4cm}|p{1cm}}
         \multicolumn{2}{c|}{\bf India (366)} & \multicolumn{2}{c|}{\bf UK (183)} & \multicolumn{2}{c}{\bf USA (1333)} \\
         \hline
         Keywords/phrases & \#(\%) of policies & Keywords/phrases & \#(\%) of policies& Keywords/phrases & \#(\%) of policies  \\
         \hline
         \hline
         {\bf biometric data} & 31(8.5)  & {\bf PHI} & 41(22.4) & {\bf PHI} & 479(35.9) \\
         {\bf medical history} & 35(9.5) & {\bf patient information} & 71(38.7) &{\bf personal health information} & 111(8.3)\\
         {\bf medical records} & 37(10.1)  & {\bf health records} & 53(28.9) &{\bf personally identifiable information} & 332(24.9) \\
         &&&&{\bf health information} & 624(46.7)\\
         \hline
         {\bf cookie} & 222(60.6) &  {\bf cookie} & 98(53.5) & {\bf cookie} & 397(29.7)\\
         {\bf logs} & 30(8.1) & {\bf logs} & 12(6.5) &  {\bf logs} & 57(4.2)\\
         \hline
         {\bf covid} & 0(0) &{\bf covid} & 60(32.7) & {\bf covid} & 84(6.3)\\
         {\bf pandemic} & 0(0) & {\bf pandemic} & 13(7.1) & {\bf pandemic} & 16(1.2) \\
         \hline
         {\bf Data Disclosure Practices} & 228(62.3) & {\bf Data Disclosure Practices} & 128(69.9) &{\bf Data Disclosure Practices} & 899(67.4)\\
    \end{tabular}
    \caption{Number and percentage of policies discussing various attributes and data practices for each country.}
    \label{tab:dist_traits}
\end{table*}
\fi

\if{0}
\begin{table*}
    \centering
    \begin{tabular}{l|c}
         Kewords/phrases & Number(Perentage) of policies  \\
         \hline
         \multicolumn{2}{c}{\bf India (366)} \\
         \hline
         {\bf biometric data} & 31 \\
         {\bf medical history} & 35 \\
         {\bf medical records} & 37 \\
         {\bf cookie} & 222\\
         {\bf logs} & 30\\
         {\bf covid} & 0 \\
         {\bf pandemic} & 0 \\
         {\bf Data Disclosure Practices} & 228\\
         \hline
         \multicolumn{2}{c}{\bf UK (183)} \\
         \hline
         {\bf PHI} & 41\\
         {\bf patient information} & 71\\
         {\bf health records} & 53\\
         {\bf cookie} & 98\\
         {\bf logs} & 12\\        
         {\bf covid} & 60 \\
         {\bf pandemic} & 13 \\
         {\bf Data Disclosure Practices} & 128\\
         \hline
         \multicolumn{2}{c}{\bf USA (1333)} \\
         \hline
         {\bf PHI} & 479 \\
         {\bf personal health information} & 111 \\
         {\bf health information} & 624 \\
         {\bf personally identifiable information} & 332 \\  
         {\bf cookie} & 397\\
         {\bf logs} & 57\\
         {\bf covid} & 84 \\
         {\bf pandemic} & 16 \\
         {\bf Data Disclosure Practices} & 899\\
         \hline         
    \end{tabular}
    \caption{Number and percentage of policies discussing various attributes and data practices for each country.}
    \noteng{Reduce one row}
    \label{tab:dist_traits}
\end{table*}
\fi

\subsection{Linguistic Properties of Privacy Policy Text}
We first investigated the distribution of the number of words, sentences, and Flesch Reading Ease Score---the result is shown in Table~\ref{tab:stats_country}. We did this analysis to understand the variation in the \textit{linguistic} complexity of the policies across countries and check for trivial (essential empty) policies that we might have \textit{accidentally} collected.The Indian privacy policies are on average 1061.99 words and 36.60 sentences long, whereas policies from the USA and the UK are 1557.88 and 1783.80 words long and are made up of 58.10 and 61.0 sentences~\footnote{Sentences obtained using NLTK sent\_tokenize module}. The word and sentence distribution was statistically significantly different (using Mann-Whitney U test) with a medium effect size (using Cohen's D) between India and US (p < 0.0001, Cohen's D = 0.49), and India and UK (p < 0.0001, Cohen's D = 0.43). The Flesch Reading Ease scores of all policies were between 50 to 60 which indicate that the documents are uniformly fairly difficult to read (and non-empty)---perhaps due to potentially complex privacy policies. 

\begin{table}[h]
    \centering
    \resizebox{\columnwidth}{!}{
    
    \begin{tabular}{|l|c|c|c|c|c|c|}
    \hline    
    Country (\#Policies) & \multicolumn{2}{c|}{\#Words} &\multicolumn{2}{c|}{\#Sentences}&\multicolumn{2}{c|}{Readability Score} \\
       & Mean & Median & Mean & Median & Mean & Median \\
       \hline
       {\bf USA (1333)} & 1557.88 & 1097.0 & 58.10 & 42.0 & 51.15 &51.66  \\
       \hline
       {\bf UK (183)} & 1783.80 & 1040.0 & 61.0 & 42.0 &55.88 & 55.73\\
       \hline       
       {\bf India (366)} & 1061.99 & 727.0 & 36.60 & 28.5 &52.47 &52.58 \\
       \hline
    \end{tabular}}
    \caption{Distribution of Number of Words, Sentences and Flesch Reading Ease score for each country}
    \label{tab:stats_country}
\end{table}

\if{0}
\begin{table*}[t]
    \centering 
    \begin{tabular}{p{2cm}|p{4.3cm}|p{4.3cm}|p{4.3cm}}
        \textbf{Category} & {\bf India (366)} & {\bf UK (183)} & {\bf USA (1333)} \\
        \hline
        \textbf{Information Collected} & biometric data(31, 8.5\%); medical history (35, 9.5\%); \textbf{medical records (37, 10.1\%)}
                                       & PHI (41, 22.4\%); health records (53, 28.9\%); \textbf{patient information (71, 38.7\%)}
                                       & PHI (479, 35.9\%); \textbf{health information (624, 46.7\%)} \\ \hline
        
        \textbf{Cookies and Logs}      & Cookies (222, 60.6\%); Logs (30, 8.1\%)
                                       & Cookies (98, 53.5\%); Logs (12, 6.5\%)
                                       & Cookies (397, 29.7\%); Logs (57, 4.2\%) \\ \hline
        
        \textbf{COVID-19}              & covid (0, 0\%); pandemic (0, 0\%)
                                       & covid (60, 32.7\%); pandemic (13, 7.1\%)
                                       & covid (84, 6.3\%); pandemic (16, 1.2\%)\\ \hline
                                       
        \textbf{DDP}                   & \textbf{share (227, 62\%)}; disclose (186, 50.8\%); sell (145, 39.7\%); pass (113, 30.1\%)
                                       & \textbf{share (118, 64.5\%)}; disclose (62, 33.8\%); sell (19, 10.4\%); pass (71, 38.8\%)
                                       & share (708, 53.1\%); \textbf{disclose (713, 53.5\%)}; sell (366, 27.5\%); pass(289, 21.7\%) \\ \hline
        
    \end{tabular}
    \caption{Attributes as discussed in policies along with the count of policies they occur in and percentages. (\textbf{DDP}: Data Disclosure Practice)}
    \label{tab:dist_traits}
\end{table*}

\begin{table*}
    \centering
    \begin{tabular}{p{5cm}|p{5cm}|p{5cm}}
    \textbf{India} & \textbf{UK} & \textbf{USA} \\
    \hline
     We will not disclose or sell any of your personal information, including your name, address, age, sex or medical history to any third party without your permission. 
     & Existing law which allows confidential patient information to be used and shared appropriately and lawfully in a public health emergency is being used during this outbreak 
     & Except as set forth above, you will be notified when PII may be shared with third parties, and will be able to prevent the sharing of this information.\\
     \hline
     We do not sell, trade, or rent Users personal identification information to others 
     & ... share personal/confidential patient information ... in disease surveillance for the purposes of protecting public health, ... and managing the outbreak 
     &  ... disclose personal information ... when required by law  ... Cooperate with the investigations of purported unlawful activities and conform ... - Protect and defend the rights ... - Identify persons ... misusing our Website or its related properties.\\
    \hline
    \end{tabular}
    \caption{Two most frequent data disclosure practices as observed for each country.\new{Can be pushed into Appendix}}
    
    \label{tab:snippets_freq}
\end{table*}
\fi

\subsection{Data Collection and Disclosure Practices Revealed in the Privacy Policies}

Next we did a simple lexical analysis to check (i). the type of information  (using keywords) mentioned in the collected policies, and (ii). the data practices mentioned in the policy text about disclosure of sensitive healthcare data. Table ~\ref{tab:dist_traits} presents a summary of our results.  %

\subsubsection{Type of Information Mentioned in Policies}

{\bf USA and UK.} We find for USA, out of 1,333 policies 479 (35.9\%) mention `PHI' (short for Personal Health Information), and 624 (46.7\%) policies mention medical records, whereas 397 (29.7\%) policies mention `cookie', and only 57 (4.2\%) policies mention `logs'. For UK, out of 183 policies, 41 (22.4\%) policies mention `PHI', 71 (38.7\%) policies mention medical records, whereas 98 (53.5\%) policies mention `cookies' and only 12(6.5\%) policies mention `logs'. 

\noindent\textbf{India.} The Indian IT Rules, 2011~\cite{IT_ACT_2011} identifies `medical records and history' as one of the healthcare specific sensitive information. We checked (using keywords) if Indian policies talk about this information. We find that out of 366 policies, 35 (10.1\%) policies mention `medical history', 37 (10.1\%) policies mention  `medical records'. We also looked into how many policies talk about online data (cookies and logs) and found that 222 (60.6\%) policies mention `cookie', and 30 (8.1\%) mentions `logs' in them. Interestingly, potentially with different healthcare laws in place for USA and UK, we see that the mention of attributes relevant to patient information is much more frequent (46.7\% and 38.7\% respectively) than India (10.1\%) (Blocks 1 and 2 of Table~\ref{tab:dist_traits}).

\begin{table}[t]
    \centering 
    \resizebox{\columnwidth}{!}{
    \begin{tabular}{p{1.5cm}|p{2cm}|p{2cm}|p{2cm}}
        \textbf{Keywords} & {\bf India (\#,\%)} & {\bf UK (\#,\%)} & {\bf USA (\#,\%)} \\
        \hline
        \multicolumn{4}{c}{\textbf{Healthcare Specific Information}} \\ \hline
        medical records & 37, 10.1\% & 71, 38.7\% & 624, 46.7\% \\ 
        PHI             & 0, 0\%        & 41, 22.4\% &  479, 35.9\% \\ \hline
        
        \multicolumn{4}{c}{\textbf{Online Information}} \\ \hline
        cookies & 222, 60.6\% & 98, 53.5\% & 397, 29.7\% \\
        logs & 30, 8.1\% & 12, 6.5\%  & 57, 4.2\% \\ \hline
        
        \multicolumn{4}{c}{\textbf{COVID-19 specific}} \\ \hline
        covid & 0, 0\% & 60, 32.7\% & 84, 6.3\%\\
        pandemic & 0, 0\%  & 13, 7.1\% & 16, 1.2\% \\ \hline
        
        \multicolumn{4}{c}{\textbf{Data Disclosure Practices}} \\ \hline                      
        share     & 227, 62\% & 118, 64.5\% & 708, 53.1\% \\ 
        sell      & 145, 39.7\% & 19, 10.4\% & 366, 27.5\% \\
        disclose  & 186, 50.8\% & 62, 33.8\%& 713, 53.5\%\\
        pass      & 113, 30.1\%& 71, 38.8\% & 289, 21.7\%\\ \hline
        
    \end{tabular}}
    \caption{Attributes as discussed in policies along with the count of policies they occur in and percentages.}
    \label{tab:dist_traits}
\end{table}

\subsubsection{\bf Privacy Policy and Pandemic} 
Interestingly with the outbreak of COVID during our study, we found that the USA and UK have updated policies that also take into account disclosure practices during COVID-period, whereas India has no policy with a specific clause relating to data handling related to pandemic times (Block 3 of Table~\ref{tab:dist_traits}). Specifically, 84 policies (6.3\%) in the USA and 60 (32.7\%) policies in the UK mention `covid', and 16 (1.2\%) in USA and 13(7.3\%) in UK mention `pandemic', whereas in India no policies mentioned `pandemic' or `covid'. 

\subsubsection{\bf Data Disclosure Practices}

In our final preliminary analysis of policy text, we identify how many policies actually mention the data disclosure practices (to disclose users' information) of these organizations. Specifically, we looked for sentences containing keywords {\em share, disclose, sell, provide, and pass}~\cite{ddp-kewords}. We find that 62.3\% Indian policies mention such data disclosure practices, whereas 53.5\% and 64.5\% policies from the USA and UK mention such practices (Block 4 of Table~\ref{tab:dist_traits}). Thus the fraction of policies that discuss such practices are almost similar (the most frequent practices are in Table~\ref{tab:snippets_freq} of Appendix~\ref{append:ddp}). 

However, we noted that the data disclosure practices (i.e., the sentences) from USA and UK are detailed and diligently mentioned under what conditions and to whom is the data being disclosed. In contrast, in India, we noted that the policies just mentioned `We do not sell, trade or rent users personal identification information to others', where {\em others} seems indefinite---it doesn't explicitly mention the third parties and organizations (either govt. or private), and what {\em information } covered by these clauses. A potential reason is the right to `Accounting of Disclosures', which is exclusive to USA (mentioned in 245 policies)---consequently, the USA hospitals are required to give the patient an account of the disclosures it has made of the patient's data over a specific time frame (e.g. 6 months, 1 year). Interestingly, although both USA and UK policies contained COVID-related data disclosure, most frequent data disclosure practices in UK policies are related to COVID, which is not the case for USA. In Section~\ref{sec:RQ3} and~\ref{sec:RQ4} We further explore these data practices and their legal compliance.

Next, after our data collection and preliminary analysis, we audit these collected policies using next phases of our workflow (Figure~\ref{fig:work_flow})---starting with micro level auditing.

\if{0}
\subsubsection{\bf India.}
The Indian IT Act 2011, has identified 8 attributes under Sensitive Personal Information (SPI).
Out of these, (i).`physical, physiological, and mental health condition', (ii).`medical records and history' and (iii).`Biometric information' discuss health data.
We look into how many of the policies actually talk about the above three attributes in an automated fashion.
We find that out of 366 policies, 35 policies mention `medical history', 37 policies mention  `medical records', and 31 policies mention `biometric data'  in their text. We also looked into how many policies talk about cookies and logs of user data and found that 222 policies mention `cookie', and 30 mentions `logs' in them. 
\subsubsection{\bf UK and USA} In the UK, we find that out of 183 policies,  41 policies mentions `PHI' (short for Personal Health Information), 71 policies mentions `patient information', and 53 policies mentions `health records', whereas 98 policies mentions `cookies' and only 12 policies mention `logs' in their text. In the USA, we find that out of 1333 policies 479 mention `PHI', and 624 policies mention `health information', whereas 397 policies mention `cookie', and only 57 policies mention `logs'. This brings up the idea that with healthcare laws in place for the USA and UK, we see that the mention of attributes relevant to patient information is much more frequent (46.7\% and 38.7\%) than what is observed for India(10.1\%)(Row 1 of Table~\ref{tab:dist_traits}).
\fi

\fi
\subsection{Obtaining Clean Privacy Policy Text}\label{sec:RQ2}

Since we collected policies from websites, our policies might be a mixture of policies dedicated to website and cookie and policies specific to hospitals' handling of sensitive information of its users (as observed in Table \ref{tab:dist_traits}).  Thus in our data cleaning phase, we aim to obtain a clean set of hospital policies which we will analyze in the further stages of the workflow. We leverage a simple intuition: Multiple privacy policies of hospitals operating in same country (regulated by same law interpreted in a similar way), would have \textit{similar} data practices and similar policy text---these similar policies can be analyzed together in a group, and a set of clean policies can thus be obtained. So, we take a graph clustering-based approach to group policies with similar text and in extension similar data practices.

\subsubsection{Clustering of Privacy Policies}
In our clustering phase, we leverage a popular graph community detection algorithm Louvain~\cite{Louvain-2008} on policies from a country. It utilizes semantic similarity at the policy level (obtained using Word Mover Distance~\cite{kusner2015word})  above a particular threshold as edge weights, to cluster together the policies that are very similar in nature. We now discuss in brief our clustering strategy.

\noindent{\bf Building graph of privacy policies.} The nodes in our graph are privacy policies and the weights on the edges are the similarity scores based on Word Mover Distance~\cite{kusner2015word}. Specifically, we obtain the Word Mover Distance score for every node pair (i.e. Privacy Policy) in the graph, we next normalize the scores thus obtained to scale them between 0 and 1,  we then apply a $1-normalized score$ step to take this score from distance domain to similarity domain. %
We assigned an edge between a pair of nodes if the corresponding similarity is greater than some threshold. 
This was necessary, as simply putting all possible edges with similarity weights led to all the policies collapsing into a single community. We next discuss how we identify thresholds for each country. %

\noindent{\bf Selecting edge-thresholds for each country.} To decide on the threshold to assign edges for the community detection, we looked into several community metrics~\cite{leskovec2010empirical} that determined the quality of generated communities. We used modularity, normalized cut-ratio, and conductance. Additionally, we also used coverage, which is computed as the fraction of policies covered at a threshold. 

To determine the thresholds, we choose the regions of higher modularity,  lower cut Ratio, and lower conductance, and moderate coverage. The distribution of scores for each metric across similarity scores is shown in Figures~\ref{fig:India_Comm_Thr} (India),~\ref{fig:UK_Comm_Thr} (UK), and ~\ref{fig:USA_Comm_Thr} (USA) with threshold highlighted in red  (See Appendix~\ref{sec:appendix_comm}).

\begin{table}[h]
    \centering
    \resizebox{\columnwidth}{!}{
    \begin{tabular}{|c|l|l|l|l|l|l|l|}
         \hline
         Country & \#Thres- &\#Policies & \#Comm- & \multicolumn{4}{l|}{Size of Communities} \\
         &hold& Covered &unities& >50& 11-50 & 6-10 &2-5 \\
         
         \hline
         \textbf{USA}   & 0.44 & 1110 & 4 & 2 & 0 & 2 & 0 \\
         \textbf{UK}    & 0.45 & 161 & 3  & 2 & 0 & 1 & 0\\
         \textbf{India} & 0.49 & 212 & 8 & 2 & 3 & 1 & 2 \\
\hline    \end{tabular}}
    \caption{Privacy policy communities for USA, UK and India. }
    \label{tab:comm_dist}
\end{table}

\noindent \textbf{Sizes of communities obtained.} After applying the graph-based community detection algorithm, the community distribution for each country is shown in Table~\ref{tab:comm_dist}. Next we characterize these communities in order to obtain a set of cleaned privacy policies.

\subsubsection{Characterizing Communities to Obtain Clean \\ Policies} \label{sec:charec_comm}

To perform a detailed characterization study for each of the countries, we selected the most populous communities covering in total at least 80\% of the policies of the country.  The  communities thus selected for each country are as follows: (a). {\bf India:}  Total 8 communities (\textbf{I-1} to \textbf{I-6}) with the respective sizes of 79, 66, 24, 19, 12, 8 (b). {\bf UK:} A total of 2 communities of sizes 82 (\textbf{UK-1}) and 73 (\textbf{UK-2}) (c). {\bf USA:} A total of 2 communities of sizes of 765 (\textbf{US-1}) and 327 (\textbf{US-2}). The other communities are much smaller and sparser.

\noindent{\bf Coding privacy policies.} In the coding phase two researchers coded whether a privacy policy discusses  specific data practices using Open Coding~\cite{Vollstedt2019}. Our initial codebook was inspired by OPP-115 codes~\cite{OPP_115}. Then a researcher went over the dataset and augmented the codebook with additional data practices. Next two researchers used this codebook to independently code the dataset. Cohen's kappa between two sets of codes was 0.8, showing significant agreement. Finally, the two coders met to discuss and resolve small number of disagreements, and assigned final codes.

\noindent \textbf{Final codebook:} There are six categories namely, (a). Online Data Collection ({\bf OD}), (b). First Party Collection ({\bf 1C}), (c). First Party Usage ({\bf 1U}), (d). Third Party Disclosure ({\bf 3D}), (e). Data Retention ({\bf DR}), and (f). User Rights (edit, access, and delete) ({\bf UR}) across which the annotation task was carried out. \textbf{OD} covers practices specific to collection and processing of user's online Non-PII information (such as cookies, logs, IP address and so on). \textbf{1C and 1U} cover practices specific to collection and processing of user's (sensitive) healthcare information. \textbf{3D} covers practices specific to disclosure of information to third parties and to entities outside the healthcare organizations. \textbf{UR} covers practices discussing users' rights over their information being held by the healthcare organizations.

\begin{table}[h]
\footnotesize
\resizebox{\columnwidth}{!}{
\begin{tabular}{|l|l|l|l|l|l|l|l|}
\hline         
Comm            &OD     & 1C    & 1U    & 3D    & DR    & UR    & \textbf{Avg} \\ \hline
     \multicolumn{8}{|c|}{\textbf{USA}} \\  \hline
\textbf{US-1}   &0.04   & 0.43  & 1     & 1     & 0     & 1     & \textbf{0.69} \\ \hline
US-2            &0.97   & 0.03  & 0.03  & 1     & 0     & 0.06  & 0.22 \\ \hline

     \multicolumn{8}{|c|}{\textbf{UK}} \\ \hline
\textbf{UK-1}   &0.32   & 0.86  & 0.96  & 0.96  & 0.78  & 0.78  & \textbf{0.87} \\ \hline
UK-2            &0.92   & 0.38  & 0.28  & 0.52  & 0.34  & 0.16  & 0.34 \\ \hline

 \multicolumn{8}{|c|}{\textbf{INDIA}} \\ \hline
\textbf{I-1}    & 0.96  & 0.78  & 0.53  & 0.97     & 0.50  & 0.61  &  \textbf{0.68} \\ \hline
I-2             & 1     & 0.06  & 0.06  & 0.92  & 0.18  & 0.30  & 0.30 \\ \hline
I-3             & 0.63  & 0.08  & 0.08  & 0.92  & 0.04  & 0.04  & 0.23 \\ \hline
I-4             & 1     & 0     & 0     & 0     & 0     & 0.74  & 0.14 \\ \hline
I-5             & 1     & 0     & 0     & 0     & 0.83  & 0.91  & 0.35 \\ \hline
I-6             & 1     & 0     & 0     & 0.75  & 0     & 0.75  & 0.3 \\ \hline
\end{tabular}}
    \caption{Scores for communities from each country for a data practice category. Categories are shortened for brevity.}
\label{tab:annot_scores}
\end{table}

\noindent \textbf{Scores:} Using this codebook, researchers coded random 110 policies from USA (77 from US-1, and 33 from US-2), random 100 from UK (50 from UK-1 and UK-2 each) and 208 policies from India (spanning all 6 communities from I-1 to I-6). Overall 418 policies across 3 different countries were coded. We tabulate the final codes and assign a score to each community under a category as the fraction of policies actually discussing that category. Ideally, a `clean' community should have a majority of policies discussing all the categories. Average scores for selected communities, across each data practice is in Table~\ref{tab:annot_scores} (excluding $OD$). `clean' communities for each country (with highest average score) are in bold.

\vspace{1mm}
\noindent{\bf Discussion.} We now discuss insights (using the final coded privacy policies) about the privacy policy communities we detected for each country.

    \noindent \textbf{(i) USA.} We see that \textbf{US-1} performs much better than \textbf{US-2} across all the categories  {(See Table~\ref{tab:annot_scores}, Under USA Row US-1 and US-2)} except for OD. Looking into the text from the {\bf US-1} community, we found that policies discuss in-depth health information usage of by the first party ({\bf 1U},1). %
    The salient difference between these two communities is that {\bf US-2} contains policies discussing website and cookie privacy policy (\textbf{OD}, 0.97) whereas \textbf{US-1} community discusses more detailed healthcare privacy policy (\textbf{OD}, 0.06 and \textbf{1U}, 1).
    
    \noindent \textbf{(ii) UK.} Like USA, we see that \textbf{UK-1} performs much better than \textbf{UK-2} across all the categories  {(See Table~\ref{tab:annot_scores}, Under UK Row UK-1 and UK-2)}. %
    We found that (similar to USA) policies from \textbf{UK-2} are mostly website and cookie privacy policies ({\bf OD}, 0.92) and have low scores across categories.

    \noindent \textbf{(iii) India.} We observe that on average the policies for the {\bf I-1} community  {(See Table~\ref{tab:annot_scores}, Under India, Rows for I-1)}, discusses about most of the  accepted data practices. We find that the policies present in this community comprises the hospital privacy policies discussing the handling of user-health data({\bf 1C}, 0.78) and practices related to third parties({\bf 3D}, 0.97). For the rest of the communities, we found the focus mainly to be on online data ({\bf OD}>0.5) with almost no focus on handling of sensitive personal medical information ({\bf 1C, 1U}$\leq$0.08).

\vspace{1mm}
\noindent{\bf Summary.} Our  cleaned data revealed that the countries with strict healthcare-specific laws in place (UK and USA) for a significant period, resulted in a few big communities, where the difference was observed primarily in the nature of policies, some being online website privacy and cookie policies, whereas others being hospital privacy policies. But for India, with the lack of a descriptive, concise healthcare-specific law, we have a lot of small communities which exhibit different traits as discussed. We found that (a) policies from India were more focused on discussing  Online Data (\textbf{OD}>0.5 for all), whereas with USA and UK that was not the case. (b) policies for USA had no mention of how the hospital retains Personal Health Information (PHI) of the user  (\textbf{DR},0) and little mention of how it collects this information  (\textbf{1C},0.43). %

After identifying clean, hospital-specific policies  we now delve into our analysis part of the workflow focusing on the policies from clean clusters from each country namely \textbf{US-1} containing 765 policies,\textbf{ UK-1} with 82 policies, and \textbf{I-1} with 79 policies. We note that the clean set also retains the representativeness and diversity of initial data (Sec.\ref{sec:representative}) as shown in Table~\ref{tab:facility_desc}. The state distribution for USA and India also remained same in clean data.

\if{0}

\vspace{1mm}
\noindent{\bf USA.}
We considered three extra categories in line with OPP-115~\cite{OPP_115} besides the six common categories. These are  (a). User`s Choice ({\bf UC}), (b). Data Security ({\bf DS}), and (c). Policy Change ({\bf PC}). 
\if{0}
\begin{figure}[h]
\begin{minipage}[t]{0.5\linewidth}
    \includegraphics[width=\linewidth]{Chart/Word_CDF_Comm_USA.eps}
    \label{f1}
\end{minipage}%
\begin{minipage}[t]{0.5\linewidth}
    \includegraphics[width=\linewidth]{Chart/Sent_CDF_Comm_USA.eps}
    \label{f2}
\end{minipage}
\caption{CDF distributions for the USA communities {\bf (Left)} \#Words {\bf (Right)} \#Sentences}
\label{fig:US_Comm_dist}
\end{figure}
\fi
As opposed to the \textbf{first} community, the {\bf second} community performs much better across all the categories  {(See Table~\ref{tab:annot_scores}, Under USA Row US-1 and US-2)}. Looking into the text from the {\bf second} community, we found that policies discuss in-depth health information collection and usage of data by the first party ({\bf 1C} and {\bf 1U}, 1 and 1). Almost all the policies discuss the Users' rights to data ({\bf UR}, 1) and choices the users have over their data ({\bf UC}, 0.98). The salient difference between these two communities is that the {\bf first} community contains policies discussing website and cookie privacy policy (\textbf{OD}, 1) whereas the second community discusses more detailed healthcare privacy policy (\textbf{OD}, 0.04).

\vspace{1mm}
\noindent{\bf UK.}
We considered three extra categories in line with provisions of DPA 2018~\cite{DPA-2018} and OPP-115~\cite{OPP_115} besides the six  agreed categories. The categories are  (a). User`s Choice ({\bf  UC}), (b). Data Security ({\bf DS}), and (c). Data Protection Officer ({\bf DPO}). We find that on average policies from the {\bf first} community  {(See Table~\ref{tab:annot_scores}, Under UK, Row UK-1)} are more extensive about the data practices. %
Majority of the constituent policies  talk in general about hospital privacy policies, mentioning in detail the type of personal information collected along with its purpose ({\bf 1C}, 0.86), and the usage ({\bf 1U}, 0.96) and disclosure of such personal sensitive information ({\bf 3D}, 0.96).
For the {\bf second} community (See Table~\ref{tab:annot_scores}, Under UK, Row UK-2), the scores across each category are much lower than the scores for the first community. %
The first major difference is observed across {\bf UR}, 0.16 for the second community whereas 0.78 for the first community.
The second major difference is observed across {\bf DPO},  0.26 for the second community whereas 0.80 for the first community.
We checked the texts from the second community and found that (similar to USA) these policies are mostly website and cookie privacy policies ({\bf OD}, 0.92) and hence suffer from such low scores across categories. 

\vspace{1mm}

\noindent{\bf India.}
We considered three extra categories in line with provisions of IT Rules, 2011\cite{IT_ACT_2011} besides the six agreed categories. These are (a). Legal Disclosure ({\bf LD}), (b). Grievance Officer ({\bf GO}), and (c). Storage Standards ({\bf SS}). Since the communities in India are fairly small in size, the results are derived by considering  all the policies across all the communities. 

{\bf Top two Indian communities.} We observe that on average the policies for the {\bf I-1} community  {(See Table~\ref{tab:annot_scores}, Under India, Rows for I-1)}, talk about most (0.85) of the  accepted data practices. This is also the best performing community, as the policies mentioned in the communities cover almost every possible aspect as prescribed in the law. We find that the policies present in this community comprises the hospital privacy policies discussing the handling of user-health data({\bf 1C}, 0.94) and practices related to third parties({\bf 3D}, 0.97) and legal disclosure({\bf LD}, 0.84). {\bf I-5} community has a similar distribution across categories as we saw for the I-1 community and has a good overall score (0.76)  {(See Table~\ref{tab:annot_scores}, Under India, Row I-5)}.
However, we  found that the fraction of policies in I-1 community that discussed data retention (\textbf{DR}, 0.71) and user rights (\textbf{UR}, 0.80) were much more than those in I-5 community ((\textbf{DR}, 0.2); (\textbf{UR}, 0.1)). 

{\bf Other Indian Communities:} Policies from  {\bf I-2, I-3, I-4, and I-7} communities  mostly talk about data practices related to cookie policies and logging of user activities on the website (\textbf{OD}: 1, for most of these communities) making them more of a website privacy policy than a hospital privacy policy. Policies from these communities  {(See Table~\ref{tab:annot_scores}, Under India, Rows I-2, I-3, I-4, and I-7)} almost never discuss any data practice specific to data retention ({\bf DR}, best:0.2), and grievance officer ({\bf GO}, 0 for all). Policies in {\bf I-6} community  {(See Table~\ref{tab:annot_scores}, Under India, Row I-6)} discusses %
users' rights ({\bf UR}, 0.70), but all of them use the same phrase `All users can see, edit, or delete their personal information at any time (except they cannot change their username). Website administrators can also see and edit that information'. Further, only 1 policy out of 10 talks about third-party disclosure ({\bf 3D}, 0.10) and states: `We do not collect any personal data. Data shared by user for taking appointments is placed secured on our server which is not shared with any third parties' to convey the practice.

\fi
\section{Summarization and Thematic Analysis} \label{sec:RQ3}

In this step of the workflow, we aim to compare the data practices of privacy policies across countries (which are regulated by different laws) across two broad categories of \textit{First Party Collection and Usage}, and \textit{Third Party Disclosure}. Since each policy community consisted of \textit{similar privacy policies}, we created a representative summary (\textit{template}) for a given community which encompassed the data practices of the policies in the community---for each community, we use automated text-summarization followed by a thematic analysis to (i) preserve original practices in summary (ii) omit redundant manual effort of checking hundreds of (very similar) privacy policies from healthcare organizations (iii) ensure inclusion of most frequent and diverse data practices.

\subsection{Methodology to Create Templates}
The methodology of identifying our category-specific templates is broadly classified into two phases: (a) \textbf{Phase-I Finding Segments:} In this phase, for a community we identify the segments present in each policies across four data practice categories in an automated fashion. (b) \textbf{Phase-II Summarization:} In this phase, we summarize the segments identified from the previous phase to bring out representative sentences for each category and create templates using an extractive summarization framework. 
Note that these segments belong to a single community with policies containing very similar data practices (by construction).

\subsubsection{Phase-I: Finding Segments}\label{sec:find_segs}
We used the OPP-115 corpus~\cite{OPP_115} which has 3,792 annotated segments from 115 privacy policies. We consider annotations  across 2 different categories of data practices that aligns with our discussion in Section~\ref{sec:charec_comm} :``First party Collection/Use''({\bf 1U + 1C}) and ``Third party Disclosure'' ({\bf 3D}) %
which are also grounded from the laws of the countries we consider in this study, and have significant data support in the OPP-115 dataset (around 40\%). The rest are marked as ``Others''. We used the annotated segments from OPP-115 dataset to train 5 different text classification models to identify the particular data practice category discussed in the segment in an automated fashion. Out of these five models, the best performing was TextCNN (See Table~\ref{tab:models_results} in Appendix~\ref{sec:appendix_template}). Next for our dataset of policies, we first parse the policies using the ASDUS tool~\cite{gopinath2018supervised} to create the segments and then classify them using  TextCNN to obtain category-specific practices.

\subsubsection{Phase-II: Constructing Templates}\label{sec:det_temp}
To construct category-specific templates which comes from multiple segments, we use an unsupervised multi-document extractive summarization technique (in-line with \cite{zhao2020summpip}). First, we obtain cosine similarity for each segment pair with segments encoded using a 300-dimension fasttext~\cite{bojanowski2016enriching} model tuned on privacy policy corpus~\cite{amos2021privacy}. Then, we construct a graph of these segments with edge weight as cosine similarity and obtain importance score for each segment as PageRank~\cite{cite:pagerank} weights obtained from the above graph. Lastly, we apply MMR (Maximal Marginal Relevance)~\cite{cite:MMR} using the importance scores and inter-segment cosine similarities to identify a set of top-10 important yet diverse segments. We call this set as \textit{template} from a country for a category. We also compare this approach to various multi-document summarization baselines in Appendix \ref{sec:appendix_template}. Now we present the thematic analysis that we conduct on the \textit{templates} thus obtained, in a cross-country fashion.

\subsection{Comparing USA, UK and India Templates}
After capturing the representative data practices in \textit{Templates}, we now explore how the \textit{descriptiveness} of such practices vary across countries.  Specifically, we conduct a thematic analysis on data practices and then unpack the thematic differences across countries.

\begin{table}[h]
    \footnotesize
    \centering
    \resizebox{\columnwidth}{!}{
    \begin{tabular}{|l|l|}
    \hline
    Category  &  Themes(Countries it occurs in)\\
    \hline
    First Party Collection,  & Use of data collected (IN, USA, UK) \\
                        Use       (\textbf{1U + 1C}) & Sources of information collected (IN, UK) \\
                    
                                    & Explicit attributes collected (IN) \\
                                    & Defines SPI (IN) \\
                                    
                                    & Methods of Information storage (UK) \\
                                    & Transmission of information (UK) \\
                                    & Defines PHI (USA) \\
                                    & Regarding Authorization for Usage (USA) \\
                                    \hline
    Third Party Disclosure (\textbf{3D})          & Legal Disclosure (IN,UK,USA)\\
                                    & Purpose of Disclosure (IN, UK, USA) \\
                                    & Information Disclosed To (IN, UK, USA) \\
                                    & Restrict Disclosure (USA)\\
                                    & Limited Disclosure (USA)\\
                                    & Third-Party Responsibilities (IN)\\
                                    \hline

    \end{tabular}}
    \caption{Themes identified in the templates along with countries where the theme is identified in parenthesis. (IN: India)  }
\label{tab:short_themes}
\end{table}

\begin{table*}[t]
    \centering
    \footnotesize
    \resizebox{\textwidth}{!}{
    \begin{tabular}{|p{1.7cm}|p{7.5cm}|p{7.5cm}|}
    \hline 
{\bf Theme} & \multicolumn{1}{c|}{\bf India} & \multicolumn{1}{c|}{\bf UK} \\
    \hline
     Sources of information collected
     & Collection of information through from you when you make payment
... to complete optional online surveys
     & Information from you … face to face, during a telephone call ...via a form we have asked you to complete.\\
     \hline
     Use of Data Collected
     &… use your Personal Information to contact you and deliver information to you ... targeted to your interests, such as targeted banner advertisements
     & Purpose is for your direct care and treatment this includes to ensure safe and high-quality care for all our patients. ... for other purposes such as research.\\
    \hline
    \end{tabular}}
    \caption{First Party Collection/Use (1U + 1C) Example Practices (India and UK)}
    \label{tab:India_UK_First_Party}
\end{table*}

\begin{table*}[t]
    \centering
    \footnotesize
    \resizebox{\textwidth}{!}{
    \begin{tabular}{|p{1.65cm}|p{7.5cm}|p{7.5cm}|}
    \hline
    {\bf Theme} & \multicolumn{1}{c|}{\bf India} & \multicolumn{1}{c|}{\bf UK} \\
    \hline
     Legal Disclosure
     & … disclose any information as is necessary to satisfy or comply with any applicable law, regulation, legal process or governmental request
     & ... due to a legal requirement disclosure under a court order, ... to the Health and Safety Executive ... to the police ... to debt collecting agencies\\
     \hline
     Purpose of Disclosure
     & ... for reasons such as website hosting, data storage, software services, email services, marketing, fulfilling customer orders,providing payment services, data analytics
     & … to obtain payment for services that we provide to you ... may use and disclose medical information about you for research purposes \\
     \hline
     Information Disclosed To
     &… may also disclose or transfer End-User’s personal and other information a User provides, to a third party as part of reorganization or a sale of the assets
     & … with health and care organisations and other bodies engaged in disease surveillance, ... to the Department of Health, with authorised non-NHS authorities and organisation \\
    \hline
    \end{tabular}}
    \caption{Third Party Disclosure (3D) Example Practices (India and UK)}
    \label{tab:India_UK_Third_Party}
\end{table*}

\subsubsection{Thematic Analysis}
We manually coded these templates using Open Coding \cite{Vollstedt2019} to bring out the data collection, usage and disclosure-related themes covered in the templates for each country. In our open coding two researchers first collaboratively created a codebook to identify codes pertaining to the two key categories--First party collection/use and Third party disclosure by going over the templates. Next they independently coded the sentences from the template (cohen's kappa 0.72, substantial agreement). The codes finally met, resolved the differences in codes and assigned final codes. The themes identified are shown in Table~\ref{tab:short_themes}. We note that the UK and USA have similar themes under the same category. From our analysis of the countries' templates we find 2 distinct groups coming forth--India on one side and the UK and the USA on the other. We take the UK's template as representative of the latter group and compared India and the UK.

\noindent{\bf Comparing First Party Collection/Use.} %
In India's template we see explicit mention of the information (SPI as mentioned in clauses of IT Rules, 2011 \cite{IT_ACT_2011}) collected whereas UK's and USA's do not. India and USA explain SPI/PHI(Sensitive Personal Information and Personal Health Information), but the UK does not.  UK's template included practices of information storage and transmission. %
The themes common to both India and the UK are sources of information collected and the way information is used. Examples from the templates for these themes are shown in Table~\ref{tab:India_UK_First_Party}. In the first row, the theme identified is the `Sources of information collected'---UK's template mentions even the most specific sources of information, whereas India mentions collection through broad payment gateways or surveys. In the second row, the theme being discussed is `Use of data collected'. We discover that in India, the major use of information is for communication, whereas the UK's template mentions treatment of the patient as the primary reason.

\noindent{\bf Comparing Third-Party Disclosure.}
The themes common to both India and the UK are legal disclosure, purpose of disclosure of information to third parties, and description of which third parties get the information. We discuss these themes using  Table~\ref{tab:India_UK_Third_Party} to highlight the differences between India and UK's templates. From the first row, under the theme `Legal Disclosure', we see that unlike India, in UK there are details of exactly which legal agency will be given access to one's data. In the second row the theme is `Purpose of Disclosure'. India's template focuses on the operational side of storing/processing the data. On the other hand UK's template focuses on payment and research when it comes to reasons for disclosing data. The final row, under the theme `Information Disclosed To', illustrates that India's template stresses upon the complete transfer of data as a subset of `data sharing'. UK's template is much more explicit, naming the possible entities for data sharing. The Indian template explicitly discuss the responsibilities of third-parties for the data received from hospitals---possibly due to explicit directive in Rule 7 of IT Rules, 2011~\cite{IT_ACT_2011}. %

\noindent{\bf Uniqueness of template from USA.} In the template for USA we see %
two notable exceptions of `Restrict the Disclosure' and `Regarding Authorization for usage'. Since HIPAA \cite{hipaa-law} is a law primarily about the accountability on account of data, it focuses especially on users' rights. According to the above two categories we see user's choice of disallowing the sharing of data is respected and permission is taken for some specific data sharing (e.g. `Written authorization is required prior to using or disclosing your PHI for marketing activities' ). Under third party disclosure we get the full list of healthcare professionals with whom the user's data may be shared (e.g. `doctors, nurses, technicians, students preparing for health care related careers'). Under legal disclosure, we found a complete list of claims/proceedings which might necessitate information disclosure (e.g. `accreditation purposes, patients and residents claims, grievances or lawsuits, health care contracting relating to our operations'). Under `Limited Disclosure' the policies stated disclosing limited information of patients to their care taker in case of an incapacitated (e.g. unconscious) patient.

\noindent{\bf Summary.} Overall our analysis uncovered a strong correlation between country-specific law and privacy policy structure. More detailed and thorough legislation forces privacy policy makers to be at least as explicit as the law demands. We found that in the case of USA and UK the data practices identified discuss the permitted and authorized usage and disclosure practices as prescribed in the respective laws. However, for Indian policies things were more general because of absence of a clear, concise, descriptive law with the practices often been linked more to the operation of website and general communication and payment related discussion. This further ascertains our findings in Sec.~\ref{sec:charec_comm} where we saw very few Indian policies discussing the practices specific to sensitive medical information of users. 

\if{0}
Here we deploy various unsupervised \textit{multi-document extractive} text summarization (i.e. the original text is preserved in the summary) algorithms with an aim to obtain \textit{template} for a category, which club together all the similar segments and generates one representative template. We now describe the methods used to build the templates and metrics used to determine the efficacy of the templates thus obtained. 

\noindent {\bf Methods and Metrics} We use 3 summarization methods: two well-known strategies~\cite{louis2013automatically} namely {\bf Random-N, Avg Probability} and PageRank+MMR.
In the PageRank+MMR approach, we first process the segments for a category to build a graph with nodes as the sentence from the segments, and use cosine similarity of the sentence representation as the edge on that graph. We apply PageRank~\cite{cite:pagerank} on the said graph and extract sentences and corresponding PageRank weights. Next, we extract top-10 sentences by applying Maximal Marginal Relevance~\cite{cite:MMR} on top of the output from PageRank stage, which resulted in an important yet diverse set of sentences.

To evaluate the summaries, we used the  standard metrics employed for  unsupervised multi-document summarization framework~\cite{louis2013automatically}: JS-Divergence, Cosine Similarity, and Word Mover Distance. Additionally, we also used SUPERT~\cite{gao-etal-2020-supert} which aims to compute the semantic similarity between a summary and a pseudo-summary generated by ~\cite{gao-etal-2020-supert}. A higher SUPERT value means a good summary. %

\fi

\if{0}
\begin{table}
\centering
\resizebox{\columnwidth}{!}{
\begin{tabular}{|l|l|l|l|l|}
\hline
Baselines used  & \multicolumn{4}{c|}{Metrics Used}                  \\ \hline
                & JS   & Cos Sim & SUPERT & \multicolumn{1}{c|}{WMD} \\ \hline
\multicolumn{5}{|c|}{India Third Party}                              \\ \hline
Random-N        & 0.67 & 0.70    & 0.45   & 2.58                     \\ \hline
Avg Probability & 0.79 & 0.57    & \textbf{0.51}   & 2.56                     \\ \hline
PageRank+MMR        & \textbf{0.67} & \textbf{0.74}    & 0.45   & \textbf{2.48}                     \\ \hline
\multicolumn{5}{|c|}{India First Party}                     \\ \hline
Random-N        & 0.59 & 0.72    & 0.52   & 2.45                     \\ \hline
Avg Probability & 0.79 & 0.55    & 0.47   & 2.45                    \\ \hline
PageRank+MMR        & \textbf{0.57} & \textbf{0.76}    & \textbf{0.54}   & \textbf{2.41}                     \\ \hline
\multicolumn{5}{|c|}{UK Third Party}                                 \\ \hline
Random-N        & 0.76 & \textbf{0.66}    & 0.47  & \textbf{2.39}                     \\ \hline
Avg Probability & 0.69 & 0.57    & 0.46   & 2.45                     \\ \hline
PageRank+MMR        & \textbf{0.68} & 0.63    & \textbf{0.50}   & 2.43                     \\ \hline
\multicolumn{5}{|c|}{UK First Party}                                 \\ \hline
Random-N        & 0.76 & \textbf{0.62}    & 0.36   & 2.39                     \\ \hline
Avg Probability & 0.77 & 0.56    & 0.37   & 2.45                     \\ \hline
PageRank+MMR        & \textbf{0.73} & 0.59    & \textbf{0.44}   & \textbf{2.38}                     \\ \hline
\multicolumn{5}{|c|}{USA Third Party}                                 \\ \hline
Random-N        & 0.69 & \textbf{0.61}    & 0.47   & 2.25                     \\ \hline
Avg Probability & 0.67 & 0.56    & 0.49   & 2.19                     \\ \hline
PageRank+MMR        & \textbf{0.67} & 0.56    & \textbf{0.54}   & \textbf{2.18}                     \\ \hline
\multicolumn{5}{|c|}{USA First Party}                                 \\ \hline
Random-N        & 0.68 & 0.58    & 0.49   & 2.21                     \\ \hline
Avg Probability & 0.66 & \textbf{0.61}    & 0.50   & \textbf{2.13}                    \\ \hline
PageRank+MMR        & \textbf{0.66} & 0.56    & \textbf{0.52}   & 2.18                     \\ \hline
\end{tabular}}
\caption{Metrics reported across each category for summarization methods}
\label{tab:scores_India}
\end{table}

\noindent{\bf Results:} For each of the three countries we created two summaries using three different methods, thus totalling 18 summaries. All the summaries obtained from each of the methods are {\bf ten sentences long}. Overall there are 24 sets of comparisons, one for each of the methods for each of the 4 metrics across 6 different settings (2 categories $\times$ 3 countries) . Out of these 24 comparisons, Random-N performs better in 4 comparisons and Average Probability in 3 of them. PageRank+MMR performs at par or better than the other methods in \textbf{17 out of the 24} comparisons. All scores obtained across categories for each country are in Table~\ref{tab:scores_India} ( Appendix~\ref{sec:appendix_temp}). Thus we use the PageRank+MMR for two categories: ``First Party Collection/Use'' (\textbf{(1U + 1C)}) and ``Third Party Disclosure'' (\textbf{3D}) to generate final representative templates.

\fi

\section{Appropriateness and Compliance of Data Practices} \label{sec:RQ4}

In the final phase of our auditing workflow, we checked the legal compliance of the healthcare organization's data practices with country-specific laws. Specifically, we (a) using {\em Contextual Integrity theory}(CI)~\cite{nissenbaum2004privacy} and Vagueness theory~\cite{bhatia2016theory} checked the appropriateness of the data practices  ~\cite{shvartzshnaider2019going}, and (b) checked legal compliance of these data practices using feedback from legal experts.

\subsection{Identifying Data Practices} 
We used the following 3-step approach to identify the data practices that we further evaluated for appropriateness and legal compliance:

(i). We used the segmentation and classification approach (Sec.~\ref{sec:find_segs}) to obtain segments specific to First Party Collection/Usage(1U+1C) and Third Party Disclosure (3D). 

(ii). We ran the PageRank+MMR algorithm (Sec.~\ref{sec:det_temp}) on these segments to obtain a ranking where important yet diverse set of segments were present at the top.

(iii). We then manually selected a list of non-redundant and important segments from the output of PageRank+MMR and leverage these segments for further analysis.

\noindent We finally analyzed 69 segments from USA, 37 from UK, and 55 from India to investigate appropriateness and legal compliance of data practices with applicable laws.

\subsection{Appropriateness of Data Practices}\label{sec:appropriateness}

We leveraged the theory of {\em contextual integrity} (CI) to identify how complete the practices are in terms of the presence of CI-parameters. Recall that (Sec.~\ref{sec:rel_works}), according to CI, a data practice is an information flow constituted of five parameters related to the information: \textit{subject, attributed, sender, recipient, transmission principle}.  
Following the lines of prior work~\cite{shvartzshnaider2019going}, we check if our data practices are \textit{incomplete}. A practice is incomplete if any of the CI parameters is absent in the practice, leading to the following categories of incompleteness: missing recipient, missing sender, missing attribute, and missing transmission principle. We further examined the data practices to check if the practices are vague (in line with the \textit{vagueness taxonomy}~\cite{bhatia2016theory}). We categorize the practices with mention of keywords from one or more combinations of vague terms thus classifying them into four categories: Conditionality (C), Generalization (G), Modality (M), and Numeric Quantifier (N). Descriptions of these categories are in Table~\ref{tab:append_vague}. All the data practices were annotated by two researchers following the guidelines described in the respective studies (with a near-perfect inter-annotator agreement.) We now discuss our key observations in terms of appropriateness. %

\begin{table}[t]
    \footnotesize
    \centering
    \begin{tabular}{p{2cm}|p{5cm}}
         \textbf{Category} & \textbf{Definition}\\ \hline
         Conditionality(C) & the action to be performed is dependent upon a variable or unclear trigger \\
         Generalization(G) & the action or information types are vaguely abstracted with unclear conditions \\
         Modality(M)       & the likelihood or possibility of the action is vague or ambiguous\\
         Numerical(N)      & the action or information type has a vague quantifier \\
    \hline
    \end{tabular}%
    \caption{Explaining category of Vagueness~\cite{bhatia2016theory}}
    \label{tab:append_vague}
\end{table}

\vspace{1pt}
\noindent {\bf USA.} In the case of USA, we find that 6 out of 69 data practices discussed were incomplete. Out of these, 3 had missing recipients  and 3 had missing transmission principle. %
In terms of vagueness, we found that in USA ``modality'' was the most dominant category (50 data practices) followed by ``numeric'' (2 data practices).

\vspace{1pt}
\noindent {\bf UK.} In the case of UK, we find that 5 out of 37 data practices discussed were incomplete, all of which had missing recipients. %
Just like USA, also in UK, in terms of vagueness ``modality'' was the most dominant category (18 practices) followed by ``numeric'' (3 practices).

\vspace{1pt}
\noindent {\bf India.}  We find that 5 out of 55 data practices in India were incomplete. Out of these 2 had missing recipients and 3 had missing transmission principle %
In fact data practices in India almost never clearly mention the information being used or disclosed---they just mentioned `Personal Information' without clarifying what specific type of information is under consideration, unlike USA where examples of PHI were explicitly mentioned. In terms of Vagueness, the most dominant category was ``modality'' (23 practices), followed by ``numeric'' (2 practices), and ``generalization'' (1 practice).

\vspace{2pt}
\noindent{\bf Implications.} In all of the three countries for a few cases either transmission principle or recipient is missing. We show examples of such incomplete statements in Table \ref{tab:Incomplete} (Appendix \ref{append:incomplete_prac}). In those cases, the users are ill-informed about with whom the data is being shared or why it is being shared (or collected). In terms of vagueness, the most dominant category is ``modality''---in the case of USA and UK that is mostly because of such practices being directly copied from law which itself uses vague terms like `can', `may' etc. Our findings are in line with Kaur et al.~\cite{comprehensiveKeyWords} which presented a comprehensive keywords based analysis of online general privacy policies in Canada, USA, UK, and Germany. In the case of India, vagueness is observed because of the construct of the practices. We note that, such vagueness further denies a user the surety of knowing why exactly their information will be used (or shared). When the vagueness is `numeric' the user is also unsure about how much and how often their data will be used (or shared). Alarmingly, `Generalization', the most severe form of vagueness was encountered only in India---the data practice vaguely described what party will have the right to use and disclose user information (stating ``\textit{Generally} these contractors do not have any independent right to share this information, however certain contractors...will have rights to use and disclose the personal information...in accordance with their own privacy policies'').%

\subsection{Legal Compliance of Data Practices}

In the light of revealed incompleteness in some policies, we finally ask: are some of the data practices legally non-compliant? For USA and UK, we leveraged the clauses of the respective laws as they are quite well-defined and healthcare-specific (HIPAA and DPA). We find that semantically the data practices are very similar to law %
, whereas for India there were no specific use-cases/clauses to evaluate against. So for India we took help from legal experts (experienced in Indian IT law) to capture potential multiple interpretations of the law.

\vspace{1pt}
\subsubsection{ Legal compliance of USA privacy policies.} PART 164, SUBPART-E of HIPAA~\cite{hipaa-law} mentions several cases of usage and disclosures: {\bf [164.502]} Uses and disclosures of protected health information: General rules, {\bf [164.504]} Uses and disclosures: Organizational requirements, {\bf [164.506]} Uses and disclosures to carry out treatment, payment, or health care operations, {\bf [164.508]} Uses and disclosures for which an authorization is required, {\bf [164.510]} Uses and disclosures requiring an opportunity for the individual to agree or to object, {\bf [164.512]} Uses and disclosures for which an authorization or opportunity to agree or object is not required. Using these clauses we evaluated the compliance of the identified data practices (both complete and incomplete), and found all the practices to be compliant within the definitions of the clauses mentioned above.

\vspace{1pt}
\subsubsection{Legal compliance of UK privacy policies.} Schedule 1, Part 1, Paragraphs 2, 3, and 4 of DPA 2018~\cite{DPA-2018} mention explicitly the reasons for which data should be used/processed by a healthcare organization. These paragraphs are to be read with Art. 6 (Lawfulness of processing), Art. 9 (Processing of special categories of personal data) and Art. 11(1) of GDPR. Using the clauses in the above parts of DPA 2018 and GDPR we checked the compliance of the data practices. We found, like USA, all the UK data practices to be fully compliant with the law. %

\vspace{1mm}
\noindent \textbf{Understanding the reason for legal compliance in USA and UK data practices.} We noted (Section~\ref{sec:appropriateness}) that in USA and UK some data practices are sometimes directly copied from law. We surmised that this strategy of copying from law is a key reason behind the compliance of privacy policies in USA and UK.  However, one can argue that \textit{simply copying} a few keywords or even phrases from law might not ensure compliance---data practices are often contextual and open to interpretation. Such interpretation might require in-depth knowledge of the underlying laws (like HIPAA, GDPR, and DPA). So to capture if the strategy of copying from law in USA and UK preserves the meaning of the data practices identified in law, we conduct a WMD-based (Word Mover Distance) analysis between the data practices and the applicable legal clauses. WMD-based similarity captures not only the syntactic similarity (at word-level), but also semantic similarity (capturing context)~\cite{ma2019efficient}. Thus, we simply compute the WMD between the data practices and applicable legal clauses/provisions (the full results are in Appendix~\ref{app:wmdlaw}). %
We found that the minimum WMD for the USA and UK data practices are respectively 0.52 and 0.68---considerably lower than that of Indian data practices (0.90), signifying higher similarity between law and privacy policies for USA and UK. In fact, the WMD values for both USA and UK are statistically significantly lower than India (using U-test) with very large effect sizes: for India-USA, p\_value<0.0001 (Cohen's d: 2.07), and for India-UK, p\_value<0.0001 (Cohen's d: 3.07). %
These results indicate that for USA and UK, the data practices are copied from law with intact semantic meaning, which is not the case for India. Hence for Indian policies, we use an expert-in-the-loop strategy to evaluate the state of compliance of the practices from Indian Hospitals.

\subsubsection{Legal compliance of Indian data practices}\label{sec:India_Compliance}

\noindent Indian policies are governed by IT Rules, 2011 (more commonly known as SPDI Rules, 2011~\cite{IT_ACT_2011}). Rule 3 defines what attributes are classified as sensitive personal data or information, Rule 4 mandates \textit{body corporate} (Healthcare Organization) to provide policy for privacy and disclosure of information, Rule 5 pertains to the collection of information, Rule 6 deals with the disclosure and  Rule 7 deals with the transfer of information. %
Using this background, we conducted a small survey with four experts (formally studied and working with Indian IT law).

\vspace{1pt}

\noindent \textbf{Survey set up:} In the survey, our four experts were shown each of the data practices and asked to annotate the legal compliance of the data practice in line with IT Rules, 2011~\cite{IT_ACT_2011}. They were also required to mention the rules from these laws based on which they made their decision, along with the rationale for their judgement. Apart from judging compliance, the annotators were also asked to rate the vagueness in the practice and also identify the cause of vagueness. In total, the survey took 10 hours (spread across a week) and each of the experts were compensated INR 2000 (approx. \$26.62). For ethical reasons throughout the study, we did not collect any personally identifiable information from the legal experts. %

\vspace{1pt}

\noindent \textbf{Inter rater agreement:} After our pilot deployment with two experts, we found the inter-annotator agreement (Krippendorff's alpha) for the compliance for data practices is very low (0.018). Upon further investigation using rationale provided in survey, we identified three reasons for confusion: First, although the IT Rule 5(2)(a) mentioned data collection should be for \textit{lawful purpose}, it did not differentiate between lawful and unlawful purposes. Second, there were no explicit guidelines regarding non-Personally Identifiable information (PII).  Third, the experts sometimes \textit{implicitly} assumed consent was taken from the user while processing (or disclosing) the data. The reason is: privacy policy follows the `Notice and Consent' paradigm---a user agrees to all the practices being notified in the policy. We discussed with the experts in the pilot and resolved confusion by revisiting the interpretation of existing law in the context of healthcare with them. After the second round of deployment with the improved contextual understanding, the inter-annotator agreement was 0.78, with very small cases of disagreement, which the experts then discussed and resolved within themselves. The experts also found the data practices and the IT Rules, 2011 themselves to be far less descriptive than that of HIPAA.

\vspace{1pt}

\noindent \textbf{Results:} Our survey results indicated that out of 55 practices (derived from 48 policies), 12 practices (i.e., 21.8\%) were found to be unaligned with the IT Rules due to a total of six themes. Out of these 12, 10 practices were vague in accordance with IT Rules~\cite{IT_ACT_2011}. The primary cause of non-alignment for these 10 was in the construct of the data practice. We identified five themes after analyzing experts' responses: \textit{First,} absolute discretion is given to ``third parties'' to collect information (3 practices) (not in-line with Rule 6), \textit{Second,} not mentioning what explicit SPI is being used or disclosed (3 practices) (not in-line with Rule 4, 5 and 6), \textit{Third,} not clearly disclosing purpose for which information was collected (or disclosed) (3 practices) (not in line with Rule 4, 5 and 6), \textit{Fourth,} unclear description of third-parties to which information is disclosed (2 practices) (not in line with Rule 6 and 7), and \textit{Fifth,} upon updating policy, consent was not taken again (1 practice) (not in line with Rule 5). Remaining two practices were not vague but unaligned with the law under the sixth theme: absence of explicit consent and/or lawful purpose for which information was collected (not in line with Rule 5). %

\vspace{2mm}

\noindent \textbf{Validation using senior legal counsel:} %
Finally we took additional counsel from a senior legal professional with a survey %
to validate our findings. The legal professional has more than 10 years of experience in healthcare privacy (self-reported, independently verified by us) and is conversant in Indian IT laws as well as HIPAA and GDPR. The senior professional completely agreed with the findings (both compliance and non-compliance) with high inter-annotator agreement. The senior professional also pointed out case verdicts from Indian courts that ruled the non-alignments observed in the policies as potential violations of Rule 4 (AADHAR case~\cite{SC_aadhar}, TDB case~\cite{Kereal_HC_Case}), and Rule 5 (WhatsApp Policy case~\cite{cite:WhatsApp_Case}) of the IT Rules, 2011. Using the court verdicts, we surmised the following four non-alignments observed (from 48 policies) above as \textit{potential} violation: {\em not mentioning what explicit SPI is being used or disclosed} (found in 10 policies),  {\em not clearly disclosing the purpose for which information was collected (or disclosed)} (found in 7 policies), {\em upon updating policy, consent was not taken again} (found in 5 policies), and {\em absence of explicit consent and/or lawful purpose for which information was collected} (found in 1 policy). We present samples of data practices from each of these categories in Table \ref{tab:Samples_Violation} (Appendix \ref{appendix:samples_violation}). In fact, they mentioned---`\textit{The implementation of the SPDI Rules is woeful in some parts of the healthcare sector. Often, when we read policies of some hospitals, we do not see the mention of  a Grievance Officer, Option of withdrawal, how data retention is done.}'

\vspace{2mm}

\noindent \textbf{Impact of PDP Bill, 2022 in India.} Apart from IT Rules, 2011 which governs the current state of handling users' data, India is also seeing an upcoming data protection framework: The  Digital Personal Data Protection Act, 2022 (PDP Bill, 2022) \cite{PDP_2022}. However this framework is not enacted as a law yet. Thus, we simply discuss the data practices which would be in-line with the provisions of PDP Bill, 2022. This new data protection framework aims at affirming a Notice-and-Consent Paradigm. In Chapter 2 Section 5, the Bill quotes that processing can only be done for a lawful purpose. In Chapter 2 Section 6, the Bill discusses various clauses to ensure that the \textit{Data Fiduciary} (Hospitals) provides a clear, clean notice describing the type of personal data to be collected and the purpose for processing such information. In Chapter 2 Section 7, the Bill discusses the need for consent from \textit{Data Principal} and also highlights the right to withdraw consent without any obstruction in the services to be provided. In Chapter 2 Section 8, Bill discusses the notion of \textit{deemed consent}, in which it lists cases where the \textit{Data Principal} is deemed to have given consent. In Chapter 2 Section 9, the Bill discusses the obligations on \textit{Data Fiduciary} to store/collect/process the sensitive data. It also includes a clause to appoint a Data Protection Officer like the IT Rules, 2011. Additionally in Chapter 3, it clearly discusses the rights of the \textit{Data Principal} in a manner similar to GDPR. In comparison to the current framework of IT Rules 2011, this new data protection framework does reduce some vagueness. However, the data practices identified in this work in violation with IT Rules 2011, also remain in violation with the PDP Bill, 2022. Thus our findings remain valid even in the context of the proposed PDP Bill, 2022 in India.

\subsubsection{Implications for better compliance}  Interestingly, practices in USA and UK are in some sense \textit{copied} from the relevant parts of the law itself, ensuring compliance. On the contrary, privacy policy creators in India rely on more general IT Rules, 2011 which our experts found to be less concise and descriptive than HIPAA. In India, we observed that non-alignment (and vagueness) often came from improper drafting of sentences in policies and resulted in potential violations with the law. However, in absence of a mixed-method auditing workflow as proposed in this work, it is difficult for policy makers to identify potential violations and potentially explain the law in a uniform and concise manner. Similarly, without such a mixed-method workflow, it might be difficult for even hospitals (that might have misinterpreted the law in absence of concise use cases) to find and fix these violating data practices at scale and handle the data of their patients more responsibly.

\if{0}
Overall, we observed that main source of vagueness comes from the definition in laws.
\subsection{\bf Appropriateness and Legal Compliance -- UK and USA.} 
\subsubsection*{USA}
\if{0}
\begin{table}[h]
    \centering
    \resizebox{\columnwidth}{!}{
    \begin{tabular}{p{1.5cm}|p{4cm}|p{2cm}}
    Case                            & Example   & Explanation \\ \hline
    Missing Transmission Principle & We and our Affiliates and Service Providers may collect personal information about you on the Website and from other sources, including commercially available sources. & The purpose for which such information is used is not discussed\\ \hline
    Missing Recipient & If you are present and able to agree or object then we may only disclose your PHI if you don't object after you have been informed of your opportunity to do so (although such agreement may be reasonably inferred from the circumstances).  & To whom the information will be disclosed is not mentioned \\ \hline
    \end{tabular}}
    \caption{Incomplete examples as observed for USA data practices}
    \label{tab:USA_Incomplete}
\end{table}
\fi

\textbf{Appropriateness.} In case of USA we find that 6 out of 69 data practices discussed were incomplete. Out of these 3 had missing recipients \noteng{I think the best would be giving these examples as footnote, also it is not clear what is missing, that needs to be specified}\notegb{We are mentioning it is a case of missing recipient} and 3 had missing transmission principle (See Table~\ref{tab:Incomplete}). In terms of vagueness, we found that ``modality'' was the most dominant category occurring in 50 flows, followed by ``numeric'' occurring in 2 flows and ``generalization'' in 1 flow.

\noindent \textbf{Compliance. }PART 164, SUBPART-E of HIPAA mentions several cases of usage and disclosures: {\bf [164.502]} Uses and disclosures of protected health information: General rules, {\bf [164.504]} Uses and disclosures: Organizational requirements, {\bf [164.506]} Uses and disclosures to carry out treatment, payment, or health care operations, {\bf [164.508]} Uses and disclosures for which an authorization is required, {\bf [164.510]} Uses and disclosures requiring an opportunity for the individual to agree or to object, {\bf [164.512]} Uses and disclosures for which an authorization or opportunity to agree or object is not required. Using these clauses we evaluated the compliance of the data practices that we identified, and found all the statements to be compliant within the definitions of the clauses mentioned above.

\subsubsection*{UK}
\if{0}
\begin{table}[h]
    \centering
    \resizebox{\columnwidth}{!}{
    \begin{tabular}{p{1.5cm}|p{4cm}|p{2cm}}
    Case                            & Example   & Explanation \\ \hline
    Missing Recipient & Any information used or shared during the Covid-19 outbreak will be limited to the period of the outbreak unless there is another legal basis to use the data. & Clear description of the party recieving the information is absent\\ \hline
    \end{tabular}}
    \caption{Incomplete examples as observed for UK data practices}
    \label{tab:UK_Incomplete}
\end{table}
\fi
\noindent \textbf{Appropriateness.} In case of UK we find that 5 out of 37 data practices discussed were incomplete, all of which had missing recipients (See Table~\ref{tab:UK_Incomplete}). In terms of vagueness, we found that ``modality'' was the most dominant category occurring in 18 flows, followed by ``numeric'' occurring in 3 flows and ``generalization'' in 1 flow.

\noindent \textbf{Compliance.} Schedule 1, Part 1, Paragraphs 2, 3, and 4 of DPA 2018 mention explicitly the reasons for which data might be used/processed by a healthcare organization. These paragraphs are to be read with Art. 6 (Lawfulness of processing), Art. 9 (Processing of special categories of personal data) and Art. 11(1) of GDPR. Using the clauses in the above parts of DPA 2018 and GDPR we checked the compliance of the data practices. We found all the statements to be fully compliant with the law.

\subsection{\bf Appropriateness and Legal Compliance -- India.}
\if{0}
\begin{table}[]
    \centering
    \resizebox{\columnwidth}{!}{
    \begin{tabular}{p{1.5cm}|p{4cm}|p{2cm}}
    Case                            & Example   & Explanation \\ \hline
    Missing Transmission Principle  &    We may collect any and all personal information you provide to us, like your name, mobile phone number, email address . . . feedback, and any other information you provide us. & The purpose for which information is collected is not mentioned in the text \\ \hline
    Missing Recipients & You acknowledge that some countries where we may transfer Your Personal Information may not have data protection laws .... SIMS will place contractual obligations on the transferee which will oblige the transferee to adhere to the provisions of this Privacy Policy. & A clear description of entities to whom the information is disclosed is absent in the statement.\\ \hline
    \end{tabular}}
    \caption{Incomplete examples as observed for Indian data practices}
    \label{tab:IN_Incomplete}
\end{table}
\fi
\noindent \textbf{Appropriateness.} In case of India we find that 5 out of 56 data practices discussed were incomplete. Out of these 2 had missing recipients and 3 had missing transmission principle (Samples of such cases are shown in Table~\ref{tab:Incomplete}). We also observed that statements in India almost never clearly mention the information being used or disclosed. The practices just mentioned `Personal Information' without clarifying what specific type of information is under consideration, unlike USA where examples of PHI were observed in sentences. In terms of Vagueness, we found that ``modality'' was the most dominant category occurring in 23 flows, followed by ``numeric'' occurring in 2 flows, and ``conditionality'' and ``generalization'' in 1 and 1 flow each. 

\noindent \textbf{Legal Compliance.} In case of India however things are not straight-forward \noteng{what does not straight-forward mean??} \notegb{Basically the law is not very descriptive as HIPAA, so we cannot do simple lookup and say this is compliant and this is not.}since there is no healthcare specific law in place. From the text of the privacy policies we found that these policies are governed by IT Rules 2011 (more commonly knows as SPDI Rules, 2011). Rule 3 of this law defines what attributes are classified as sensitive personal data or Information, Rule 4 of the law mandates \textit{body corporate}(Healthcare Organization) to provide policy for privacy and disclosure of information, Rule 5 pertains to Collection of Information, Rule 6 and 7 discuss under what circumstances can the information be disclosed or transferred. In case of Indian Law, as opposed to HIPAA, there are not well defined use cases in the body of the law using which one can identify legal compliance. \noteng{So is the alignment not possible to do by you} \notegb{Yes and that's why we take legal expertise.} So to establish the alignment of the data practices we conducted an annotation task with the help of 4 law undergraduates. \noteng{undergraduates or graduates, were they conversant with Indian IT law??}\notegb{The law students are 3rd year LLBs, hence I have written undergraduates, and yeah they are conversant with IT Law} They were shown the data practices and  asked to annotate the legal compliance of the data practice in line with IT Rules 2011~\cite{IT_ACT_2011}. They were also required to mention the rules from these laws based on which they made their decision, along with the rationale for their judgement. Apart from judging compliance, the annotators were also asked to rate the vagueness of the statement and also identify the cause of vagueness.

After first round of annotations, we computed the inter-annotator agreement. We found it to be 0.018 which was very low. The annotators then sat together and discussed the causes of conflicts, looking into the rationale that was used for making the judgement. We identified the following main reasons behind the  confusion: (a). Reading the data practices as a standalone statement might not be sufficient, as looking it together  the privacy policy might give a better overall picture, b) The IT Rule 5(2)(a), very loosely states that information can be collected for a lawful purpose linked with the functioning of business entity, but it doesn't mention any specific guidelines to differentiate lawful purposes from unlawful purposes, thus causing the ambiguity in the alignment of the statement, (c). Whenever these practices talked about the aggregated information or had mention of non-PII, there were disagreements since such information is not covered in the ambit of any rules of IT Rules. We also found that \new{there was inherent bias of consent} \noteng{what is this? not clear}\notegb{Even though the clause does not mention that we will take consent the annotators assumed that consent is there. This is because of the fact that being present in privacy policy means consent should be taken.} being taken since providing the privacy policy itself constitutes as an act of consent because of paradigm of `Notice and Choice' by design. \new{The annotators also seemed to connect different statements to bring out the judgements in case there were some discrepancies in the statements.} After the first round, the annotators then discussed and resolved such confusions by looking into the interpretation of the law. As for point (a) they were instructed to consider the statements as standalone, complete data practice since they represented a complete information flow. 

After conducting the second round of annotations, the inter-annotator agreement went up to 0.78, with very small cases of disagreement, which the annotators then discussed and resolved within themselves. The annotators also found the data practices and the law itself to be far less descriptive than that of HIPAA. Overall, we observed that out of 56 practices, 15 were found to be not in compliance. We find that out of these 15 statements, 13 statements were vague in accordance with IT Rules. We find the primary cause of such vagueness and non-compliance to be in the construct of the statement. We identified following themes after analyzing annotators responses: (a). absolute discretion is given to "third parties" to collect information, (b). not mentioning what explicit SPI is being used or disclosed, (c). does not disclose the purpose for which information was collected (or disclosed), (d). upon updating policy, consent was not taken again, and (e). information being used for the purpose it was not collected. The remaining two were not vague however the causes we found was (a). absence of explicit consent and lawful purpose for which information was collected, and (b). absence of clear description of the third parties to which personal information was shared. We also found 4 statements to be vague in accordance with law, yet were compliant. The phrases that we found to be problematic are: `including but not limited to', `information that might not be required to carry out the task', and `assumed consent'. There were also 4 cases where law itself was not applicable since the statement under consideration discusses about collecting, using, or disclosing Non-PII (which is not covered in ambit of IT Rules 2011). 

\subsection{Implications}
Overall, we observed that main source of vagueness comes from the definition in laws. We find that practices from US and UK are in some sense copied from the relevant parts of the law itself, which in turn uses such vague terms like \textit{may} thereby leading to such higher fractions of Modality vagueness. On contrary, policy makers in India with absence of any healthcare-specific law rely on more general IT Rules. In India, we observed that  vagueness and non-compliance often came from improper drafting of sentences. Annotators also mentioned that certain purposes for which information was collected does not seem to be attached with the functioning of hospitals. This in turn make things difficult as opposed to HIPAA which clearly mentions the use-cases for lawfully handling PHI, IT Rules just mention that 'purpose should be legal and connected to functioning of Body Corporate', without describing clearly what such purposes should be and what to discard as unlawful.

\if{0}
\subsection{\bf Appropriateness of Data Practices.} 
 
Using the theory of Contextual Integrity, we identify how complete the practices are in terms of the parameters present in the practices. The theory of Contextual Integrity identifies a data practice as an information flow constituted of five parameters: Subject of the information, attributes covered in the information flow, Sender of the information, Recipient of the information and the transmission principle followed for handling the information. 

Following the lines of~\cite{shvartzshnaider2019going}, we evaluate the data practices to identify incompleteness in these practices. A practice is incomplete if it has any of these parameter(s) is missing, leading to following categories: missing Recipient, missing sender, missing attribute and missing transmission principle. We also do a follow-up to identify how many of these data practices are vague using the vagueness taxonomy~\cite{bhatia2016theory}. We categorize the practice with mention of keywords from one or more combinations of vague terms thus classifying into 4 categories: Conditionality(C), Generalization(G), Modality(M), and Numeric Quantifier (N). We now discuss our observations for each country.
\vspace{1mm}

\noindent \textbf{USA.} In case of USA we find that 10 out of 69 (14.4\%) data practices discussed were incomplete. Out of these 8 had missing Recipients (\textit{If we agree to the requested restriction, we may not use or disclose your PHI in violation of that restriction unless it is needed to provide emergency treatment.}) and 2 had missing transmission principle (\textit{We and our Affiliates and Service Providers may collect personal information about you on the Website and from other sources, including commercially available sources.}). In terms of Vagueness, we found that `Modality' was the most dominant category occurring in 40 flows, followed by `Conditional' occurring in 17 flows, and Numeric and Generalization in 4 and 3 flows each.
\vspace{1mm}

\noindent \textbf{UK.} In case of UK we find that 6 out of 37 (16.2\%) data practices discussed were incomplete, all of which had missing Recipients (\textit{Any information used or shared during the Covid-19 outbreak will be limited to the period of the outbreak unless there is another legal basis to use the data.}) In terms of Vagueness, we found that `Modality' was the most dominant category occurring in 18 flows, followed by `Conditional' occurring in 7 flows, and Numeric and Generalization in 3 and 1 flows each.
\vspace{1mm}

\noindent \textbf{India.} In case of India we find that 12 out of 59 (20.3\%) data practices discussed were incomplete. Out of these 5 had missing Recipients (\textit{Institute will only disclose your personal information when it, in its sole discretion, deems it necessary in order to protect its rights or the rights of others, to prevent harm to persons or property, to fight fraud and credit risk, or to enforce or apply the Terms of Use.}),  4 had missing transmission principle (\textit{Users may be asked for, as appropriate, name, email address, phone number. Users may, however, visit our Site anonymously.}), and 3 had missing attributes (\textit{We or our Partners may also use such information in an aggregated or non-personally identifiable form for research, statistical analysis and business intelligence purposes}). In terms of Vagueness, we found that `Modality' was the most dominant category occurring in 24 flows, followed by `Conditional' occurring in 7 flows, and Numeric and Generalization in 4 and 3 flows each.

Overall, we observed that main source of vagueness comes from the definition in laws. We find that practices from US and UK are in some sense copied from the relevant parts of the law itself, which in turn uses such vague terms like may, might etc thereby leading to such higher fractions of Modality vagueness. On contrary, policy makers in India with absence of any healthcare-specific law rely on more general IT Rules.

\subsection{\bf Legal Compliance of US and UK Practices}
The main idea here is to establish the practices aligning with the law for any country({\it compliant}), and those deviating from normal course of law ({\it non-compliant}). This is achieved by checking for clauses in the laws and checking for their mention in the policy. 

\noindent{\bf USA.} PART 164, SUBPART-E of HIPAA mentions several cases of usage and disclosures: {\bf [164.502]} Uses and disclosures of protected health information: General rules, {\bf [164.504]} Uses and disclosures: Organizational requirements, {\bf [164.506]} Uses and disclosures to carry out treatment, payment, or health care operations, {\bf [164.508]} Uses and disclosures for which an authorization is required, {\bf [164.510]} Uses and disclosures requiring an opportunity for the individual to agree or to object, {\bf [164.512]} Uses and disclosures for which an authorization or opportunity to agree or object is not required. Using these clauses we evaluated the compliance of the data practices that we identified, and found all the statements to be compliant within the definitions of the clauses mentioned above.

\subsection{\bf Legal Compliance of Indian Practices}
\noindent {\bf India.} In case of India however things are not straight-forward since there is no healthcare specific law in place. From the text of the privacy policies we found that these policies are governed by IT Rules 2011 (more commonly knows as SPDI Rules, 2011). Rule 3 of this law defines what attributes are classified as sensitive personal data or Information, Rule 4 of the law mandates \textit{body corporate}(Healthcare Organization) to provide policy for privacy and disclosure of information, Rule 5 pertains to Collection of Information, Rule 6 and 7 discuss under what circumstances can the information be disclosed or transferred. In case of Indian Law, as opposed to HIPAA, there are not well defined use cases in the body of the law using which one can identify legal compliance. So to establish the alignment of the data practices we conducted an annotation task with the help of 4 law undergraduates. They were shown the data practices and were asked to annotate the legal compliance of the data practice in line with IT Rules 2011~\cite{IT_ACT_2011}. They were also required to mention the rules from these laws based on which they made their decision, along with the rationale for their judgement.

The annotators were divided into 2 groups of 2 people each. They were given 29 and 30 data practices chosen randomly to annotate. After first round of annotations, we computed the inter-annotator agreement. We found it to be 0.018 which was very low. The annotators then sat together and discussed the causes of conflicts, looking into the rationale that was used for making the judgement. We identified the following main reasons for confusions: (a).  Reading the data practices as a standalone statement might not be sufficient, as looking at it together with the privacy policy might give a better overall picture, (b). The IT Rule 5(2)(a), very loosely states that information can be collected for a lawful purpose linked with the functioning of business entity, but it doesn't mention any specific guidelines to differentiate lawful purposes from unlawful purposes, thus causing the ambiguity in the alignment of the statement, (c). Whenever these practices talked about the aggregated information or had mention of non-PII, there were disagreements since such information is not covered in the ambit of any rules of IT Rules. We also found that there was an inherent bias of consent \noteng{not sure what do you mean by inherent bias??} being taken since providing the Privacy Policy itself constitutes as an act of consent because of paradigm of `Notice and Choice' by design. \noteb{The annotators also seemed to connect statements to bring out the judgements in case there were discrepancies.} After the first round, the annotators then discussed and resolved such confusions by looking into the interpretation of the law. As for point (a) they were instructed to consider the statements as standalone, \noteb{complete data practice} since they represented a complete information flow. 

After conducting the second round of annotations, the inter-annotator agreement went up to 0.78, with very small cases of disagreement, which the annotators then discussed and resolved within themselves. The annotators also found the data practices and the law itself to be far less restrictive than that of HIPAA. Overall, we observed that out of 58 practices, only 3 were found to be not in compliance. In one statement, it was due to mention of disclosure without any prior notice ({\it You agree and confirm that we do not rent, sell, or share Personal Information about you with other people (save with your consent) or non-affiliated companies except to transfer/disclose Personal Information about you to trusted partners, may or may not be for gain, to promote certain products/services for commercial purposes, \textbf{without any prior notice to you.}}), in one case it was due to the discrepancy in the formation of the statement ({\it The hospital may collect and process relevant information about you via forms on the website (www.newmillenniumhospital.in ); in addition to forms given as hard copies form or by any other means. This information is stored with us for varied purposes which \textbf{may or may not be explicitly shared at the outset}}), and lastly it was because of mention of phrase `including but not limited to' ({\it New Millennium Multispecialty Hospital will collect, record, store and use your personal data for the following specified purposes including but not limited to:- Providing you with information ... Carrying out our obligations arising from any transactions or agreements ... Ensuring that our website is presented in the most effective manner for you and your computer}), as it is not in line with Rule 5(3) and 5(5) which states that users must be made aware of all the purposes for which information is collected and it should be used only for the purpose it was collected.  
\fi
\fi
\section{Limitations}
We identify two limitations of our study. First, the number of unique privacy policies is comparatively low when compared to the total number of hospitals in a country. However, we have tried our best and used established methods of identifying and analyzing privacy policies. Support from a government or regulators will possibly improve our dataset. However, even with our best-effort analysis, we identified interesting findings including legal non-compliance. Second, our segmentation-summarization based approach might miss certain important data practices(or segments) which could not be identified correctly either in the segmentation or summarization phase. Thus, we only identified relatively clearly mentioned data practices. In spite of the limitation, our audit uncovered problems even with these relatively clearly mentioned data practices, underlining the value of our research.

\section{Conclusion and Future Work}

The significance of our work lies in building up a novel workflow to audit the privacy policies of the healthcare organizations of a country or a region. The application of this workflow to the three countries brings out interesting insights---although we find that all the three countries have some incompleteness in data practices, we find that the policies in India are much less specific than the other two countries. We conclude (after consultation with legal experts) that a primary reason behind this is the absence of concise use cases and descriptive rules in the Information Technology Rules, 2011 \cite{IT_ACT_2011} for regulating health data practices. Overall, during audit we found \textit{unclear} statements, vague (and non-compliant) data practices in Indian healthcare organizations. However, there are subtle but significant differences even between UK and USA privacy policies---in USA there is much more stress on users' consent for usage and disclosure but less stress on the mode of data collection and retention; these may have arisen due to differences in the law of the two  countries. 
We strongly believe upon presenting this useful mixed-method auditing workflow, a natural key future avenue of exploration is to provide concrete data-driven suggestions to the healthcare organization to create better privacy policies as well as help regulators in removing ambiguity from existing legal policies. To that end, the present work would act as a solid foundation. %

\bibliographystyle{plain}
\bibliography{audit-health}

\begin{thebibliography}{10}

\bibitem{ahmad-etal-2020-policyqa}
Wasi Ahmad, Jianfeng Chi, Yuan Tian, and Kai-Wei Chang.
\newblock {P}olicy{QA}: A reading comprehension dataset for privacy policies.
\newblock In {\em EMNLP Findings}, pages 743--749, 2020.
\newblock \url{https://aclanthology.org/2020.findings-emnlp.66}.

\bibitem{amos2021privacy}
Ryan Amos, Gunes Acar, Elena Lucherini, Mihir Kshirsagar, Arvind Narayanan, and
  Jonathan Mayer.
\newblock Privacy policies over time: Curation and analysis of a
  million-document dataset.
\newblock In {\em Proceedings of the Web Conference 2021}, pages 2165--2176,
  2021.

\bibitem{ddp-kewords}
A.I. Anton, J.B. Earp, and A.~Reese.
\newblock Analyzing website privacy requirements using a privacy goal taxonomy.
\newblock In {\em IEEE RE}, pages 23--31, 2002.
\newblock \url{https://ieeexplore.ieee.org/document/1048502}.

\bibitem{apthorpe2018discovering}
Noah Apthorpe, Yan Shvartzshnaider, Arunesh Mathur, Dillon Reisman, and Nick
  Feamster.
\newblock {Discovering smart home internet of things privacy norms using
  contextual integrity}.
\newblock {\em ACM IMWUT}, pages 1--23, 2018.
\newblock \url{https://dl.acm.org/doi/10.1145/3214262}.

\bibitem{apthorpe2019evaluating}
Noah Apthorpe, Sarah Varghese, and Nick Feamster.
\newblock {Evaluating the Contextual Integrity of Privacy Regulation: Parents'
  IoT Toy Privacy Norms Versus COPPA}.
\newblock In {\em Usenix Security}, pages 123--140, 2019.
\newblock
  \url{https://www.usenix.org/conference/usenixsecurity19/presentation/apthorpe}.

\bibitem{bannihatti2020finding}
Vinayshekhar Bannihatti~Kumar, Roger Iyengar, Namita Nisal, Yuanyuan Feng, Hana
  Habib, Peter Story, Sushain Cherivirala, Margaret Hagan, Lorrie Cranor,
  Shomir Wilson, et~al.
\newblock Finding a choice in a haystack: automatic extraction of opt-out
  statements from privacy policy text.
\newblock In {\em ACM WWW'20}, pages 1943--1954, 2020.
\newblock \url{https://doi.org/10.1145/3366423.3380262}.

\bibitem{barbaresi-2021-trafilatura}
Adrien Barbaresi.
\newblock {Trafilatura: A Web Scraping Library and Command-Line Tool for Text
  Discovery and Extraction}.
\newblock In {\em ACL-IJCNLP}, pages 122--131, 2021.
\newblock \url{https://aclanthology.org/2021.acl-demo.15}.

\bibitem{GDPRContentAnalysis}
Nastaran Bateni, Jasmin Kaur, Rozita Dara, and Fei Song.
\newblock Content analysis of privacy policies before and after gdpr.
\newblock In {\em {2022 19th Annual International Conference on Privacy,
  Security \& Trust (PST)}}, pages 1--9, 2022.

\bibitem{bhatia2016theory}
Jaspreet Bhatia, Travis~D Breaux, Joel~R Reidenberg, and Thomas~B Norton.
\newblock A theory of vagueness and privacy risk perception.
\newblock In {\em IEEE (RE) Conference}, pages 26--35, 2016.
\newblock \url{https://ieeexplore.ieee.org/document/7765508}.

\bibitem{Louvain-2008}
Vincent~D Blondel, Jean-Loup Guillaume, Renaud Lambiotte, and Etienne Lefebvre.
\newblock Fast unfolding of communities in large networks.
\newblock {\em Journal of Statistical Mechanics: Theory and Experiment},
  2008(10):P10008, 2008.
\newblock \url{http://dx.doi.org/10.1088/1742-5468/2008/10/P10008}.

\bibitem{bojanowski2016enriching}
Piotr Bojanowski, Edouard Grave, Armand Joulin, and Tomas Mikolov.
\newblock Enriching word vectors with subword information.
\newblock {\em arXiv preprint arXiv:1607.04606}, 2016.

\bibitem{cite:MMR}
Jaime Carbonell and Jade Goldstein.
\newblock {The use of MMR, diversity-based reranking for reordering documents
  and producing summaries}.
\newblock In {\em ACM SIGIR}, pages 335--336, 1998.
\newblock \url{https://doi.org/10.1145/290941.291025}.

\bibitem{hipaa-law}
{Center for Disease Control and Prevention}.
\newblock {Health Insurance Portability and Accountability Act of 1996
  (HIPAA)}.
\newblock \url{https://www.cdc.gov/phlp/publications/topic/hipaa.html}, 1996.

\bibitem{cite:WhatsApp_Case}
{Competition Commission of India}.
\newblock {Updated Terms Of Service And ... vs Whatsapp Llc}.
\newblock \url{https://indiankanoon.org/doc/99533020/}, 2022.

\bibitem{gao-etal-2020-supert}
Yang Gao, Wei Zhao, and Steffen Eger.
\newblock {SUPERT}: Towards new frontiers in unsupervised evaluation metrics
  for multi-document summarization.
\newblock In {\em ACL}, pages 1347--1354, 2020.
\newblock \url{https://aclanthology.org/2020.acl-main.124}.

\bibitem{goncalves-2015-qualityoflife}
Joaquim Gon{\c c}alves, Br{\'\i}gida~M{\'o}nica Faria, Lu{\'\i}s~Paulo Reis,
  Victor Carvalho, and {\'A}lvaro Rocha.
\newblock {Data mining and electronic devices applied to quality of life
  related to health data}.
\newblock In {\em CISTI}, pages 1--4, 2015.
\newblock \url{https://ieeexplore.ieee.org/document/7170627}.

\bibitem{gopinathautomatic}
Abhijith Athreya~Mysore Gopinath, Vinayshekhar~Bannihatti Kumar, Shomir Wilson,
  and Norman Sadeh.
\newblock {Automatic Section Title Generation to Improve the Readability of
  Privacy Policies}.
\newblock In {\em USENIX SOUPS}, 2020.
\newblock
  \url{https://www.usenix.org/conference/soups2020/presentation/gopinath}.

\bibitem{gopinath2018supervised}
Abhijith Athreya~Mysore Gopinath, Shomir Wilson, and Norman Sadeh.
\newblock Supervised and unsupervised methods for robust separation of section
  titles and prose text in web documents.
\newblock In {\em EMNLP}, pages 850--855, 2018.
\newblock \url{https://aclanthology.org/D18-1099/}.

\bibitem{comprehensiveKeyWords}
Jasmin Kaur, Rozita~A. Dara, Charlie Obimbo, Fei Song, and Karen Menard.
\newblock A comprehensive keyword analysis of online privacy policies.
\newblock {\em Information Security Journal: A Global Perspective},
  27(5-6):260--275, 2018.

\bibitem{kenneally2012menlo}
Erin Kenneally and David Dittrich.
\newblock The menlo report: Ethical principles guiding information and
  communication technology research.
\newblock {\em Available at SSRN 2445102}, 2012.
\newblock \url{https://papers.ssrn.com/sol3/papers.cfm?abstract_id=2445102}.

\bibitem{Kereal_HC_Case}
{Kerala High Court}.
\newblock Kerela {H}igh {C}ourt: {S}uo {M}otu vs {T}ravancore {D}evaswom
  {B}oard - {T}{B}{D}.
\newblock \url{https://indiankanoon.org/doc/142754318/}, 2022.

\bibitem{keymanesh_toward_2020}
Moniba Keymanesh, Micha Elsner, and Srinivasan Parthasarathy.
\newblock Toward {Domain}-{Guided} {Controllable} {Summarization} of {Privacy}
  {Policies}.
\newblock In {\em NLLP}, pages 18--24, 2020.
\newblock \url{http://ceur-ws.org/Vol-2645/\#paper3}.

\bibitem{kusner2015word}
Matt Kusner, Yu~Sun, Nicholas Kolkin, and Kilian Weinberger.
\newblock From word embeddings to document distances.
\newblock In {\em ICML}, pages 957--966, 2015.
\newblock \url{https://proceedings.mlr.press/v37/kusnerb15.html}.

\bibitem{DPA-2018}
UK~Legislation.
\newblock {Data Protection Act of 2018 (DPA18)}.
\newblock \url{https://www.legislation.gov.uk/ukpga/2018/12/contents/enacted},
  2018.

\bibitem{leskovec2010empirical}
Jure Leskovec, Kevin~J Lang, and Michael Mahoney.
\newblock Empirical comparison of algorithms for network community detection.
\newblock In {\em WWW}, pages 631--640, 2010.
\newblock \url{https://dl.acm.org/doi/10.1145/1772690.1772755}.

\bibitem{liao2020measuring}
Song Liao, Christin Wilson, Long Cheng, Hongxin Hu, and Huixing Deng.
\newblock Measuring the effectiveness of privacy policies for voice assistant
  applications.
\newblock In {\em ACM ACSAC}, page 856–869, 2020.
\newblock \url{https://doi.org/10.1145/3427228.3427250}.

\bibitem{libert2018automated}
Timothy Libert.
\newblock An automated approach to auditing disclosure of third-party data
  collection in website privacy policies.
\newblock In {\em WWW}, pages 207--216, 2018.
\newblock \url{https://doi.org/10.1145/3178876.3186087}.

\bibitem{liu2014step}
Fei Liu, Rohan Ramanath, Norman Sadeh, and Noah~A Smith.
\newblock {A step towards usable privacy policy: Automatic alignment of privacy
  statements}.
\newblock In {\em ACL-COLING}, pages 884--894, 2014.
\newblock \url{https://aclanthology.org/C14-1084/}.

\bibitem{louis2013automatically}
Annie Louis and Ani Nenkova.
\newblock Automatically assessing machine summary content without a gold
  standard.
\newblock {\em Computational Linguistics}, 39(2):267--300, 2013.
\newblock \url{https://aclanthology.org/J13-2002/}.

\bibitem{ma2019efficient}
Yinglong Ma, Peng Zhang, and Jiangang Ma.
\newblock An ontology driven knowledge block summarization approach for chinese
  judgment document classification.
\newblock {\em IEEE Access}, 6:71327--71338, 2018.

\bibitem{massey-analysis}
Aaron Massey, Jacob Eisenstein, Annie Antón, and Peter Swire.
\newblock Automated text mining for requirements analysis of policy documents.
\newblock In {\em IEEE RE 2013}, pages 4--13, 2013.
\newblock \url{https://ieeexplore.ieee.org/document/6636700}.

\bibitem{IN-dir}
Medindia.
\newblock {Directory for Indian Hospitals}.
\newblock
  \url{https://www.medindia.net/patients/hospital_search/hospital_list.asp},
  2001.

\bibitem{10.1007/978-3-642-39371-6_8}
Gabriele Meiselwitz.
\newblock Readability assessment of policies and procedures of social
  networking sites.
\newblock In {\em OCSC}, pages 67--75, 2013.
\newblock \url{https://link.springer.com/chapter/10.1007/978-3-642-39371-6_8}.

\bibitem{MM_Deletion_Privacy}
Mohsen Minaei, Mainack Mondal, and Aniket Kate.
\newblock {"My Friend Wanted to Talk About It and {I} Didn't": Understanding
  Perceptions of Deletion Privacy in Social Platforms}.
\newblock {\em CoRR}, 2020.
\newblock \url{https://arxiv.org/abs/2008.11317}.

\bibitem{IT_ACT_2011}
{Ministry of communications \& Information technology}.
\newblock {Information Technology (Reasonable security practices and procedures
  and sensitive personal data or information) Rules, 2011}, 2011.
\newblock
  \url{https://www.meity.gov.in/writereaddata/files/GSR313E_10511(1)_0.pdf}.

\bibitem{PDP_2022}
Government of~India Ministry~of Electronics \& Information Technology~\.
\newblock The digital personal data protection bill, 2022, 2018.

\bibitem{DISHA_ACT_2018}
{Ministry of Health and Family Welfare}.
\newblock {Digital lnformation Security in Healthcare, act (DISHA)}, 2018.
\newblock \url{https://www.nhp.gov.in/NHPfiles/R_4179_1521627488625_0.pdf}.

\bibitem{nissenbaum2004privacy}
Helen Nissenbaum.
\newblock Privacy as contextual integrity.
\newblock {\em Wash. L. Rev.}, 79:119, 2004.
\newblock \url{https://digitalcommons.law.uw.edu/wlr/vol79/iss1/10/}.

\bibitem{Google-dir}
Nv7-Github.
\newblock {GoogleSearch}.
\newblock \url{https://github.com/Nv7-GitHub/googlesearch}, 2020.

\bibitem{oltramari2018privonto}
Alessandro Oltramari, Dhivya Piraviperumal, Florian Schaub, Shomir Wilson,
  Sushain Cherivirala, Thomas~B Norton, N~Cameron Russell, Peter Story, Joel
  Reidenberg, and Norman Sadeh.
\newblock {PrivOnto: A semantic framework for the analysis of privacy
  policies}.
\newblock {\em Semantic Web}, 9(2), 2018.
\newblock \url{https://dl.acm.org/doi/abs/10.3233/SW-170283}.

\bibitem{cite:pagerank}
Lawrence Page, Sergey Brin, Rajeev Motwani, and Terry Winograd.
\newblock The pagerank citation ranking: Bringing order to the web.
\newblock Technical report, Stanford InfoLab, 1999.
\newblock \url{http://ilpubs.stanford.edu:8090/422/}.

\bibitem{gdpr-article35-2021}
European Parliament and Council of~the European~Union.
\newblock {GDPR}.
\newblock \url{https://eur-lex.europa.eu/eli/reg/2016/679/oj}, 2016.

\bibitem{ravichander_breaking_2021}
Abhilasha Ravichander, Alan~W Black, Thomas Norton, Shomir Wilson, and Norman
  Sadeh.
\newblock Breaking {Down} {Walls} of {Text}: {How} {Can} {NLP} {Benefit}
  {Consumer} {Privacy}?
\newblock In {\em ACL-IJCNLP}, pages 4125--4140, 2021.
\newblock \url{https://aclanthology.org/2021.acl-long.319}.

\bibitem{ravichander-etal-2019-question}
Abhilasha Ravichander, Alan~W Black, Shomir Wilson, Thomas Norton, and Norman
  Sadeh.
\newblock Question answering for privacy policies: Combining computational and
  legal perspectives.
\newblock In {\em EMNLP-IJCNLP}, pages 4947--4958, 2019.
\newblock \url{https://aclanthology.org/D19-1500}.

\bibitem{sathyendra_helping_nodate}
Kanthashree~Mysore Sathyendra, Abhilasha Ravichander, Peter~Garth Story, Alan~W
  Black, and Norman Sadeh.
\newblock Helping users understand privacy notices with automated query
  answering functionality: An exploratory study.
\newblock Technical report, Technical report, Carnegie Mellon University, 2017.
\newblock \url{https://usableprivacy.org/static/files/CMU-ISR-17-114R.pdf}.

\bibitem{sathyendra2017identifying}
Kanthashree~Mysore Sathyendra, Shomir Wilson, Florian Schaub, Sebastian
  Zimmeck, and Norman Sadeh.
\newblock {Identifying the Provision of Choices in Privacy Policy Text}.
\newblock In {\em ACL-EMNLP}, pages 2774--2779, 2017.
\newblock \url{https://aclanthology.org/D17-1294/}.

\bibitem{shvartzshnaider2019going}
Yan Shvartzshnaider, Noah Apthorpe, Nick Feamster, and Helen Nissenbaum.
\newblock Going against the (appropriate) flow: a contextual integrity approach
  to privacy policy analysis.
\newblock In {\em HCOMP}, pages 162--170, 2019.
\newblock \url{https://ojs.aaai.org/index.php/HCOMP/article/view/5266}.

\bibitem{srinath-etal-2021-privacy}
Mukund Srinath, Shomir Wilson, and C~Lee Giles.
\newblock Privacy at scale: Introducing the {P}riva{S}eer corpus of web privacy
  policies.
\newblock In {\em ACL-ICNLP}, pages 6829--6839, 2021.
\newblock \url{https://aclanthology.org/2021.acl-long.532}.

\bibitem{SC_aadhar}
{Supreme Court of India}.
\newblock Justice {K.S.}{P}uttaswamy(retd) vs {U}nion of {I}ndia.
\newblock \url{https://indiankanoon.org/doc/127517806/}, 2022.

\bibitem{tekieh-2015-healthdatamining}
Mohammad~Hossein Tekieh and Bijan Raahemi.
\newblock {Importance of Data Mining in Healthcare: A Survey}.
\newblock In {\em ASONAM}, pages 1057--1062, 2015.
\newblock \url{https://ieeexplore.ieee.org/document/7403678}.

\bibitem{USA-dir}
{The Centers for Medicare \& Medicaid Services}.
\newblock {Hospital General Information}.
\newblock \url{https://data.cms.gov/provider-data/dataset/xubh-q36u}, 2022.

\bibitem{tomuro_automatic_2016}
Noriko Tomuro, Steven Lytinen, and Kurt Hornsburg.
\newblock Automatic {Summarization} of {Privacy} {Policies} using {Ensemble}
  {Learning}.
\newblock In {\em CODASPY}, pages 133--135, 2016.
\newblock \url{https://dl.acm.org/doi/10.1145/2857705.2857741}.

\bibitem{UK-dir}
NHS UK.
\newblock {Authorities and Trusts}.
\newblock
  \url{https://www.nhs.uk/servicedirectories/pages/nhstrustlisting.aspx}, 2018.

\bibitem{Vollstedt2019}
Maike Vollstedt and Sebastian Rezat.
\newblock {\em An Introduction to Grounded Theory with a Special Focus on Axial
  Coding and the Coding Paradigm}, pages 81--100.
\newblock Springer International Publishing, 2019.
\newblock \url{https://doi.org/10.1007/978-3-030-15636-7_4}.

\bibitem{OPP_115}
Shomir Wilson, Florian Schaub, Aswarth~Abhilash Dara, Frederick Liu, Sushain
  Cherivirala, Pedro~Giovanni Leon, Mads~Schaarup Andersen, Sebastian Zimmeck,
  Kanthashree~Mysore Sathyendra, N~Cameron Russell, et~al.
\newblock The creation and analysis of a website privacy policy corpus.
\newblock In {\em ACL 2016}, pages 1330--1340, 2016.
\newblock \url{https://aclanthology.org/P16-1126/}.

\bibitem{xue-2018-health-data}
Baohong Xue and Ting Liu.
\newblock {Physical Health Data Mining of College Students Based on DRF
  Algorithm}.
\newblock {\em Wireless Personal Communications}, 102(4), 2018.
\newblock \url{https://link.springer.com/article/10.1007/s11277-018-5410-5}.

\bibitem{zaeem2018privacycheck}
Razieh~Nokhbeh Zaeem, Rachel~L German, and K~Suzanne Barber.
\newblock Privacycheck: Automatic summarization of privacy policies using data
  mining.
\newblock {\em ACM TOIT}, 18(4):1--18, 2018.

\bibitem{zhao2020summpip}
Jinming Zhao, Ming Liu, Longxiang Gao, Yuan Jin, Lan Du, He~Zhao, He~Zhang, and
  Gholamreza Haffari.
\newblock Summpip: Unsupervised multi-document summarization with sentence
  graph compression.
\newblock In {\em Proceedings of the 43rd international acm sigir conference on
  research and development in information retrieval}, pages 1949--1952, 2020.

\bibitem{zimmeck2019maps}
Sebastian Zimmeck, Peter Story, Daniel Smullen, Abhilasha Ravichander, Ziqi
  Wang, Joel Reidenberg, N~Cameron Russell, and Norman Sadeh.
\newblock {MAPS: Scaling privacy compliance analysis to a million apps}.
\newblock {\em PETS}, 2019(3):66--86, 2019.
\newblock \url{https://usableprivacy.org/static/files/popets-2019-maps.pdf}.

\bibitem{zimmeck2017automated}
Sebastian Zimmeck, Ziqi Wang, Lieyong Zou, Roger Iyengar, Bin Liu, Florian
  Schaub, Shomir Wilson, Norman Sadeh, Steven Bellovin, and Joel Reidenberg.
\newblock Automated analysis of privacy requirements for mobile apps.
\newblock In {\em 2016 AAAI Fall Symposium Series}, 2016.
\newblock
  \url{https://www.aaai.org/ocs/index.php/FSS/FSS16/paper/viewPaper/14113}.

\end{thebibliography}

\appendix
\section*{Appendix}
\section{Data Disclosure Practices}\label{append:ddp}

Two most frequent data practices observed for India, UK, and USA is shown in Table~\ref{tab:snippets_freq}.

\section{Community Metric Distribution}\label{sec:appendix_comm}

\begin{figure*}[h]
	\centering
	\begin{minipage}{0.5\columnwidth}
		\centering
		\includegraphics[width=0.8\textwidth]{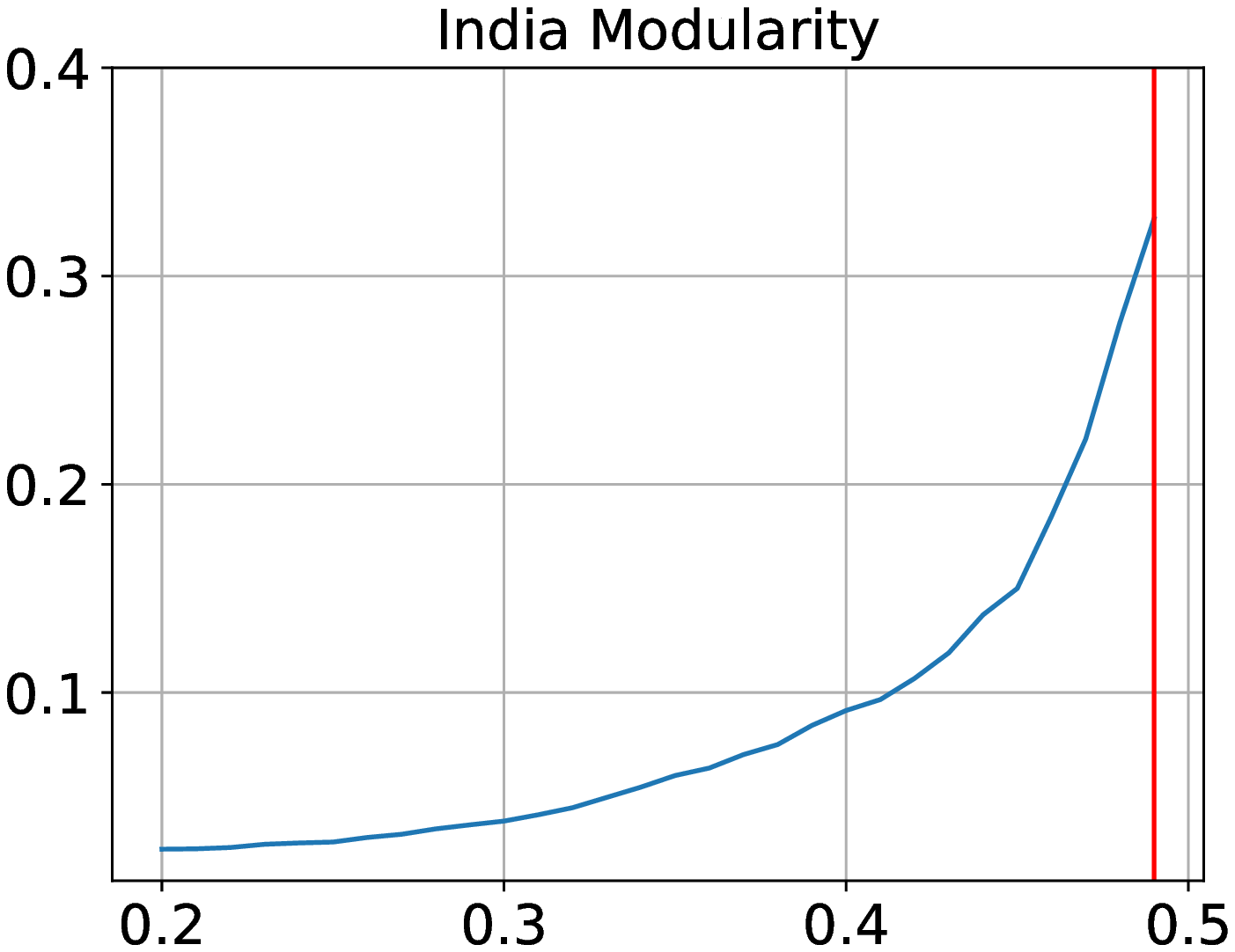}
		\label{label1}
	\end{minipage}%
	\begin{minipage}{0.5\columnwidth}
		\centering
		\includegraphics[width=0.8\textwidth]{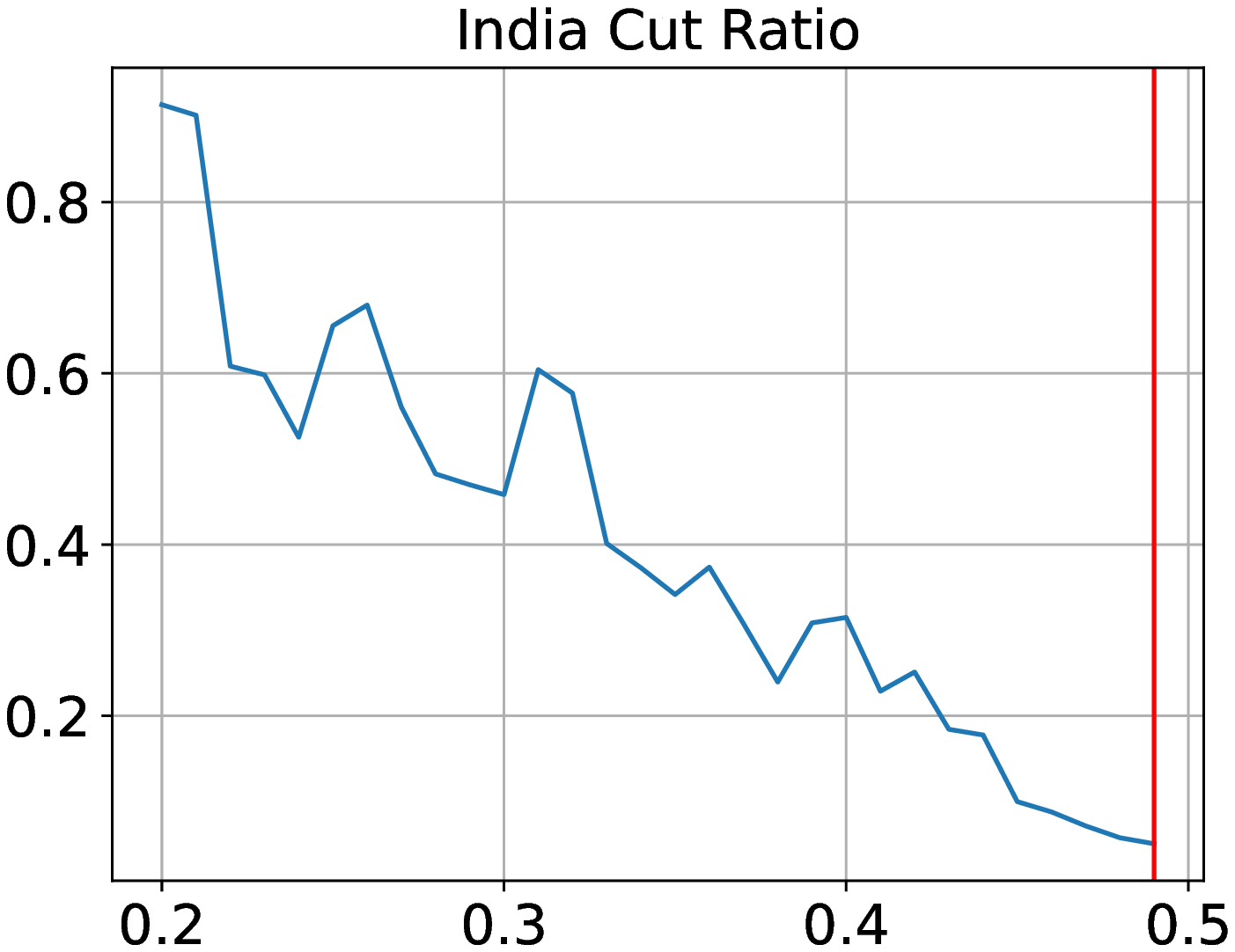}
		\label{label_1a}
	\end{minipage}
	\begin{minipage}{0.5\columnwidth}
		\centering
		\includegraphics[width=0.8\textwidth]{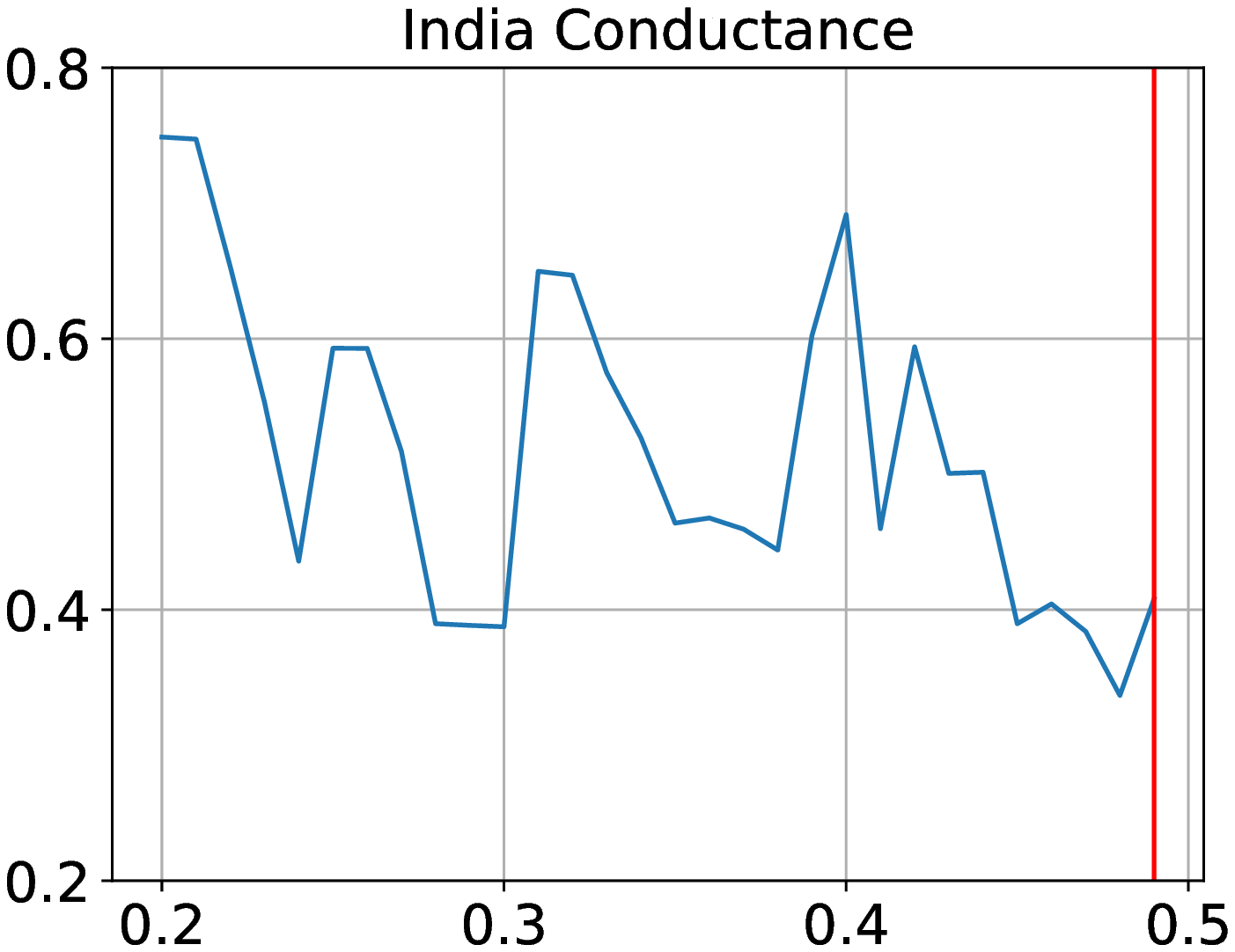}
		\label{label_1b}
	\end{minipage}
	\begin{minipage}{0.5\columnwidth}
		\centering
		\includegraphics[width=0.8\textwidth]{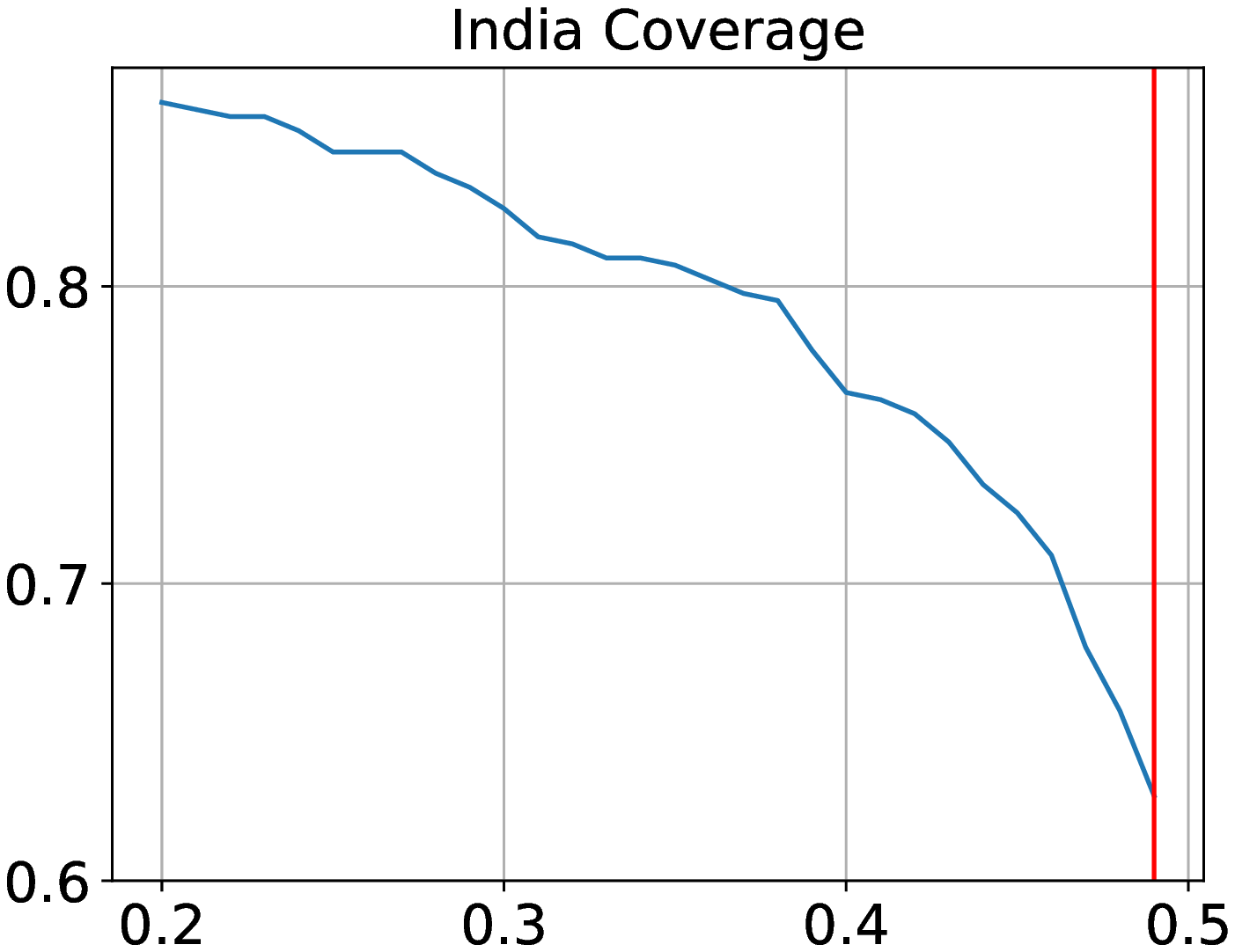}
		\label{label_1c}
	\end{minipage}	
	\caption{Community Metric Plots: Modularity, Cut Ratio, Conductance and Coverage for India}
	\label{fig:India_Comm_Thr}
\end{figure*}

\begin{figure*}[h]
	\centering
	\begin{minipage}{0.5\columnwidth}
		\centering
		\includegraphics[width=0.8\textwidth]{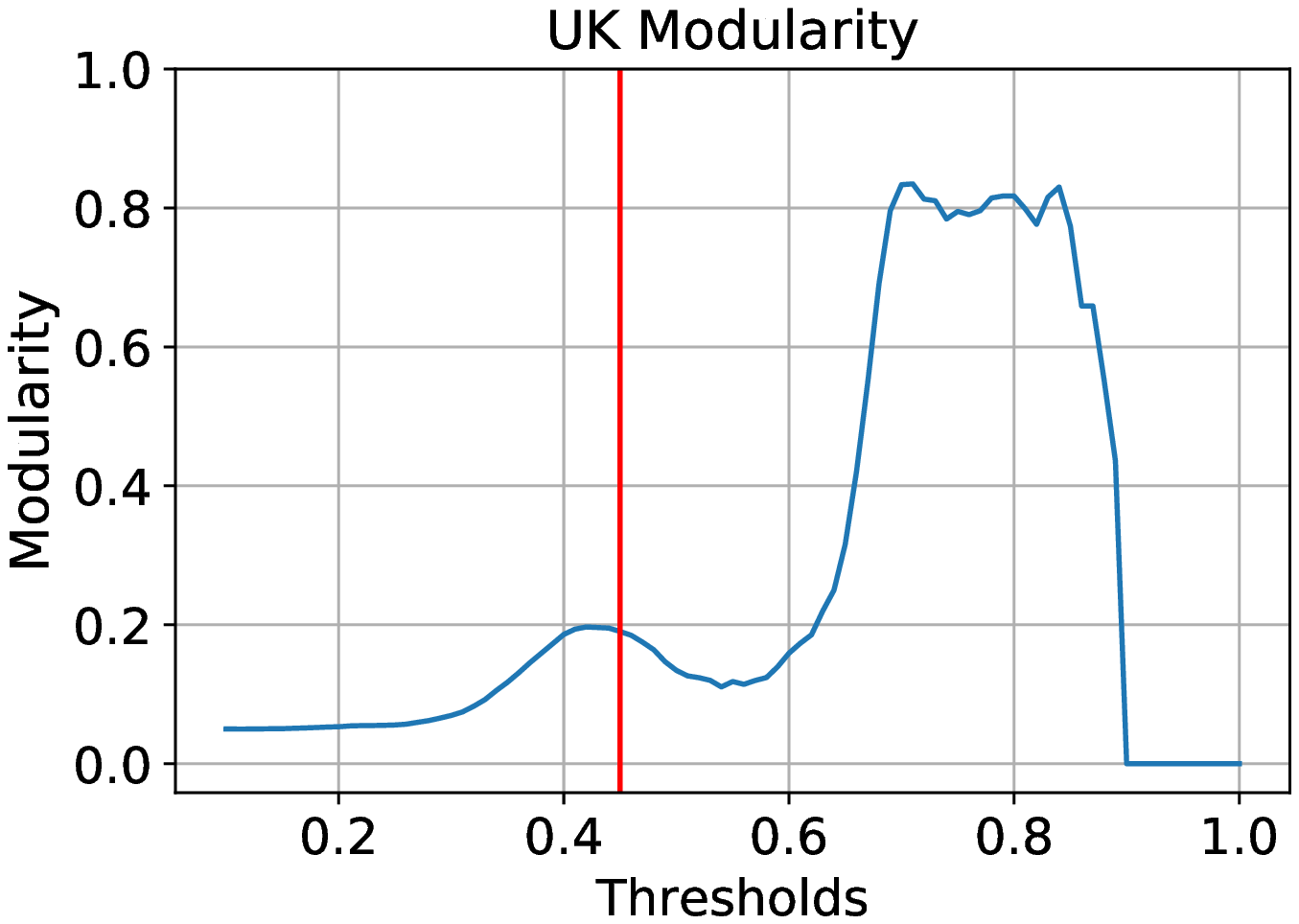}
		\label{label2}
	\end{minipage}%
	\begin{minipage}{0.5\columnwidth}
		\centering
		\includegraphics[width=0.8\textwidth]{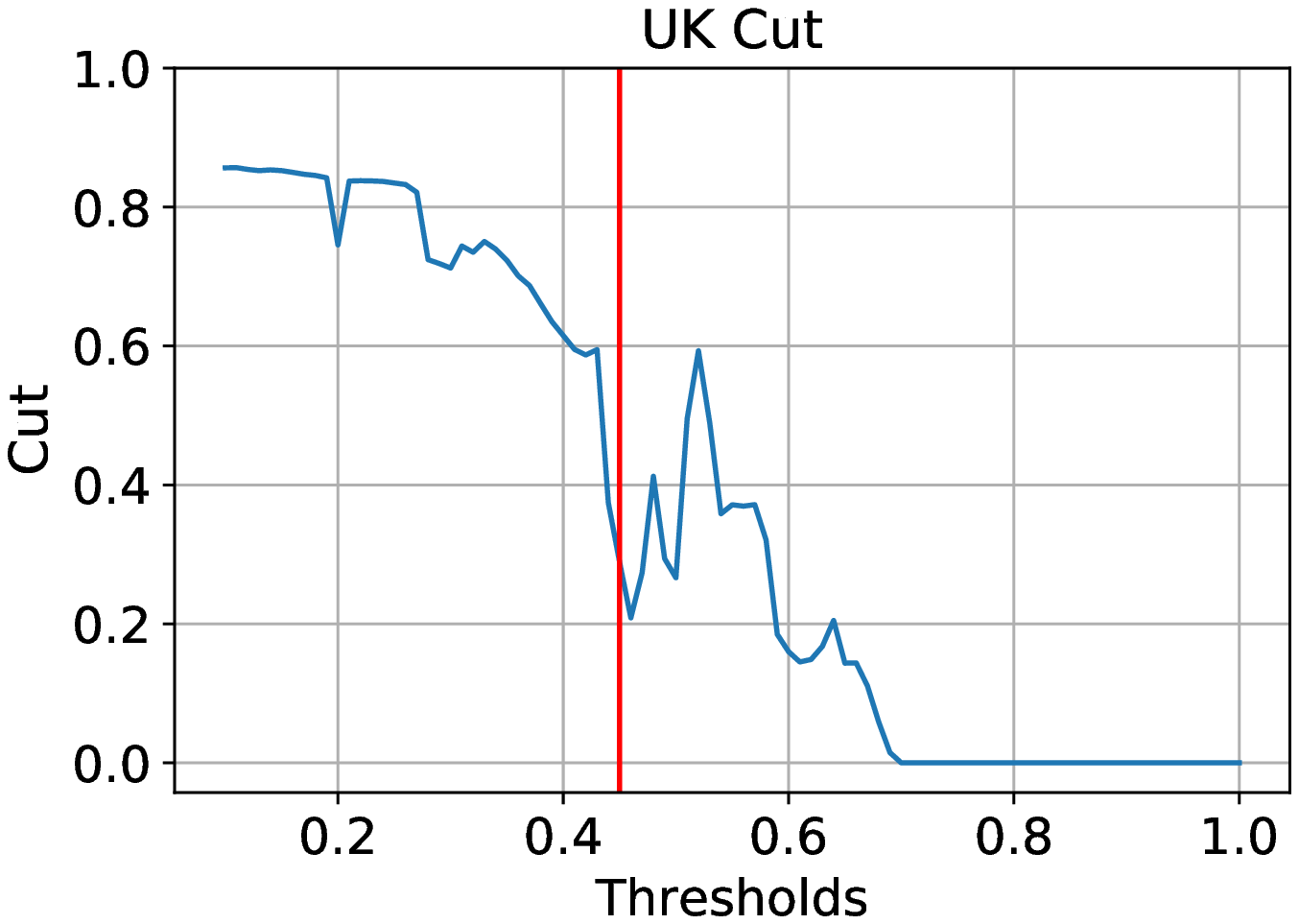}
		\label{label2_a}
	\end{minipage}
	\begin{minipage}{0.5\columnwidth}
		\centering
		\includegraphics[width=0.8\textwidth]{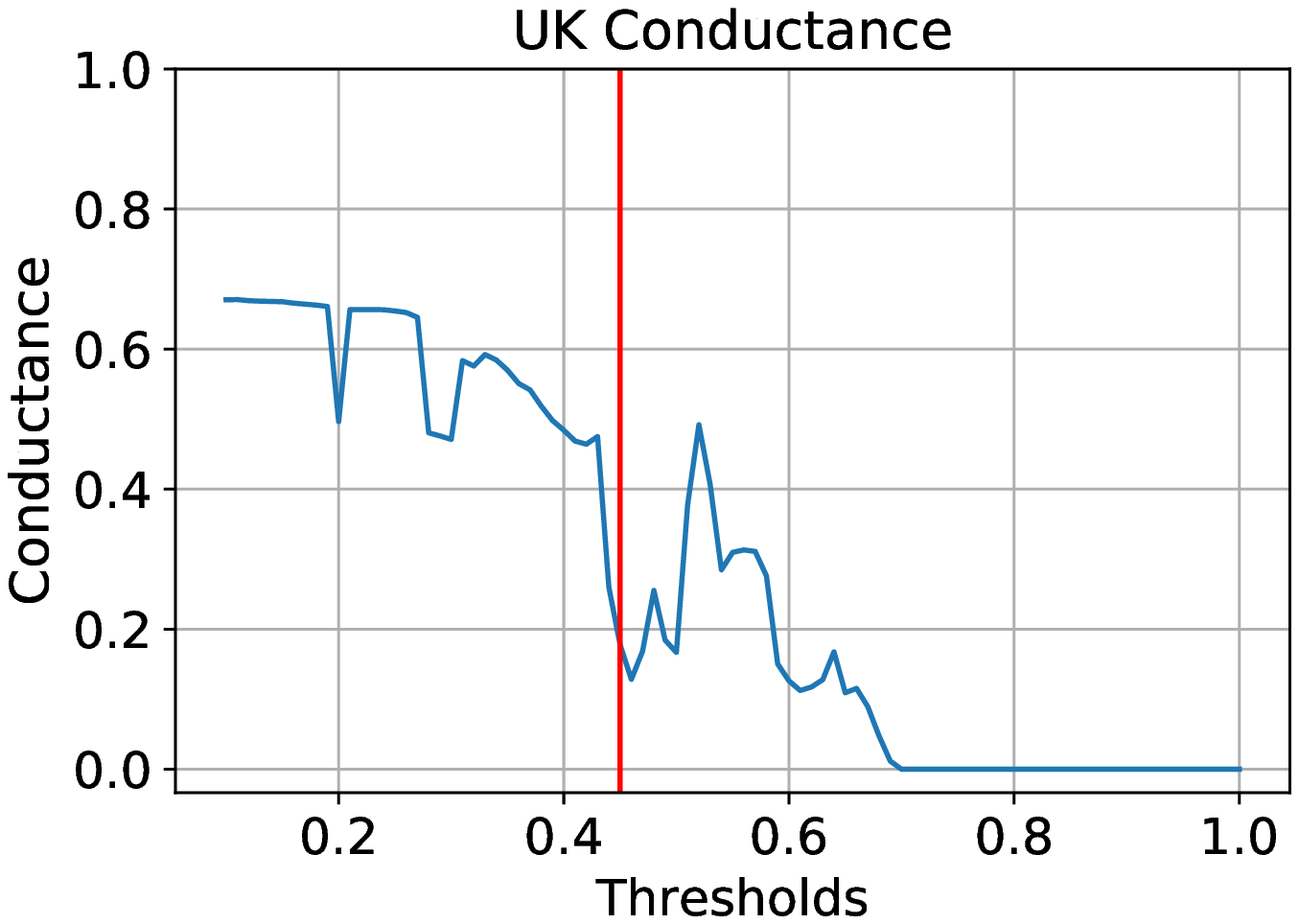}
		\label{label2_b}
	\end{minipage}
	\begin{minipage}{0.5\columnwidth}
		\centering
		\includegraphics[width=0.8\textwidth]{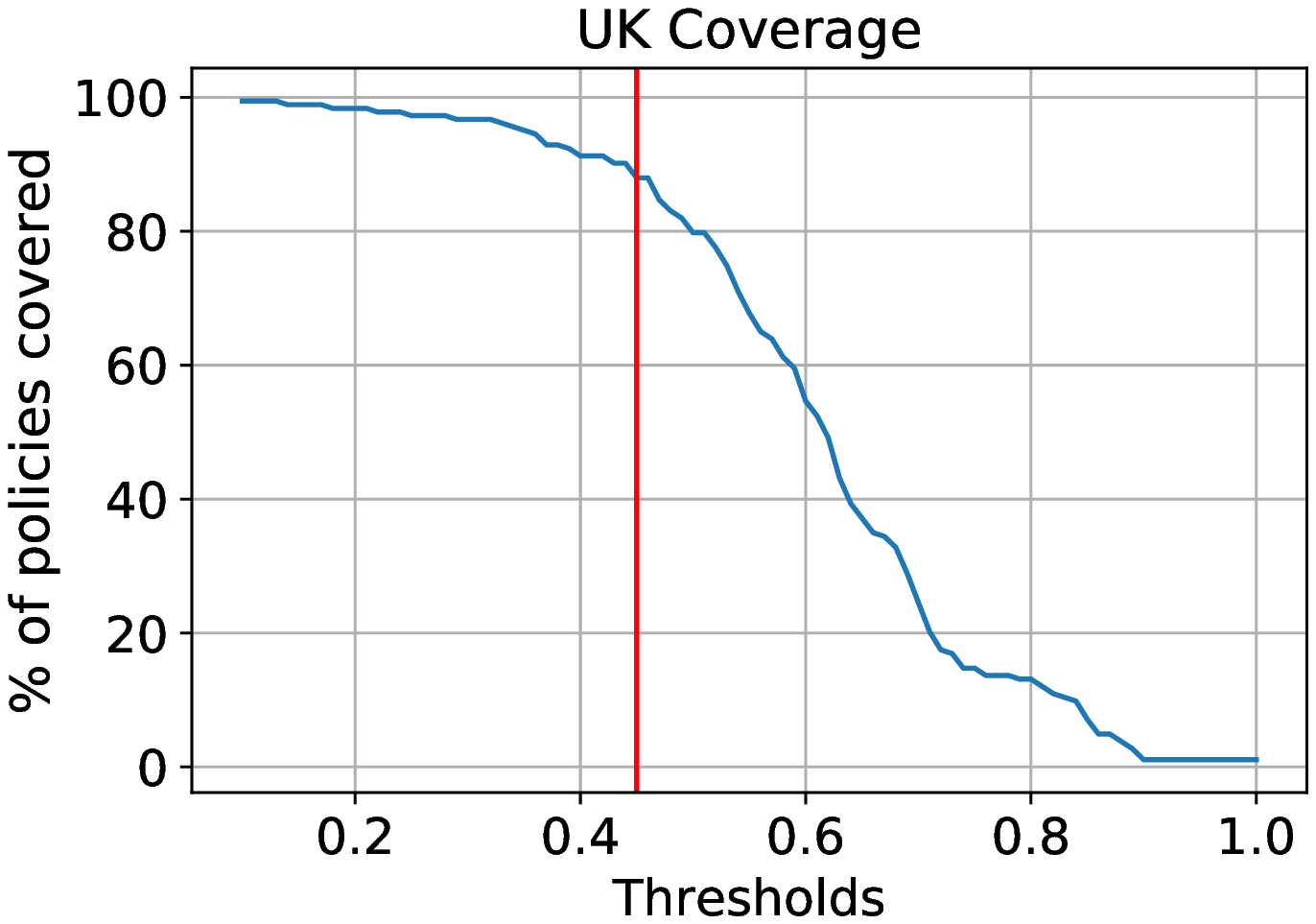}
		\label{label2_c}
	\end{minipage}	
	\caption{Community Metric Plots: Modularity, Cut Ratio, Conductance and Coverage for UK}
	\label{fig:UK_Comm_Thr}
\end{figure*}

\begin{figure*}[h]
	\centering
	\begin{minipage}{0.5\columnwidth}
		\centering
		\includegraphics[width=0.8\textwidth]{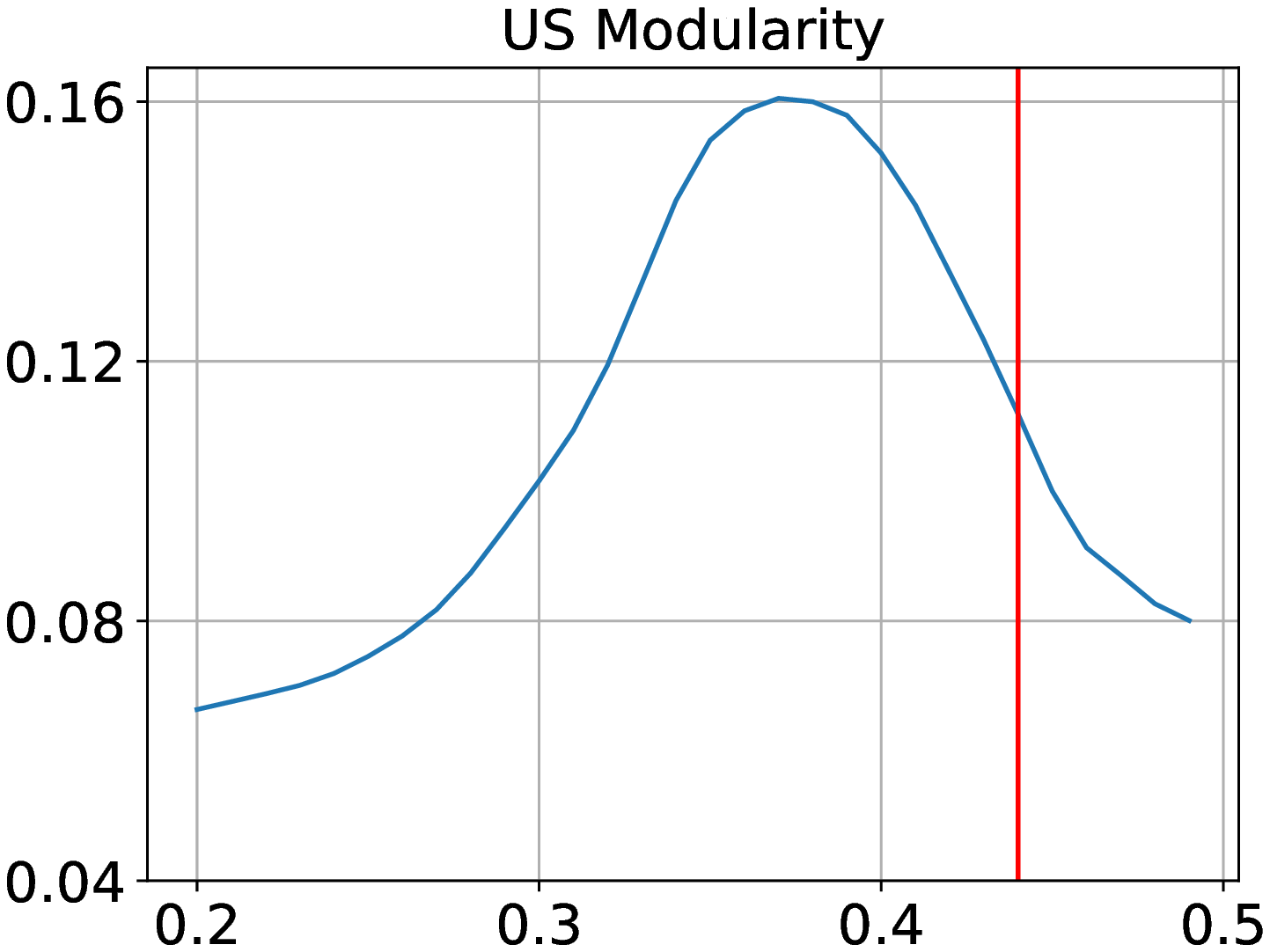}
		\label{label3}
	\end{minipage}%
	\begin{minipage}{0.5\columnwidth}
		\centering
		\includegraphics[width=0.8\textwidth]{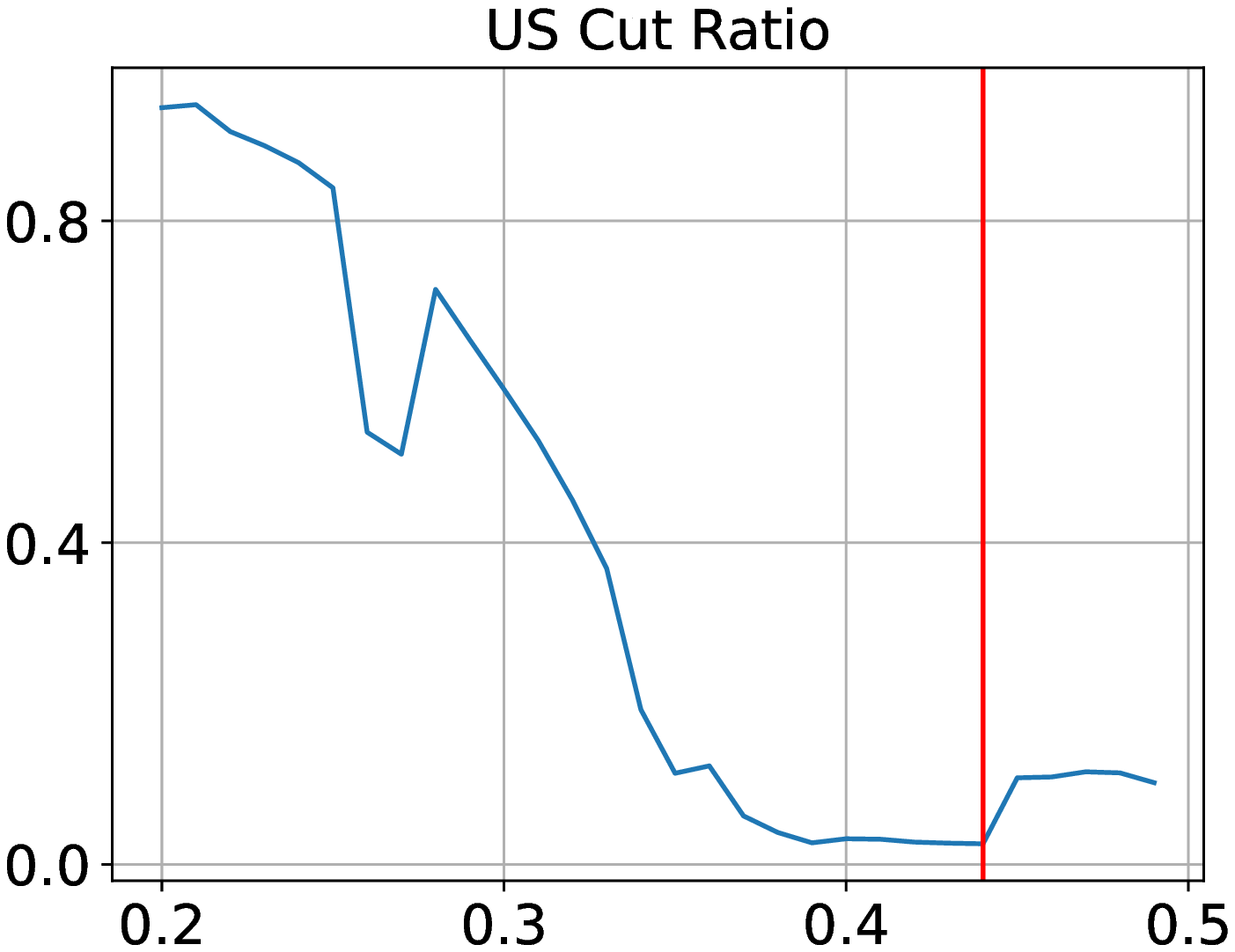}
		\label{label3_a}
	\end{minipage}
	\begin{minipage}{0.5\columnwidth}
		\centering
		\includegraphics[width=0.8\textwidth]{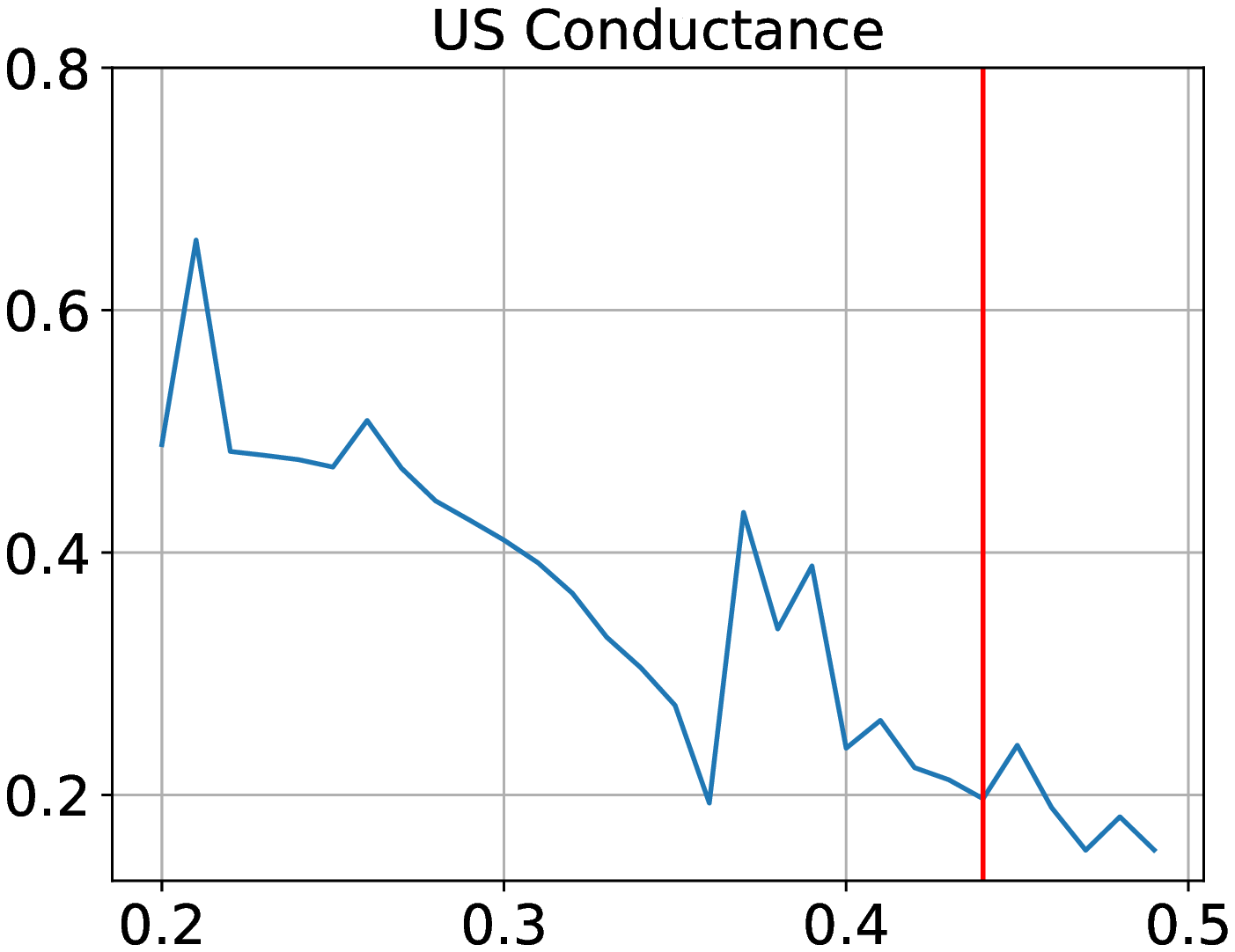}
		\label{label3_b}
	\end{minipage}
	\begin{minipage}{0.5\columnwidth}
		\centering
		\includegraphics[width=0.8\textwidth]{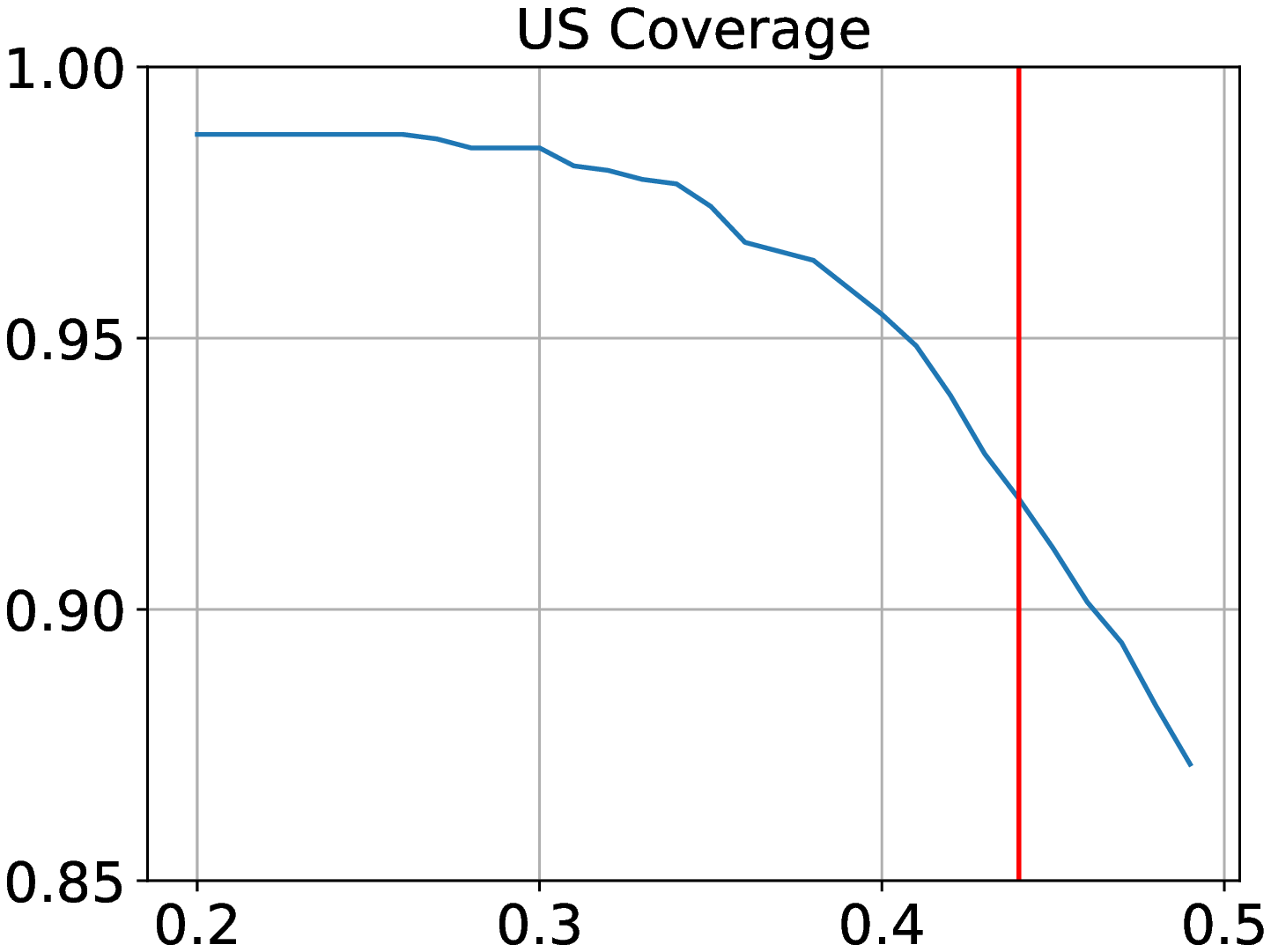}
		\label{label3_c}
	\end{minipage}	
	\caption{Community Metric Plots: Modularity, Cut Ratio, Conductance and Coverage for USA}
	\label{fig:USA_Comm_Thr}
\end{figure*}

Here we show the plots for each of the country reporting the metrics across thresholds. For India the plots are in Figure \ref{fig:India_Comm_Thr}. For the UK in Figure \ref{fig:UK_Comm_Thr}, for USA in Figure \ref{fig:USA_Comm_Thr}.

\noindent \textbf{India}: In Figure~\ref{fig:India_Comm_Thr}, we observe that at value 0.58 for India, we have 58\% coverage. After 0.58 the Cut Ratio, Conductance and Modularity just stabilizes. So we choose this point as our threshold for obtaining communities.

\noindent \textbf{UK}: In Figure~\ref{fig:UK_Comm_Thr}, we observe two peaks in modularity one at 0.45 and other at 0.70. A similar trend is observed for conductance and cut ratio, where local minimas occur at these two points. At 0.70, we see that coverage is at 20\%, which is very less as compared to coverage at 0.45 (>80\%). Hence we pick this as threshold, since we don't want to make conclusions based on a very small fraction of policies.

\noindent \textbf{USA}: In Figure~\ref{fig:USA_Comm_Thr}, just like the case with UK, we observe peaks in modularity one at 0.38. But for rest of the metrics we see peek at 0.44. Hence we pick it as the threshold as there is not a huge drop in modularity with very less conductance and cut ration.

\section{Analysis of Templates}\label{sec:appendix_template}
Here we show our results discussing best performing classification model OPP-115 dataset in Table~\ref{tab:models_results}. Other than the summarization approach discused in Sec.~\ref{sec:det_temp} [\textbf{PageRank+MMR}], we use 2 baseline summarization methods: two well-known strategies~\cite{louis2013automatically} namely {\bf Random-N, Avg Probability}. To evaluate the summaries, we used the  standard metrics employed for  unsupervised multi-document summarization framework~\cite{louis2013automatically}: JS-Divergence, Cosine Similarity, and Word Mover Distance. Additionally, we also used SUPERT~\cite{gao-etal-2020-supert} which aims to compute the semantic similarity between a summary and a pseudo-summary generated by ~\cite{gao-etal-2020-supert}. A higher SUPERT value means a good summary. The template evaluation scores are in Table~\ref{tab:scores_India}. A full discussion of themes along with sample statements is shown in Table~\ref{tab:themtic_analysis}.

\section{Incomplete Data Practices}\label{append:incomplete_prac}
We discuss the cases of incomplete complete data practices observed in USA, UK, and India along with the reasoning in Table~\ref{tab:Incomplete}.

\section{Examples of Data Practices} \label{appendix:samples_violation}
Here we show a data practice from each of the category of potential violation we discussed in Section \ref{sec:India_Compliance} in Table \ref{tab:Samples_Violation}.

\section{WMD analysis for checking alignment of data practices and law across countries}\label{app:wmdlaw}

\begin{table}[!t]
    \footnotesize
    \centering
    \begin{tabular}{c|c}
        \textbf{Country} &  \textbf{WMD} (Min/Median/Max) \\ \hline
        USA	    & 0.5185/0.8621/1.0274 \\
        UK	    & 0.6789/0.8523/0.9085 \\
        India   & 0.8986/1.0368/1.1856

    \end{tabular}
    \caption{WMD Analysis for countries: USA, UK and India. We computed the WMD on the data practices with the applicable provisions from the governing laws.}
    \label{tab:WMD_Scores}
\end{table}

We computed WMD between data practices from privacy policies and applicable legal regulations and report minimum/median/maximum WMD across data practices in a country. Our WMD analysis is shown in Table \ref{tab:WMD_Scores}. Note that for data practices in India, the WMD is high, signifying that the privacy policiy data practices might not be copied from the law.

\clearpage

\begin{table*}[h]
    \footnotesize
    \centering
    \begin{tabular}{p{5cm}|p{5cm}|p{5cm}}
    \textbf{India} & \textbf{UK} & \textbf{USA} \\
    \hline
     We will not disclose or sell any of your personal information, including your name, address, age, sex or medical history to any third party without your permission. 
     & Existing law which allows confidential patient information to be used and shared appropriately and lawfully in a public health emergency is being used during this outbreak 
     & Except as set forth above, you will be notified when PII may be shared with third parties, and will be able to prevent the sharing of this information.\\
     \hline
     We do not sell, trade, or rent Users personal identification information to others 
     & ... share personal/confidential patient information ... in disease surveillance for the purposes of protecting public health, ... and managing the outbreak 
     &  ... disclose personal information ... when required by law  ... Cooperate with the investigations of purported unlawful activities and conform ... - Protect and defend the rights ... - Identify persons ... misusing our Website or its related properties.\\
    \hline
    \end{tabular}
    \caption{Two most frequent data disclosure practices as observed for each country.}
    
    \label{tab:snippets_freq}
\end{table*}

\clearpage

\begin{table}[h]
\begin{tabular}{|l|l|l|l|c|}
\hline
\textbf{Model Name}                      & \multicolumn{3}{c|}{\textbf{Macro}} & \textbf{Accuracy}    \\ \hline
                                         & P          & R         & F1        &  \\ \hline
FastText                                 & .41        & .41       & .41       & .66                  \\ \hline
\textbf{TextCNN} & .72        & .67       & .68       & .75                  \\ \hline
RCNN                                     & .46        & .46       & .46       & .71                  \\ \hline
TextBiRNN                                & .58        & .55       & .56       & .68                  \\ \hline
RCNNVariant                              & .68        & .61       & .62       & .74                  \\ \hline
TextRNN                                  & .51        & .50       & .50       & .62                 \\ \hline
\end{tabular}
\caption{Macro and Accuracy Scores as obtained from sklearn's classification\_report tool}
\label{tab:models_results}
\end{table}

\begin{table}[h]
\footnotesize
\resizebox{\columnwidth}{!}{
\begin{tabular}{|p{3cm}|p{1.125cm}|p{1.25cm}|p{1.125cm}|p{1.125cm}|}
\hline
Methods  & \multicolumn{4}{c|}{Metrics Used}                  \\ \hline
                & JS   & Cos Sim & SUPERT & \multicolumn{1}{c|}{WMD} \\ \hline

\multicolumn{5}{|c|}{USA Third Party}                                 \\ \hline
Random-N        & 0.69 & \textbf{0.61}    & 0.47   & 2.25                     \\ \hline
Avg Probability & 0.67 & 0.56    & 0.49   & 2.19                     \\ \hline
PageRank+MMR        & \textbf{0.67} & 0.56    & \textbf{0.54}   & \textbf{2.18}                     \\ \hline
\multicolumn{5}{|c|}{USA First Party}                                 \\ \hline
Random-N        & 0.68 & 0.58    & 0.49   & 2.21                     \\ \hline
Avg Probability & 0.66 & \textbf{0.61}    & 0.50   & \textbf{2.13}                    \\ \hline
PageRank+MMR        & \textbf{0.66} & 0.56    & \textbf{0.52}   & 2.18                     \\ \hline

\multicolumn{5}{|c|}{UK Third Party}                                 \\ \hline
Random-N        & 0.76 & \textbf{0.66}    & 0.47  & \textbf{2.39}                     \\ \hline
Avg Probability & 0.69 & 0.57    & 0.46   & 2.45                     \\ \hline
PageRank+MMR        & \textbf{0.68} & 0.63    & \textbf{0.50}   & 2.43                     \\ \hline
\multicolumn{5}{|c|}{UK First Party}                                 \\ \hline
Random-N        & 0.76 & \textbf{0.62}    & 0.36   & 2.39                     \\ \hline
Avg Probability & 0.77 & 0.56    & 0.37   & 2.45                     \\ \hline
PageRank+MMR        & \textbf{0.73} & 0.59    & \textbf{0.44}   & \textbf{2.38}                     \\ \hline

\multicolumn{5}{|c|}{India Third Party}                              \\ \hline
Random-N        & 0.67 & 0.70    & 0.45   & 2.58                     \\ \hline
Avg Probability & 0.79 & 0.57    & \textbf{0.51}   & 2.56                     \\ \hline
PageRank+MMR        & \textbf{0.67} & \textbf{0.74}    & 0.45   & \textbf{2.48}                     \\ \hline
\multicolumn{5}{|c|}{India First Party}                     \\ \hline
Random-N        & 0.59 & 0.72    & 0.52   & 2.45                     \\ \hline
Avg Probability & 0.79 & 0.55    & 0.47   & 2.45                    \\ \hline
PageRank+MMR        & \textbf{0.57} & \textbf{0.76}    & \textbf{0.54}   & \textbf{2.41}                     \\ \hline
\end{tabular}}
\caption{Metrics reported across each category for summarization methods}
\label{tab:scores_India}
\end{table}

\clearpage

\begin{table*}
    \centering
    \tiny
    
    \begin{tabular}{p{3cm}|p{4cm}|p{7cm}}
    
         Category & Themes & Sample Sentences \\
         \hline
         \multicolumn{3}{c}{India} \\
         \hline
         First Party Collection/Use & Explicit attributes collected & information collected ... a. Contact data ... your mobile number, email address and phone number b. Demographic data ... your gender, nationality, date of birth, postal address and pin code \\
                                    & Explains what is SPI  & Personal information ... mean any information that relates to a natural person, which, either directly or indirectly, in combination with other information available or likely to be available with a body corporate, is capable of identifying such person \\
                                    & Source of information collection
 & We collect personally identifiable information (Email Id., Name, Contact number, etc.) from you when you make payment
... to complete optional online surveys. ... ask you for contact information and demographic information (like zip code, age, or income level)
\\
                                    & Use of data collected & ... use your Personal Information to contact you and deliver information to you ... targeted to your interests, such as targeted banner advertisements
 \\ \hline
         Third Party Disclosure & Explicit use of information collected & To send or facilitate communications to you, send you communications we think will be of interest to you, including information about products\\
                                    & Legal Disclosure  & disclose any information as is necessary to satisfy or comply with any applicable law, regulation, legal process or governmental request\\
                                    & Specific law regulations & Regulation 4 of the Information Technology (Reasonable Security Practices and Procedures and Sensitive Personal Information) Rules, 2011 \\
                                    & Purpose of Disclosure  & ... for reasons such as website hosting, data storage, software services, email services, marketing, fulfilling customer orders, providing payment services, data analytics, providing customer services, and conducting surveys  \\
                        & Information Disclosed To & ... may also disclose or transfer End-User's personal and other information a User provides, to a third party as part of reorganization or a sale of the assets \\
                        
                        & Responsibilities of Third Parties & will place contractual obligations on the transferee which will oblige the transferee to adhere to the provisions of this Privacy Policy \\

        \hline
         \multicolumn{3}{c}{UK} \\
         \hline
         First Party Collection/Use & Sources of Information collection & ... from you and this is likely to be done either face to face, during a telephone call, or via email.
Your information ... a referral from your GP or other healthcare organisation, or ... via a form we have asked you to complete. \\
                                    & Purpose of information collection  & ... for your direct care and treatment this includes to ensure safe and high-quality care for all our patients. ... for other purposes such as research.
 \\
                                    & Methods of information storage & The information may be written down on paper, held on a computer, or a mixture of both
 \\
                                    & Transmission of Information & ... may be transferred from this Trust to other NHS Trusts to support the safe, efficient and effective transfer of staff information
 \\ \hline        
         Third Party Disclosure & Legal
         Disclosure & ... we are under a duty to share your information, due to a legal requirement
 disclosure under a court order, ... to the Health and Safety Executive ... to the police ... to debt collecting agencies  
 \\
                                & Purpose of disclosure  & ... to pay a supplier, our legal basis is that its use is necessary for the purposes of our legitimate interests as a buyer of goods and services
\\
                                & Information Disclosed To  & with health and care organisations and other bodies engaged in disease surveillance, ... to the Department of Health, with authorised non-NHS authorities and organisations \\
        \hline
         \multicolumn{3}{c}{USA} \\
         \hline
         First Party Collection/Use &  Explicit use of information & to conduct our standard internal operations, including proper administration of records, evaluation of the quality of Treatment,payment purposes of those health professionals and facilities\\
                                    & Defines PHI  & PHI) is information that you provide us or that we create or receive about your health care. PHI includes a patient’s name, age, race, sex, ... patient’s physical or mental health in the past, present, or future, and to the care, treatment, services and payment for care needed by a patient because of his or her health. \\
                                    & Regarding Authorization for usage
 & Written authorization is required prior to using or disclosing your PHI for marketing activities that are supported by payments from third parties\\ \hline
         Third Party Disclosure & Purpose for disclosure
  & We may use and disclose your health information to obtain payment for services that we provide to you
... may use and disclose medical information about you for research purposes
We may disclose medical information about you to another health care provider or covered entity ... under certain circumstances
\\
                                & Information Disclosed To
 &  ... to doctors, nurses, technicians, students preparing for health care related careers, or other personnel who are involved in your care or treatment, including your primary care physician. ... for an insurance company to pay for your treatment, we must submit a bill that identifies you, your diagnosis, and the treatment provided to you
\\
                                & Legal Disclosure
     & ... used or disclosed to comply with laws and regulations, accreditation purposes, patients and residents claims, grievances or lawsuits, health care contracting relating to our operations, legal services, ... business management and administration, underwriting and other insurance activities.
\\
                                & Restrict Disclosure  &  You may ask for restrictions on how your health information is used or to whom your health information is disclosed: ...  While we will consider all requests for restrictions, we are not required to agree to your request
\\                                
         & Limited Disclosure & If you are unavailable, incapacitated, or . . . we determine that a limited disclosure may be in your best interest, we may share limited personal health information with such individuals without your approval \\\hline
        
    \end{tabular}
    \caption{Thematic Analysis of Templates}
    \label{tab:themtic_analysis}
\end{table*}

\clearpage
\begin{table*}[]
    \footnotesize
    \centering
    
    \begin{tabular}{p{1.5cm}|p{9cm}|p{3.5cm}}
    
    Case                            & Example   & Explanation \\ \hline
     \multicolumn{3}{c}{USA}  \\
    \hline
    Missing Transmission Principle & We and our Affiliates and Service Providers may collect personal information about you on the Website and from other sources, including commercially available sources. & The purpose for which such information is used is not discussed\\ \hline
    Missing Recipient & If you are present and able to agree or object then we may only disclose your PHI if you don't object after you have been informed of your opportunity to do so (although such agreement may be reasonably inferred from the circumstances).  & To whom the information will be disclosed is not mentioned \\ \hline
    \multicolumn{3}{c}{UK}  \\
    \hline
    Missing Recipient & Any information used or shared during the Covid-19 outbreak will be limited to the period of the outbreak unless there is another legal basis to use the data. & Clear description of the party recieving the information is absent\\ \hline
    
    \multicolumn{3}{c}{India}  \\
    \hline
    Missing Transmission Principle  &    We may collect any and all personal information you provide to us, like your name, mobile phone number, email address . . . feedback, and any other information you provide us. & The purpose for which information is collected is not mentioned in the text \\ \hline
    Missing Recipients & You acknowledge that some countries where we may transfer Your Personal Information may not have data protection laws .... SIMS will place contractual obligations on the transferee which will oblige the transferee to adhere to the provisions of this Privacy Policy. & A clear description of entities to whom the information is disclosed is absent in the statement.\\ \hline
    
    \end{tabular}
    \caption{Incomplete Information Flows examples}
    \label{tab:Incomplete}
\end{table*}

\begin{table*}
    \footnotesize
    \centering
    \begin{tabular}{p{3cm}|p{7cm}|p{4cm}}
         Category of \textit{Potential Violation} & Sample Data Practice & Explanation  \\ \hline

         \textit{Not mentioning what explicit SPI is being used or disclosed} & When you use our Website, we collect and store your Personal Information which is provided by you from time to time. Our primary goal in doing so is to provide you a safe, efficient, smooth and customized experience. This allows us to provide Services and Features that most likely meet your needs, and to customize our Website to make your experience safer and easier. & It violates Rule 4(1)(ii) because of not clearly mentioning the type of personal data that is been collected. \\ \hline

         \textit{Not clearly disclosing the purpose for which information was collected (or disclosed)} & Institution will be free to use, collect and disclose information that is freely available in the public domain without your consent. & There is a need to comply with Rule 5(3)(a) and (b) i.e. tell that information will be collected/used and further the purpose of use. \\ \hline

         \textit{Upon updating policy, consent not taken again} & By choosing the Opt-In option on the Website and after that, by providing us your personal information or availing services of hospital or by making use of the facilities provided by the Website, it is agreed by you that you have freely consented to the collection, storage, processing, disclosure and transfer of Your personal information following the provisions of this Privacy Policy and any amendments thereof. & For each and every amendment in policy, there is a fresh requirement to comply with IT Rules 2011,  and importantly Rule 5 as a whole.\\ \hline

         \textit{Absence of explicit consent and/or lawful purpose for which information was collected} & We may collect any and all personal information you provide to us, like your name, mobile phone number, email address, delivery address, order, invoicing details, materials or data you may sync through other applications (e.g. photos, contact lists), information in relation to creating an account, phone numbers you insert into your contacts or to send a message, feedback, and any other information you provide us. &  It does not complies with Rule 5 generally. Specifically with Rule 5(1) as no explicit consent is taken, and Rule 5(2) as it does not specify any lawful purpose.\\ \hline

    \end{tabular}
    \caption{Sample Statements from each category of potential violations as discussed in Section \ref{sec:India_Compliance}.}
    \label{tab:Samples_Violation}
\end{table*}

\if{0}
\section{Incomplete Data Practices}\label{append:incomplete_prac}
We discuss the cases of incomplete complete data practices observed in USA, UK, and India along with the reasoning in Table~\ref{tab:Incomplete}
\begin{table*}[]
    \centering
    \tiny
    \begin{tabular}{p{3cm}|p{4cm}|p{7cm}}
         Category & Themes & Sample Sentences \\
         \hline
         \multicolumn{3}{c}{India} \\
         \hline
         First Party Collection/Use & Explicit attributes collected & information collected ... a. Contact data ... your mobile number, email address and phone number b. Demographic data ... your gender, nationality, date of birth, postal address and pin code \\
                                    & Explains what is SPI  & Personal information ... mean any information that relates to a natural person, which, either directly or indirectly, in combination with other information available or likely to be available with a body corporate, is capable of identifying such person \\
                                    & Source of information collection
 & We collect personally identifiable information (Email Id., Name, Contact number, etc.) from you when you make payment
... to complete optional online surveys. ... ask you for contact information and demographic information (like zip code, age, or income level)
\\
                                    & Use of data collected & ... use your Personal Information to contact you and deliver information to you ... targeted to your interests, such as targeted banner advertisements
 \\ \hline
         Third Party Disclosure & Explicit use of information collected & To send or facilitate communications to you, send you communications we think will be of interest to you, including information about products\\
                                    & Legal Disclosure  & disclose any information as is necessary to satisfy or comply with any applicable law, regulation, legal process or governmental request\\
                                    & Specific law regulations & Regulation 4 of the Information Technology (Reasonable Security Practices and Procedures and Sensitive Personal Information) Rules, 2011 \\
                                    & Purpose of Disclosure  & ... for reasons such as website hosting, data storage, software services, email services, marketing, fulfilling customer orders, providing payment services, data analytics, providing customer services, and conducting surveys  \\
                        & Information Disclosed To & ... may also disclose or transfer End-User's personal and other information a User provides, to a third party as part of reorganization or a sale of the assets \\
                        
                        & Responsibilities of Third Parties & will place contractual obligations on the transferee which will oblige the transferee to adhere to the provisions of this Privacy Policy \\

        \hline
         \multicolumn{3}{c}{UK} \\
         \hline
         First Party Collection/Use & Sources of Information collection & ... from you and this is likely to be done either face to face, during a telephone call, or via email.
Your information ... a referral from your GP or other healthcare organisation, or ... via a form we have asked you to complete. \\
                                    & Purpose of information collection  & ... for your direct care and treatment this includes to ensure safe and high-quality care for all our patients. ... for other purposes such as research.
 \\
                                    & Methods of information storage & The information may be written down on paper, held on a computer, or a mixture of both
 \\
                                    & Transmission of Information & ... may be transferred from this Trust to other NHS Trusts to support the safe, efficient and effective transfer of staff information
 \\ \hline        
         Third Party Disclosure & Legal
         Disclosure & ... we are under a duty to share your information, due to a legal requirement
 disclosure under a court order, ... to the Health and Safety Executive ... to the police ... to debt collecting agencies  
 \\
                                & Purpose of disclosure  & ... to pay a supplier, our legal basis is that its use is necessary for the purposes of our legitimate interests as a buyer of goods and services
\\
                                & Information Disclosed To  & with health and care organisations and other bodies engaged in disease surveillance, ... to the Department of Health, with authorised non-NHS authorities and organisations \\
        \hline
         \multicolumn{3}{c}{USA} \\
         \hline
         First Party Collection/Use &  Explicit use of information & to conduct our standard internal operations, including proper administration of records, evaluation of the quality of Treatment,payment purposes of those health professionals and facilities\\
                                    & Defines PHI  & PHI) is information that you provide us or that we create or receive about your health care. PHI includes a patient’s name, age, race, sex, ... patient’s physical or mental health in the past, present, or future, and to the care, treatment, services and payment for care needed by a patient because of his or her health. \\
                                    & Regarding Authorization for usage
 & Written authorization is required prior to using or disclosing your PHI for marketing activities that are supported by payments from third parties\\ \hline
         Third Party Disclosure & Purpose for disclosure
  & We may use and disclose your health information to obtain payment for services that we provide to you
... may use and disclose medical information about you for research purposes
We may disclose medical information about you to another health care provider or covered entity ... under certain circumstances
\\
                                & Information Disclosed To
 &  ... to doctors, nurses, technicians, students preparing for health care related careers, or other personnel who are involved in your care or treatment, including your primary care physician. ... for an insurance company to pay for your treatment, we must submit a bill that identifies you, your diagnosis, and the treatment provided to you
\\
                                & Legal Disclosure
     & ... used or disclosed to comply with laws and regulations, accreditation purposes, patients and residents claims, grievances or lawsuits, health care contracting relating to our operations, legal services, ... business management and administration, underwriting and other insurance activities.
\\
                                & Restrict Disclosure  &  You may ask for restrictions on how your health information is used or to whom your health information is disclosed: ...  While we will consider all requests for restrictions, we are not required to agree to your request
\\                                
         & Limited Disclosure & If you are unavailable, incapacitated, or . . . we determine that a limited disclosure may be in your best interest, we may share limited personal health information with such individuals without your approval \\\hline
         
    \end{tabular}
    \caption{Thematic Analysis of Templates}
    \label{tab:themtic_analysis}
\end{table*}

\if{0}
\begin{table*}[t]
    \centering
    \resizebox{\textwidth}{!}{
     \begin{tabular}{c|l}
         Category & Definition\\ \hline
         Conditionality(C) & the action to be performed is dependent upon a variable or unclear trigger \\
         Generalization(G) & the action or information types are vaguely abstracted with unclear conditions \\
         Modality(M)       & the likelihood or possibility of the action is vague or ambiguous\\
         Numerical(N)      & the action or information type has a vague quantifier \\
    \end{tabular}}
    \caption{Explaining category of Vagueness~\cite{bhatia2016theory}}
    \label{tab:append_vague}
\end{table*}
\fi

\begin{table*}[]
    \centering
    \resizebox{\textwidth}{!}{
    \begin{tabular}{p{1.5cm}|p{9.5cm}|p{4cm}}
    
    Case                            & Example   & Explanation \\ \hline
     \multicolumn{3}{c}{USA}  \\
    \hline
    Missing Transmission Principle & We and our Affiliates and Service Providers may collect personal information about you on the Website and from other sources, including commercially available sources. & The purpose for which such information is used is not discussed\\ \hline
    Missing Recipient & If you are present and able to agree or object then we may only disclose your PHI if you don't object after you have been informed of your opportunity to do so (although such agreement may be reasonably inferred from the circumstances).  & To whom the information will be disclosed is not mentioned \\ \hline
    \multicolumn{3}{c}{UK}  \\
    \hline
    Missing Recipient & Any information used or shared during the Covid-19 outbreak will be limited to the period of the outbreak unless there is another legal basis to use the data. & Clear description of the party recieving the information is absent\\ \hline
    
    \multicolumn{3}{c}{India}  \\
    \hline
    Missing Transmission Principle  &    We may collect any and all personal information you provide to us, like your name, mobile phone number, email address . . . feedback, and any other information you provide us. & The purpose for which information is collected is not mentioned in the text \\ \hline
    Missing Recipients & You acknowledge that some countries where we may transfer Your Personal Information may not have data protection laws .... SIMS will place contractual obligations on the transferee which will oblige the transferee to adhere to the provisions of this Privacy Policy. & A clear description of entities to whom the information is disclosed is absent in the statement.\\ \hline
    
    \end{tabular}}
    \caption{Incomplete Information Flows examples}
    \label{tab:Incomplete}
\end{table*}

\clearpage
\fi

\if{0}
\begin{table*}[!t]
    \centering
    \begin{tabular}{p{5cm}|p{5cm}|p{5cm}}
    \textbf{India} & \textbf{UK} & \textbf{USA} \\
    \hline
     We will not disclose or sell any of your personal information, including your name, address, age, sex or medical history to any third party without your permission. 
     & Existing law which allows confidential patient information to be used and shared appropriately and lawfully in a public health emergency is being used during this outbreak 
     & Except as set forth above, you will be notified when PII may be shared with third parties, and will be able to prevent the sharing of this information.\\
     \hline
     We do not sell, trade, or rent Users personal identification information to others 
     & ... share personal/confidential patient information ... in disease surveillance for the purposes of protecting public health, ... and managing the outbreak 
     &  ... disclose personal information ... when required by law  ... Cooperate with the investigations of purported unlawful activities and conform ... - Protect and defend the rights ... - Identify persons ... misusing our Website or its related properties.\\
    \hline
    \end{tabular}
    \caption{Two most frequent data disclosure practices as observed for each country.}
    
    \label{tab:snippets_freq}
\end{table*}

\section{Data Disclosure Practices}\label{append:ddp}

Two most frequent data practices observed for India, UK, and USA is shown in Table~\ref{tab:snippets_freq}.

\section{Explanation of Categories}\label{sec:appendix_summ}
The categories used in the Characterization study (See Section~\ref{sec:charec_comm} Table~\ref{tab:annot_scores}) is described in Table~\ref{tab:cat_desc}.

\begin{table*}[]
    \resizebox{\textwidth}{!}{
    \begin{tabular}{|p{1cm}|p {14cm}|}
        
         \hline
         Label & Explanation  \\ \hline
        \textbf{OD} & Online Data Collection (Cookies, Server logs, IP Address, Non-PII etc.)\\ \hline
        \textbf{1C} & First Party collection(Mentions attributes collected) \\ \hline
        \textbf{1U} & First party Usage(Explicitly states how information will be used) \\ \hline
        \textbf{3D} & Third party disclosure(Explicitly states with whom data is shared) \\ \hline
        \textbf{DR} & Data Retention(For how long data is retained) \\ \hline
        \textbf{UR} & User's Rights of Accessing, Deleting or Correcting data \\ \hline
        \textbf{UC} & User's Choice(Right to opt out of data collection, right to restrict information sharing) \\ \hline
        \textbf{LD} & Legal Disclosure(Mentions disclosures to government/legal bodies will be made) \\ \hline
        \textbf{GO} & Grievance Officer(Mention details of Grievance officer) \\ \hline
        \textbf{DS} & Data Security(Mentions appropriate steps taken to ensure security of data) \\ \hline
        \textbf{SS} & Storage Standard (Mentions usage of reasonable security measures for safegaurd of data in accordance with IT Rules 2011 \cite{IT_ACT_2011})\\ \hline
        
        \textbf{DPO} & Data Protection Officer(Contact details of DPO should be mentioned) \\ \hline
        \textbf{PC} & Policy Change, Should have clause regarding changes to the policy and notification of it thereof \\ \hline
    \end{tabular}}
    \caption{Description of Categories used in the Annotation Study}
    \label{tab:cat_desc}
\end{table*}
\fi

\if{0}
\section{Vagueness Categories}\label{sec:vagueness}

The description of Vagueness categories as described in ~\cite{bhatia2016theory} is available in Table~\ref{tab:append_vague} 
\fi

\end{document}